\documentclass[12pt]{article}
\pdfoutput=1
\usepackage{times}
\usepackage{natbib}
 \bibpunct{(}{)}{;}{a}{,}{,}
\usepackage{graphicx}
\usepackage{amsmath,amssymb,amsthm}
\usepackage{bm}
\usepackage{rotating}
\usepackage[lined,ruled]{algorithm2e}
\usepackage{multirow}
\usepackage{rotating}
\usepackage{multicol}
\usepackage{titlesec}
\usepackage{latexsym}
\usepackage[pdftex,colorlinks=true,linkcolor=blue,citecolor=blue,urlcolor=blue,bookmarks=false,pdfpagemode=None]{hyperref}
\usepackage{url}
\usepackage{verbatim}
\usepackage{fancyhdr}
\usepackage{setspace}
\usepackage{paralist}
\usepackage{boxedminipage}
\usepackage{lineno}
\usepackage[top=1in, bottom=1in, left=1in, right=1in]{geometry}
\usepackage{lscape}
\usepackage{dcolumn}

\usepackage[usenames,dvipsnames]{color}

\newcolumntype{.}{D{.}{.}{-1}}

\newcommand{\bv}{\begin{array}}

 \usepackage{mathtools}

\newcommand{\nn}{\nonumber}

\newtheorem{assumption}{Assumption}[section]

\newtheorem{theorem}{Theorem}[section]

\newtheorem{lemma}{Lemma}[section]
\newtheorem{corollary}{Corollary}[section]

\newcommand{\Reals}[1]{\mathbb{R}^{#1}}

\renewcommand{\b}[1]{\boldsymbol{#1}} 		 
\newcommand{\m}[1]{\uppercase{#1}}		

\newcommand{\Ex}[1]{\mathbb{E}\left (#1 \right )} 		 
\newcommand{\ExCond}[2]{\mathbb{E}\left(#1 \mid #2\right)} 		 
\newcommand{\Var}[1]{\mathrm{Var}\left(#1\right)}   		 
\newcommand{\VarCond}[2]{\mathrm{Var}\left(#1\mid#2\right)}  		 
\newcommand{\Cov}[2]{\mathrm{Cov}\left(#1, #2\right)} 		 

\newcommand{\bigO}[1]{\mathrm{O}(#1)}
\newcommand{\littleO}[1]{\mathrm{o}(#1)}

\newcommand{\sgdM}[1]{(I - \gamma_{#1}\Fisher{\thetastar})}
\newcommand{\impM}[1]{(I + \gamma_{#1} \Fisher{\thetastar})^{-1}}
\newcommand{\Pin}[2]{\m{P}_{#1}^{#2}}
\newcommand{\Qin}[2]{\m{Q}_{#1}^{#2}}
\newcommand{\eig}[1]{\mathrm{eig}(\m{#1})}
\newcommand{\lmax}{\maxEigF}
\newcommand{\lmin}{\minEigF}

\newcommand{\MainAssumptions}{
\small
The explicit SGD procedure in Eq.~\eqref{def:sgd:explicit} 
and the implicit SGD procedure in Eq.~\eqref{def:sgd:implicit} 
operate under a combination of the following assumptions.
\begin{enumerate}[(a)]
\item \label{A:gamma_n}
	The learning rate sequence $\{\gamma_n \}$ 
	is defined as $\gamma_n = \gamma_1 n^{-\gamma}$,
	where $\gamma_1>0$ is the learning parameter, and $\gamma \in (0.5, 1]$.

\item 
\label{A:linear_model}
For the log-likelihood $\log f(Y; X, \theta)$
there exists function $\ell$ such that 
 $\log f(Y; X, \theta) \equiv \ell(X^\intercal \theta; Y)$, which depends 
 on $\theta$ only through the natural parameter $X^\intercal\theta$.
 
\item
 \label{A:h_lip} 
 Function $\ell$ is concave, twice differentiable almost surely w.r.t. natural parameter $X^\intercal\theta$ and Lipschitz with constant $L_0$ w.r.t. $\theta$.

\item \label{A:h_strong}
The observed Fisher information matrix $\Fisherobs{n}{\theta} = -\nabla^2 \ell(X_n^\intercal \theta; Y_n)$ has non-vanishing trace, i.e., 
there exists constant $\FisherMin>0$ such that 
$\mathrm{trace}(\Fisherobs{n}{\theta}) \ge \FisherMin$ almost surely, for all $\theta$.
The Fisher information matrix, $\Fisher{\thetastar} =\Ex{\Fisherobs{n}{\thetastar}}$, has minimum eigenvalue $\minEigF >0$ 
and maximum eigenvalue $\maxEigF < \infty$. Typical regularity conditions hold \citep[Theorem 5.1, p.463]{lehmann1998theory}.

\item \label{A:Cn}
Every condition matrix $ C_n $ is a fixed positive-definite matrix,
 such that $C_n = C + \bigO{\gamma_n}$, where $C\succ 0$ and symmetric, 
 and $C$ commutes with $\Fisher{\thetastar}$.
For every $C_n$, $\min \mathrm{eig}(C_n) = \minEigC >0$, and
$\max \mathrm{eig}(C_n) = \maxEigC < \infty$.

\item \label{A:errors_Lindeberg}
Let $\Xi_n = \ExCond{\nabla \log f(Y_n; X_n, \thetastar)\nabla \log f(Y_n; X_n, \thetastar)^\intercal}{\mathcal{F}_{n-1}}$, then $||\Xi_n- \Xi|| = \bigO{1}$  for all $n$, and $||\Xi_n - \Xi|| \to 0$, for a symmetric positive-definite $\Xi$.
Let $\sigma_{n, s}^2 = \Ex{\mathbb{I}_{||\xi_n(\thetastar)||^2 \ge s/\gamma_n}||\xi_n(\thetastar)||^2}$, then for all $s>0$, $\sum_{i=1}^n \sigma_{i,s}^2 = \littleO{n}$ if $\gamma=1$, and $\sigma_{n, s}^2 = \littleO{1}$ otherwise.
\end{enumerate}
}

\newcommand{\TheoremMSE}{
Let  $\delta_n = \Ex{||\thetaim{n}-\thetastar||^2}$. Suppose that \assumeMains\assumeGn,\assumeLinear,\assumeLip,
\assumeStrong, and \assumeCn\ hold. 
Then, there exist constants $n_0 >0$ and 
$\kappa = 1+2\gamma_1 \mu \minEigC\minEigF$ for some $\mu \in (0, 1]$ such that,
\begin{align}
\delta_n \le 
\frac{4 L_0^2 \maxEigC^2 \gamma_1 \kappa}{\mu\minEigF\minEigC} n^{-\gamma} + \exp\left(-\log \kappa \cdot \phi_\gamma(n) \right) [\delta_0 +\kappa^{n_0}\Gamma^2]\nn,
\end{align}
where $\Gamma^2=4L_0^2\maxEigC^2 \sum_i \gamma_i^2 < \infty$, 
and $\phi_\gamma(n) = n^{1-\gamma}$ if $\gamma<1$, and $\phi_\gamma(n) = \log n$ if $\gamma=1$.
}

\newcommand{\TheoremVariance}{Consider SGD procedures in Eq.~\eqref{def:sgd:explicit} and in Eq.~\eqref{def:sgd:implicit}, and suppose that \assumeMains\assumeGn,\assumeLip,\assumeStrong,\assumeCn\ hold, where $\gamma=1$, 
and that $2\gamma_1 C \Fisher{\thetastar} \succ I$.
The asymptotic variance of the explicit SGD estimator in Eq.~\eqref{def:sgd:explicit} satisfies
\begin{align}
n \Var{\thetasgd{n}}  \to \VarAsymp.\nn
\end{align}
The asymptotic variance of the implicit SGD estimator in Eq.~\eqref{def:sgd:implicit} satisfies
\begin{align}
	n \Var{\thetaimpl{n}}  \to \VarAsymp.\nn
\end{align}
}

\newcommand{\TheoremAveraging}{
Consider the SGD procedure defined in Eq.~\eqref{def:aisgd} and suppose \assumeMains\assumeGn,\assumeLip,\assumeStrong, and \assumeCn\ hold, where $\gamma \in [0.5, 1)$. 
Then, $\thetaBar{n}$ converges to $\thetastar$ in probability and is asymptotically efficient, i.e., 
\begin{align}
n \Var{\thetaBar{n}}  \to \Fisher{\thetastar}^{-1}.\nn
\end{align}
}

\newcommand{\TheoremNormality}{
Suppose that \assumeMains\assumeGn,\assumeLip,\assumeStrong,\assumeCn,\assumeLind\ hold. 
Then, the iterate $\thetaim{n}$ of implicit SGD in Eq.~\eqref{def:sgd:implicit} is asymptotically normal, 
such that
\begin{align}
n^{\gamma/2} (\thetaim{n}-\thetastar) \to \mathcal{N}_p(0, \Sigma),\nn
\end{align}
where $\Sigma = \VarAsymp$.
}

\newcommand{\LemmaStability}{
Let $\lmax = \max \mathrm{eig}(\Fisher{\thetastar})$,
and suppose $\gamma_n = \gamma_1/n$ and 
$\gamma_1\lmax > 1$.
Then, the maximum eigenvalue of $\Pin{1}{n}$ satisfies
	\begin{align}
	\max_{n>0} \max \mathrm{eig}(\Pin{1}{n}) =  
			\Theta(2^{\gamma_1 \lmax}/\sqrt{\gamma_1 \lmax}).\nn
	\end{align}
For the implicit method,
		\begin{align}
	\max_{n>0} \max \mathrm{eig}(\Qin{1}{n}) = \bigO{1}.\nn
	\end{align}
}

\newcommand{\LinearTheorem}{
Suppose \assumeMain\assumeLinear\ holds, {then the gradient of the log-likelihood is a scaled version of covariate $X$, i.e., for every $\theta\in\Reals{p}$ there is a
scalar $\lambda \in \Reals{}$ such that
$$
\nabla \log f(Y; X, \theta) = \lambda X.
$$
}
Thus, the gradient in the implicit update in Eq.~\eqref{def:sgd:implicit}  is a scaled version of the gradient 
calculated at the previous iterate, i.e.,
\begin{align}
\label{algo:implicit:scale}
\nabla \log f(Y_n; X_n, \thetaim{n}) = 
	 \lambda_n \nabla \log f(Y_n; X_n, \thetaim{n-1}),
\end{align}
where the scalar $\lambda_n$ satisfies
\begin{align}
\label{algo:implicit:fp}
\lambda_n \ell'(X_n^\intercal \thetaim{n-1}; Y_n) = 
 \ell'\left(X_n^\intercal \thetaim{n-1} + \gamma_n \lambda_n  \ell'(X_n^\intercal \thetaim{n-1}; Y_n) X_n^\intercal C_n X_n;  Y_n \right).
\end{align}
}

\newcommand{\thetastar}{\theta_\star}
\newcommand{\thetamle}{\theta_N^{\mathrm{mle}}}
\newcommand{\vthetastar}{\thetastar}
\newcommand{\gradsgd}{\nabla \log f(Y_n; X_n, \thetasgd{n-1})}

\newcommand{\linearScore}[2]{\ell'(X_{#1}^\intercal #2; Y_{#1}) X_{#1}}
\newcommand{\loglik}[2]{\log f(Y_{#1}; X_{#1}, #2)}

\newcommand{\aisgd}{averaged implicit SGD}

\newcommand{\asgd}{averaged explicit SGD}
\newcommand{\implicit}{implicit SGD}
\newcommand{\e}{\mathrm{e}}
\newcommand{\adagrad}{AdaGrad}
\newcommand{\proxsvrg}{Prox-SVRG}
\newcommand{\proxsag}{Prox-SAG}

\newcommand{\biglm}{\texttt{biglm}}
\newcommand{\thetaBar}[1]{\overline{\thetaim{n}}}

\renewcommand{\and}{\text{ }\mathrm{ \& }\text{ }}
\newcommand{\thetaimPlus}[1]{\theta_n^{\mathrm{im}, \diamond}}

\newcommand{\mF}[1]{\mathcal{F}_{#1}}
\newcommand{\assumeMain}{Assumption \ref{main_assumptions}}
\newcommand{\assumeMains}{Assumptions \ref{main_assumptions}}
\newcommand{\assumeGn}{\eqref{A:gamma_n}}
\newcommand{\assumeLip}{\eqref{A:h_lip}}
\newcommand{\assumeStrong}{\eqref{A:h_strong}}

\newcommand{\assumeCn}{\eqref{A:Cn}}

\newcommand{\assumeLind}{\eqref{A:errors_Lindeberg}}
\newcommand{\assumeLinear}{\eqref{A:linear_model}}

\newcommand{\maxEigC}{\overline{\lambda_c}}
\newcommand{\minEigC}{\underline{\lambda_c}}
\newcommand{\maxEigF}{\overline{\lambda_f}}
\newcommand{\minEigF}{\underline{\lambda_f}}

\newcommand{\commentEq}[1]{
\hspace{10px}\text{ \footnotesize [{\em #1}]}}
\newtheorem{innercustomthm}{Theorem}
\newenvironment{customthm}[1]
  {\renewcommand\theinnercustomthm{#1}\innercustomthm}
  {\endinnercustomthm}

\newcommand{\vtheta}{\theta}

\newcommand{\vthetan}{\theta_n}

\newcommand{\vthetanprev}{\theta_{n-1}}

\newcommand{\thetaimpl}[1]{\vtheta_{#1}^{\mathrm{im}}}
\newcommand{\thetasgd}[1]{\vtheta_{#1}^{\mathrm{sgd}}}

\newcommand{\x}{\b{x}}

\newcommand{\xn}{\b{x}_n}

\newcommand{\glmscale}{\psi}

\newcommand{\Fisher}[1]{\mathcal{I}(#1)}
\newcommand{\Fisherobs}[2]{\hat{\mathcal{I}}_{#1}(#2)}
\newcommand{\FisherMin}{b}

\newcommand{\thetastarx}[1]{\thetastar^\intercal X_{#1}}

\newcommand{\xtheta}{\thetax{}{}}

\newcommand{\Vx}{\m{V}_{\vphi}}

 \newcommand{\VarAsymp}
 {\gamma_1^2 \left (2\gamma_1 C \Fisher{\thetastar} - I \right)^{-1} C \Fisher{\thetastar} C}

\newcommand{\Mn}[2]{\m{#1}_{#2}}  
\newcommand{\mt}[2]{\Mn{#1}{#2}}

\newcommand{\I}{\m{I}}

\newcommand{\thetaim}[1]{\vtheta_{#1}^{\mathrm{im}}}

\newcommand{\linearMap}[2]{\mathbb{L}_{#1}\left\{ #2 \right \}}
\newcommand{\invLinearMap}[2]{\mathbb{L}_{#1}^{-1}\left\{#2\right\}}
\newcommand{\linearMapOne}[1]{\mathbb{L}_{#1}}
\newcommand{\linearMapIdentityOne}{\mathbb{I}}
\newcommand{\linearMapIdentity}[1]{\mathbb{I}\left \{ #1 \right \}}
\newcommand{\invLinearMapOne}[1]{\mathbb{L}_{#1}^{-1}}
\newcommand{\linearMapNull}[1]{\mathbb{L}_0 \left \{#1\right \}}
\newcommand{\linearMapNullOne}{\mathbb{L}_0}

\usepackage{algorithmic}

\begin{document}

\title{Asymptotic and finite-sample properties of estimators based on stochastic gradients\protect\thanks{Panagiotis (Panos) Toulis is an Assistant Professor of Econometrics and Statistics 
at University of Chicago, Booth School of Business~(\href{mailto:panos.toulis@chicagobooth.edu}{panos.toulis@chicagobooth.edu}).~Edoardo M.~Airoldi is an Associate Professor of Statistics at Harvard University (\href{mailto:airoldi@fas.harvard.edu}{airoldi@fas.harvard.edu}).
%
%
The authors wish to thank Joe Blitzstein, Leon Bottou, Bob Carpenter, David Dunson, Andrew Gelman, Brian Kulis, Xiao-Li Meng, Natesh Pillai and Neil Shephard for useful comments and discussion. We are grateful to four reviewers, the Associate Editor, and the Editor for offering suggestions that improved the paper. This research was sponsored, in part, by NSF CAREER award IIS-1149662, ARO MURI award W911NF-11-1-0036, and ONR YIP award N00014-14-1-0485.}}

\author{
 Panos Toulis and Edoardo M. Airoldi\\
 University of Chicago and Harvard University}
\date{}

\maketitle


\newpage

\begin{abstract}
Stochastic gradient descent procedures have gained popularity for parameter estimation from large data sets. However, their statistical properties are not well understood, in theory. And in practice, avoiding numerical instability requires careful tuning of key parameters.
Here, we introduce implicit stochastic gradient descent procedures, which involve parameter updates that are implicitly defined. Intuitively, implicit updates shrink standard stochastic gradient descent updates.
The amount of shrinkage depends on the observed Fisher information matrix, which does not need to be explicitly computed; thus, implicit procedures increase stability without increasing the computational burden.
Our theoretical analysis provides the first full characterization of the asymptotic behavior of both standard and implicit stochastic gradient descent-based estimators, including finite-sample error bounds. 
Importantly, analytical expressions for the variances of these stochastic gradient-based estimators reveal their exact  loss of efficiency.
We also develop new algorithms to compute implicit stochastic gradient descent-based estimators for generalized linear models, Cox proportional hazards,  M-estimators, in practice, and perform extensive experiments. 
Our results suggest that implicit stochastic gradient descent procedures are poised to become a workhorse for approximate inference from large data sets.

\vfill
\noindent {\bf Keywords}: Stochastic Approximation; Implicit Updates; Asymptotic Variance; Generalized Linear Models; Cox Proportional Hazards; M-estimation; Maximum Likelihood; Statistical Efficiency; Numerical Stability\end{abstract}

\newpage
\singlespacing
\small
\tableofcontents
\normalsize
\onehalfspacing
\newpage


\section{Introduction}
\label{section:introduction}

Parameter estimation by optimization of an objective function is a fundamental idea in statistics and machine learning \citep{fisher1922,lehmann1998theory,Hastie:2011fk}. 
However, classical procedures, 
such as Fisher scoring, the EM algorithm or iteratively reweighted least squares \citep{Fisher:1925vn,Demp:Lair:Rubi:1977,green1984iteratively}, do not scale to modern data sets with millions of data points and hundreds or thousands of parameters \citep{Council:2013kx}.
%

%
In particular, suppose we want to estimate the true parameter $\thetastar \in \Reals{p}$
of a distribution $f$ from $N$ i.i.d. data points $(X_i, Y_i)$, such that  conditional on covariate $X_i \in \Reals{p}$ 
outcome $Y_i \in \Reals{d}$ is distributed according to $f(Y_i;  X_i, \thetastar)$. 
%
Such estimation problems often reduce to optimization problems. For instance, the maximum likelihood estimator (MLE) is obtained by solving $\thetamle = 
\arg \max_\theta \sum_{i=1}^N \log f(Y_i; X_i, \theta)$.
Classical optimization procedures, such as Newton-Raphson or Fisher scoring, have a runtime complexity that ranges between $\bigO{Np^{1+ \epsilon}}$ and 
$\bigO{Np^{2+\epsilon}}$, in the best case and worst case respectively~\citep{lange2010numerical}.
Quasi-Newton (QN) procedures are the only viable alternative in practice because they have $\bigO{Np^2}$ complexity per iteration,  or $\bigO{Np^{1+\epsilon}}$ in certain favorable cases \citep{hennig2013quasi}. 
However, estimation from large data sets requires an even better runtime complexity 
that is roughly $\bigO{N p^{1-\epsilon}}$, i.e., linear in 
data size $N$ but sublinear in parameter dimension $p$. 
The first requirement on $N$ is generally unavoidable 
because all data points carry information from the i.i.d. assumption. Sublinearity in $p$ is therefore critical. 

Such requirements 
have recently generated interest in stochastic optimization procedures, 
especially those only relying on first-order information, i.e., 
gradients.
Perhaps the most widely popular procedure in this family is \emph{stochastic gradient descent}
(SGD), defined for $n=1, 2, \ldots $, as
\begin{align}
\label{def:sgd:explicit}
\thetasgd{n} =  \thetasgd{n-1} + \gamma_n C_n \gradsgd,
\end{align}
where $\gamma_n>0$ is the learning rate sequence,
 typically defined as $\gamma_n = \gamma_1 n^{-\gamma}$,
 $\gamma_1>0$ is the learning rate parameter, 
 $\gamma \in (.5, 1]$, and $C_n$ are $p \times p$ 
positive-definite matrices, also known as condition matrices.

Stochastic optimization procedures of this kind 
are special cases of stochastic approximation \citep{robbins1951}, 
where the estimation problem is not formulated as an optimization problem
but more generally as a characteristic equation. 
Early research considered a streaming data setting---akin to a superpopulation setting---where the characteristic 
equation is
\begin{align}
\label{eq:SA}
\ExCond{\nabla \log f(Y; X, \thetastar)}{X}=0,
\end{align}
with the expectation being over the true conditional 
distribution of outcome $Y$ given covariate $X$.
More recent research, largely in computer science and optimization, considers a finite N setting with characteristic equation
\begin{align}
\label{eq:SA2}
\Ex{\nabla \log f(Y; X, \thetamle)}=0,
\end{align}
where the expectation is over the {empirical} distribution 
of $(X, Y)$ in the finite data set.
In both settings, SGD of Eq.~\eqref{def:sgd:explicit} is well-defined: in the finite population setting of Eq.~\eqref{eq:SA2} the data point $(X_n, Y_n)$ is a random sample with
 replacement  from the finite data set; 
 in the infinite population setting of Eq.~\eqref{eq:SA} the data point $(X_n, Y_n)$ is simply the $n$th data point  in the stream.

From a computational perspective, SGD in Eq.~\eqref{def:sgd:explicit} is appealing 
because it avoids expensive matrix inversions, as in Newton-Raphson, and the log-likelihood is evaluated at a single 
data point $(X_n, Y_n)$ and not on the entire 
data set. 
From a theoretical perspective, SGD in Eq.~\eqref{def:sgd:explicit} converges, under suitable conditions,  to  $\thetasgd{\infty}$ where $\ExCond{\log f(Y; X, \thetasgd{\infty})}{X} =0$~\citep{benveniste1990adaptive, ljung1992stochastic, borkar2008stochastic}. This condition can satisfy both Eq.~\eqref{eq:SA} and Eq.~\eqref{eq:SA2}, implying that SGD can be 
used on both finite and infinite population settings.
For the rest of this paper we assume an infinite population setting,  as it is the most natural setting for stochastic approximations. 
The main difference between the streaming data setting studied in the computer science and optimization literature and the infinite population setting we consider here is that we do not condition on the observed ordering of data points, but we condition on a random ordering instead.
Moreover, most of the theoretical results presented in this paper for the infinite population case can be applied to the finite population case, where instead of estimating $\thetastar$  we estimate the MLE, or the maximum a posteriori (MAP) estimate if there is regularization.

In this paper, we introduce {\em implicit} stochastic gradient descent procedures---{\em implicit SGD} for short---defined as
\begin{align}
\label{def:sgd:implicit}
\boldsymbol{\thetaim{n}} =  \thetaim{n-1} + \gamma_n C_n \nabla\log f(Y_n; X_n, \boldsymbol{\thetaim{n}}),
\end{align}
where $\gamma_n, C_n$ are defined as in standard SGD in Eq.~\eqref{def:sgd:explicit}.
Furthermore, we provide a theoretical analysis of estimators based on stochastic gradients, for
both implicit and standard procedures.
To distinguish the two procedures, 
we will refer to standard SGD in Eq.~\eqref{def:sgd:explicit} as SGD with \emph{explicit updates}, or \emph{explicit SGD} for short,
because the next iterate $\thetasgd{n}$ can be immediately computed given $\thetasgd{n-1}$ and the data point $(X_n, Y_n)$.
In contrast, the update in Eq.~\eqref{def:sgd:implicit} is \emph{implicit} 
because the next iterate $\thetaim{n}$ appears on both sides of the equation, where the iterate was typed in boldface to emphasize the fact.

\subsection{Illustrative example}
\label{section:illustration}
\newcommand{\tnsgd}[1]{\theta_{#1}^{\mathrm{sgd}}}
\newcommand{\tnim}[1]{\theta_{#1}^{\mathrm{im}}}
Here, we motivate the main results of this paper on the comparison between implicit and explicit SGD.
Let $\vthetastar \in \Reals{}$ 
be the true parameter of a normal model with i.i.d. observations 
$Y_i  | X_i \sim \mathcal{N}( X_i\thetastar, \sigma^2)$, where the variance $\sigma^2$ is assumed known for simplicity.  The log-likelihood is 
 $\log f(Y_i; X_i, \theta) = -\frac{1}{2\sigma^2} (Y_i- X_i\theta)^2$, and the score function (i.e., gradient of log-likelihood) is given by 
$\nabla \log f(Y_i; X_i, \theta) = 
\frac{1}{\sigma^2} (Y_i - X_i \theta) X_i$. 
Let $X_i$ be distributed according to some unknown distribution 
with bounded second moment.
Assume $\gamma_n = \gamma_1/n$, for some $\gamma_1 >0$ as the learning rate. 
Then, the explicit SGD procedure in Eq.~\eqref{def:sgd:explicit} is
\begin{align}
\label{eq:sgdexample}
\thetasgd{n} & = \thetasgd{n-1}  + \gamma_n (Y_n - \thetasgd{n-1} X_n) X_n \nonumber \\
             & = (1-\gamma_n X_n^2) \thetasgd{n-1}  + \gamma_n Y_n X_n.
\end{align}
Procedure \eqref{eq:sgdexample} is 
the least mean squares filter (LMS) in signal processing, 
also known as the Widrow-Hoff algorithm \citep{widrow1960adaptive}.
The implicit SGD procedure can be derived in closed form in this problem using update in Eq.~\eqref{def:sgd:implicit} as 
\begin{align}
\label{eq:implicitexample}
\thetaim{n}  & =\thetaim{n-1}  + \gamma_n (Y_n - X_n \thetaim{n}) X_n \nonumber \\
             & = \frac{1}{1+ \gamma_n X_n^2} \thetaim{n-1} + \frac{\gamma_n}{1+\gamma_n X_n^2} Y_n X_n.
\end{align}
The procedure defined by Eq.~\eqref{eq:implicitexample} is 
known as the normalized least mean squares filter (NLMS) in signal processing \citep{nagumo1967}.

From Eq. \eqref{eq:sgdexample} we see that it is crucial for explicit SGD to have a well-specified learning rate parameter $\gamma_1$. For instance, 
if $\gamma_1 X_1^2 >> 1$ then $\thetasgd{n}$ will diverge to a value at the order of $2^{\gamma_1}/\sqrt{\gamma_1}$, before converging to the true value $\thetastar$~(see Section \ref{section:stability}, Lemma 
\ref{lemma:stability}). 
In contrast, implicit SGD is more stable to misspecification of the learning rate parameter $\gamma_1$.
For example, a very large $\gamma_1$ will not cause divergence as in explicit SGD, but it will simply put more weight 
on the $n$th observation $Y_n X_n$ than the 
previous iterate $\thetaim{n-1}$. 
Assuming for simplicity
$\thetasgd{n-1}=\thetaim{n-1}=0$, it 
also holds $\thetaim{n} = \frac{1}{1+\gamma_n X_n^2} \thetasgd{n}$,
showing that implicit SGD iterates are shrinked 
versions of explicit ones (see also Section \ref{section:discussion}).

Let $v^2 = \Ex{X^2}$, then by Theorem \ref{theorem:variance} 
the asymptotic variance of $\thetaim{n}$ (and of $\thetasgd{n}$) satisfies 
$n \mathrm{Var}(\thetaim{n}) \to \gamma_1^2 \sigma^2 v^2 / (2\gamma_1 v^2-1)$ if $2\gamma_1 v^2-1>0$.
Since $\gamma_1^2/(2\gamma_1v^2-1) \ge 1/v^2$, it is best to set $\gamma_1=1/v^2$. In this case 
$n \mathrm{Var}(\thetaim{n}) \to \sigma^2/v^2$. 
Implicit SGD can thus be optimal by setting $\gamma_n  = (\sum_{i=1}^n X_i^2)^{-1}$ in which case 
 $\thetasgd{n}$ is exactly the OLS estimator, and 
 $\thetaim{n}$ is an approximate but more stable version of the OLS estimator.
Thus, the implicit SGD estimator $\thetaim{n}$ in Eq.~\eqref{eq:implicitexample} inherits
the efficiency properties of $\thetasgd{n}$, with the added benefit 
of being stable over a wide range of learning rates.
Overall, implicit SGD is a superior form of SGD.

\subsection{Related work}
\label{section:overview:related}
Historically, the duo of explicit-implicit updates originates from the numerical 
methods introduced by Euler (ca. 1770) for solving ordinary differential equations~\citep{hoffman2001numerical}. 
The explicit SGD procedure was first proposed by Sakrison (\citeyear{sakrison1965})
as a recursive statistical estimation method and it is theoretically based on 
the stochastic approximation method of \citet{robbins1951}.
Statistical estimation with explicit SGD is a straightforward generalization 
of Sakrison's method and has recently attracted attention
in the machine learning community as a fast learning method for large-scale problems
\citep{zhang2004solving, bottou2010,  toulis2015scalable}. Applications of explicit SGD procedures in massive data problems can be found in 
many diverse areas such as large-scale machine learning \citep{zhang2004solving}, online EM algorithm \citep{cappe2009line}, image analysis and 
deep learning \citep{dean2012large} and
MCMC sampling \citep{welling2011bayesian}.

The implicit SGD procedure is less known and not well-understood.
In optimization, implicit methods have recently attracted attention 
under the guise of proximal methods, such as mirror-descent~\citep{nemirovskici83}. In fact, the implicit SGD update in Eq.~\eqref{def:sgd:implicit} can be expressed as a proximal update:
\begin{equation}
\label{intro:eq:proximal} 
\vthetan^\mathrm{im} = \arg \max_{\vtheta} \left\{-\frac{1}{2}||\vtheta - \vthetanprev^\mathrm{im}||^2 + \gamma_n 
\log f(Y_n; X_n, \theta)  \right\}.
\end{equation}
From a Bayesian perspective, $\thetaim{n}$ is
the posterior mode of a model with the standard multivariate normal $\mathcal{N}(\thetaim{n-1}, \gamma_n I)$ as the prior,
and $\log f(Y_n; X_n, \theta)$ as the log-likelihood of $\theta$
 for observation $(X_n, Y_n)$.
Arguably, the normalized least mean squares (NLMS) filter \citep{nagumo1967}, introduced in Eq.~\eqref{eq:implicitexample},
was the first statistical model that used an implicit update as in Eq.~\eqref{def:sgd:implicit},
and was shown to be consistent and robust under
excessive input noise \citep{slock1993}.
From an optimization perspective,
the update in Eq.~\eqref{intro:eq:proximal} corresponds to a stochastic version of the proximal point algorithm by \citet{rockafellar1976monotone}, which has been generalized through the idea of splitting algorithms~\citep{lions1979splitting, beck2003, singer2009efficient}; see also the 
comprehensive review of proximal methods 
in optimization by \citet{parikh2013proximal}.
Additional intuition of implicit methods 
has been provided by \citet{krakowskigeometric} and
\citet{nemirovski2009robust}, who have argued that proximal methods
can fit better in the geometry of the parameter space.
\citet{bertsekas2011incremental} derived 
the convergence rate of an implicit procedure similar to Eq.~\eqref{def:sgd:implicit} 
on a fixed data set, and compared the rates between procedures that randomly sampled data $(X_n, Y_n)$ or simply cycled through them.
\citet{toulis2014statistical} derived 
the asymptotic variance of $\thetaim{n}$ 
as estimator of $\thetastar$ in the family of generalized linear models, and
provided an algorithm to efficiently compute 
the implicit update of Eq.~\eqref{def:sgd:implicit} in such models and in the simplified setting where $C_n=I$.
%



\subsection{Contributions}
\label{section:overview}
Prior work on procedures similar to implicit SGD has considered mostly an optimization setting, in which the focus is on speed of convergence~\citep[for example]{bertsekas2011incremental}.
Instead, we focus on statistical efficiency, that is, the sampling variability of the estimator implied by implicit and explicit SGD procedures---the relevant analysis and the results of Theorem~\ref{theorem:mse2} and Theorem~\ref{theorem:variance} are novel.
Furthermore, our procedure, which we generalized in \citet{toulis2015implicitSA}, is different than typical stochastic proximal gradient procedures~\citep[for example see][]{duchi2009,rosasco2014convergence}.
In such procedures   the parameter updates are obtained by combining  a stochastic explicit update and a deterministic implicit update. In implicit SGD there is a single stochastic implicit update, which prevents numerical instability.

With regard to theoretical contributions, the asymptotic statistical efficiency of SGD procedures (both explicit and implicit) derived in Theorem~\ref{theorem:variance}  is  a key contribution of our work.
Our analysis is in fact general enough that allowed us to derive the asymptotic efficiency of other popular stochastic optimization procedures, notably of  AdaGrad~\citep{duchi2011} in Eq.~\eqref{eq:variance_adagrad} of our paper. 
The asymptotic normality of implicit SGD in Theorem~\ref{theorem:normality} is new and enables a novel comparison of explicit SGD and implicit SGD in terms of the normality of their iterates, which is also a clear point of departure from the typical optimization literature. 
The results in Section~\ref{section:stability} are also new, and formalize the advantages of implicit SGD over explicit SGD in terms of numerical stability. 
    
With regard to practical contributions, Algorithm~\ref{algo:implicit} and its variants presented in the paper are a significant extension of our earlier work beyond first-order GLMs~\citep[Algorithm 1]{toulis2014statistical}. The key contribution here is that these new algorithms make implicit SGD  as simple to implement as standard explicit SGD, whenever the fixed-point computation of the implicit update is feasible.
We provide extensive applications in Section~\ref{section:applications} and experiments in Section~\ref{section:experiments} of implicit SGD compared to explicit SGD. 
Importantly, we implemented implicit SGD through the \texttt{R} package \texttt{sgd}~\citep{tran2015stochastic} (available at \small\url{https://cran.r-project.org/web/packages/sgd/index.html}\normalsize) to compare implicit SGD with state-of-art procedures, including R's \texttt{glm()} function,  \texttt{biglm} package, the elastic net \citep[\texttt{glmnet}]{friedman2010regularization}, \adagrad\ \citep{duchi2011},
\proxsvrg\ \citep{xiao2014proximal}, and \proxsag\ \citep{schmidt2013minimizing}.

\section{Theory}
\label{section:theory}
The norm $||.||$ denotes the $L_2$ norm.
If a positive scalar sequence $a_n$ is nonincreasing and $a_n\to0$, 
we write $a_n \downarrow 0$.
For two positive scalar sequences $a_n, b_n$, equation
$b_n = \bigO{a_n}$ denotes that $b_n$ is bounded above by $a_n$, 
i.e., there exists a fixed $c>0$ such that $b_n \le c a_n$, for all $n$. 
Furthermore, $b_n = \littleO{a_n}$ denotes that $b_n / a_n \to 0$.
Similarly, for a sequence of vectors (or matrices) $X_n$,
we write $X_n = \bigO{a_n}$ to denote 
$||X_n|| = \bigO{a_n}$, and 
write $X_n = \littleO{a_n}$ to denote $||X_n|| = \littleO{a_n}$.
For two positive definite matrices $A, B$ we write  $A \prec B$ to express that $B-A$ is positive definite. 
The set of eigenvalues of a matrix $A$ 
is denoted by $\mathrm{eig}(\m{A})$; thus, $A \succ 0$ if and only if 
$\lambda>0$ for every $\lambda \in \mathrm{eig}(A)$.

\begin{assumption}
\label{main_assumptions}
\MainAssumptions
\end{assumption}
\normalsize
\emph{Remarks}.
\assumeMain\assumeGn\ is typical in stochastic approximation 
as it implies that $\sum_i \gamma_i = \infty$ and $\sum_i \gamma_i^2 < \infty$, as posited by \citet{robbins1951}.
\assumeMain\assumeLinear\ narrows our focus to 
models for which the likelihood depends on parameters $\theta$ through the linear combination $X^\intercal\theta$.
This family of models is large and includes generalized linear models, 
Cox proportional hazards models, and M-estimation. Furthermore, in Section \ref{section:discussion} we discuss a significant relaxation of  \assumeMain\assumeLinear.
\assumeMain\assumeLip\ puts a Lipschitz condition on the log-likelihood but it is used only for deriving finite-sample error bounds in Theorem~\ref{theorem:mse2}---it is possible that this condition can be relaxed.
\assumeMain\assumeStrong\ is equivalent to assuming 
strong convexity for the negative log-likelihood, which is typical for proving convergence in probability.
The assumption on the observed Fisher information is less standard
and, intuitively, it posits that a minimum of statistical 
information is received from any data point, at least for certain model parameters. Making this assumption allows us to forgo 
boundedness assumptions on the errors of stochastic gradients 
that were originally used by \citet{robbins1951},
and have since been standard.
Finally, \assumeMain\assumeLind\ posits the typical Lindeberg conditions 
that are necessary to invoke the central limit theorem and prove asymptotic normality; this assumption follows the conditions defined by \citet{fabian1968} for the normality of explicit SGD procedures.

\subsection{Finite-sample error bounds}
\label{section:non_asymptotics}
Here, we derive bounds for the errors 
$\mathbb{E}(||\thetaim{n}-\thetastar||^2)$ on a finite sample of fixed size $n$.
\begin{theorem}
\label{theorem:mse2}
\TheoremMSE
\end{theorem}

Not surprisingly, implicit SGD in Eq.~\eqref{def:sgd:implicit}
matches the asymptotic rate of explicit SGD in Eq.~\eqref{def:sgd:explicit}. 
In particular, the iterates $\thetaim{n}$ have squared error with rate $\bigO{n^{-\gamma}}$, as seen in Theorem \ref{theorem:mse2}, 
which is identical to the rate of error for the explicit iterates $\thetasgd{n}$
\citep[Theorem 22, p.244]{benveniste1990adaptive}.
One way to explain intuitively this similarity in convergence rates is to assume that both explicit and implicit
SGD are at the same estimate $\theta_0$. 
Then, using definitions in Eq.~\eqref{def:sgd:explicit} and in Eq.~\eqref{def:sgd:implicit}, a Taylor approximation of the gradient $\nabla \log f(Y_n; X_n, \thetaim{n})$ 
yields
\begin{align}
\label{eq:approx:implicit}
\Delta \thetaim{n} \approx [I + \gamma_n \hat{\mathcal{I}}_n(\theta_0)]^{-1} \Delta \thetasgd{n},
\end{align}
where $\Delta \thetaim{n} = \thetaim{n}-\theta_0$ 
and $\Delta \thetasgd{n} = \thetasgd{n}-\theta_0$.
Therefore, as $n\to\infty$, we have $\Delta \thetaim{n} \approx \Delta \thetasgd{n}$, and  the two procedures coincide.

Despite the similarity in convergence rates, the critical advantage of implicit SGD---more generally of implicit procedures---is their robustness to initial conditions and excess noise.
This can be seen in Theorem \ref{theorem:mse2} where the implicit procedure discounts the initial conditions
$\Ex{||\thetaim{0}-\thetastar||^2}$ at an exponential rate 
through the term $\exp(-\log \kappa \cdot \phi_\gamma(n))$.
Importantly, the discounting of initial conditions 
happens regardless of the specification of the learning rate.
In fact, large values of $\gamma_1$ can lead to faster discounting, and thus possibly to faster convergence, however at the expense of increased variance as implied by Theorem~\ref{theorem:variance}, which is presented in the following section.
The implicit iterates are therefore {\em unconditionally stable}, i.e., virtually any specification of the learning rate will lead to a stable discounting of the initial conditions.

In  contrast, explicit SGD is known to be very sensitive to 
the learning rate, and can numerically diverge if the rate is misspecified.
For example, \citet[Theorem 1]{moulines2011non} showed that 
there exists a term $\exp(L^2 \gamma_1^2 n^{1-2\gamma})$, 
where $L$ is a Lipschitz constant for the gradient of the log-likelihood,
amplifying the initial conditions $\mathbb{E}(||\thetasgd{0}-\thetastar||^2)$ 
of explicit SGD, which can be catastrophic 
if the learning rate parameter $\gamma_1$ is misspecified.\footnote{The Lipschitz conditions are different in the two works, 
however this does not affect our conclusions. 
Our result remains effectively unchanged if we assume Lipschitz continuity of the 
gradient $\nabla \ell$ instead of the log-likelihood $\ell$, similar to \citet{moulines2011non}; see comment after proof 
of Theorem \ref{theorem:mse2}.}
Thus, although implicit and explicit SGD have identical asymptotic performance, they are crucially different in their stability properties. This is investigated further in Section~\ref{section:stability} and in the experiments of Section \ref{section:experiments}.

\subsection{Asymptotic variance and optimal learning rates}
\label{section:asymptotics}
In the previous section we showed that $\thetaim{n} \to \thetastar$ in 
quadratic mean, i.e., the implicit SGD iterates converge to the true model parameters $\thetastar$, similar to classical results for the explicit SGD iterates $\thetasgd{n}$. 
Thus, $\thetaim{n}$ and $\thetasgd{n}$ are consistent estimators of $\thetastar$.
In the  following theorem we show that both SGD estimators have the 
same asymptotic variance. 

\begin{theorem}
\label{theorem:variance}
\TheoremVariance
\end{theorem}

\emph{Remarks}. Although the implicit SGD estimator $\thetaim{n}$ 
is significantly more stable than the explicit estimator $\thetasgd{n}$ (Theorem \ref{theorem:mse2}), 
both estimators have the same asymptotic efficiency in the limit according 
to Theorem \ref{theorem:variance}.
This implies that implicit SGD is a superior form of SGD, and should be 
preferred when the calculation of implicit updates in Eq.~\eqref{def:sgd:implicit} 
is computationally feasible. In Section \ref{section:applications} we show that this is possible in a large family of statistical models, and illustrate 
with several numerical experiments in Section \ref{section:experiments_theory}.

Asymptotic variance results in stochastic approximation similar to Theorem \ref{theorem:variance}
were first obtained by \citet{chung1954stochastic}, \citet{sacks1958}, and followed by \citet{fabian1968}, \citet{polyak1979adaptive}, and several other authors \citep[see also][Parts I, II]{ljung1992stochastic}. 
We contribute to this literature in two important ways.
First, our asymptotic variance result includes implicit SGD, 
which is a stochastic approximation procedure with implicitly defined updates, whereas other works consider only explicit updates.
Second, in our setting we estimate recursively the true parameters 
$\thetastar$ of a statistical model, and thus we can exploit the typical regularity conditions of \assumeMain\assumeStrong\ to derive the asymptotic variance of $\thetaim{n}$  (and $\thetasgd{n}$) in a simplified closed-form.
We illustrate the asymptotic variance results of Theorem \ref{theorem:variance} in Section \ref{section:experiments_variance}.

\subsubsection{Optimal learning rates}
\label{section:second_order}
\newcommand{\mle}{\vtheta_N^{\mathrm{mle}}}
\newcommand{\Jf}{\Fisher{\vthetastar}}
Crucially, the  asymptotic variance formula of Theorem \ref{theorem:variance} depends 
on the limit of the sequence $C_n$ used in the 
SGD procedures of Eq.~\eqref{def:sgd:explicit} and Eq.~\eqref{def:sgd:implicit}.
We distinguish two classes of procedures, one where $C_n=I$, known as 
{\em first-order procedures},
and a second class where $C_n$ is not trivial, known as {\em second-order procedures}.

%
In first-order procedures only gradients are used in the SGD procedures.
Inevitably, no matter how we set the learning rate parameter $\gamma_1$, first-order SGD procedures will lose statistical efficiency.
We can immediately verify this by comparing the asymptotic variance in Theorem \ref{theorem:variance} with the asymptotic variance of the maximum likelihood estimator (MLE), denoted by $\mle$, on a data set with 
$N$ data points $\{(X_n, Y_n)\}$, $n=1, 2, \ldots, N$.
Under the regularity conditions of \assumeMain\assumeStrong, the MLE is the asymptotically optimal unbiased estimator and 
$N \Var{\mle-\thetastar} \to \Fisher{\thetastar}^{-1}$.
By Theorem \ref{theorem:variance} and convergence of implicit SGD, it holds  
$N \Var{\thetaim{N}-\thetastar} \to \gamma_1^2  (2\gamma_1\Fisher{\thetastar}-I)^{-1} \Fisher{\thetastar}$, which also holds for $\thetasgd{N}$.
For any $\gamma_1>0$ we have as an identity that
\begin{align}
\label{eq:loss}
\gamma_1^2  (2\gamma_1\Fisher{\thetastar}-I)^{-1} \Fisher{\thetastar} \succeq \Fisher{\thetastar}^{-1}.
\end{align}
The proof is rather quick if we consider $\lambda_i \in \mathrm{eig}( \Fisher{\vthetastar})$ and note that 
 $\gamma_1^2 \lambda_i / (2\gamma_1 \lambda_i-1)$
 is the corresponding eigenvalue of the left-hand matrix in Ineq.~\eqref{eq:loss} and 
 $1/\lambda_i$ is the eigenvalue of $\Fisher{\thetastar}^{-1}$, and that $(2\gamma_1 \minEigF -1)>0$ implies that
$$\gamma_1^2 \lambda_i / (2\gamma_1 \lambda_i-1) \ge 1/\lambda_i,
$$
for every $\lambda_i \in \mathrm{eig}(\Fisher{\thetastar})$.
Therefore, both SGD estimators 
lose information and this loss can be quantified exactly by Ineq. \eqref{eq:loss}.
This inequality can also be leveraged to find the optimal choice for $\gamma_1$ given an appropriate objective. 
As demonstrated in the 
experiments in Section \ref{section:experiments}, this often suffices to achieve estimates that are comparable with MLE in statistical efficiency but with substantial computational gains.
One reasonable objective is to minimize the trace of the asymptotic variance matrix, i.e., to set $\gamma_1$ equal to
\begin{align}
\label{eq:optimal:rates}
\gamma_1^\star = \arg \min_{x > 1/2\minEigF} 
 \sum_i x^2  \lambda_i / (2x\lambda_i-1).
\end{align}
Eq. \eqref{eq:optimal:rates} is defined under the constraint $x > 1 / (2 \minEigF)$ because Theorem \ref{theorem:variance} requires $2 \gamma_1 \Fisher{\thetastar}-I$ to be positive definite.

Of course, the eigenvalues $\lambda_i$ are unknown in practice and need to be estimated from the data.
This problem has received significant attention recently and several methods exist \citep[see][and references within]{karoui2008spectrum}. 
We will use Eq. \eqref{eq:optimal:rates} extensively in our experiments (Section \ref{section:experiments}) in order to tune the SGD procedures.
However, we note that in first-order SGD procedures, knowing the eigenvalues $\lambda_i$ of $\Fisher{\thetastar}$ does not necessarily achieve statistical efficiency because of the spectral gap of $\Fisher{\thetastar}$, i.e., the ratio between 
its maximum eigenvalue $\maxEigF$ and minimum eigenvalue $\minEigF$; for instance, if $\minEigF=\maxEigF$, then the choice of learning rate parameter according to Eq.~\eqref{eq:optimal:rates} leads to statistically efficient first-order SGD procedures.
However, this case is not typical in practice, especially in many dimensions.

In {second-order} procedures, we assume non-trivial condition matrices $C_n$. 
Such procedures are called second-order because they usually leverage curvature information from the Fisher information 
matrix (or the Hessian of the log-likelihood). They are also known as {\em adaptive} procedures 
because they adapt their hyperparameters, i.e., learning rates $\gamma_n$ or condition matrices $C_n$, according to observed data.
For instance, let $C_n \equiv \Fisher{\thetastar}^{-1}$ and $\gamma_1=1$.
Plugging in $C_n = \Fisher{\thetastar}^{-1}$ in Theorem \ref{theorem:variance}, the normalized asymptotic variance of the SGD estimators 
is
\begin{align}
\gamma_1^2 (2\gamma_1 \Fisher{\thetastar}^{-1} \Fisher{\thetastar}-I)^{-1} \Fisher{\thetastar}^{-1} \Fisher{\thetastar}
\Fisher{\thetastar}^{-1} =  \Fisher{\thetastar}^{-1},\nn
\end{align}
which is the theoretically optimal asymptotic variance of the MLE, 
i.e., the Cram\'{e}r-Rao lower bound. 

Therefore, to achieve asymptotic efficiency, second-order
procedures need to estimate the Fisher information 
matrix at $\thetastar$.
Because $\thetastar$ is unknown one can simply use $C_n = \Fisher{\thetaim{n}}^{-1}$ (or $C_n= \Fisher{
\thetasgd{n-1}}^{-1}$) as 
an iterative estimate of $\Fisher{\thetastar}$, and the same 
optimality result holds.
This approach in second-order explicit SGD was first studied by \citet{sakrison1965}, and later by \citet[Chapter 8, Theorem 5.4]{nevelson1973stochastic}.
It was later extended by \citet{fabian1978asymptotically} and several other authors. 
Notably, \citet{amari1998natural} refers to the direction $\Fisher{\thetasgd{n-1}}^{-1} \nabla \log f(Y_n; X_n, \thetasgd{n-1})$
as the ``natural gradient" and uses information geometry arguments to prove statistical optimality.

An alternative way to implement second-order procedures 
is to use stochastic approximation to estimate $\Fisher{\thetastar}$, 
in addition to the approximation procedure estimating $\thetastar$.
For example, \citet{amari2000adaptive} proposed the following second-order procedure,
\newcommand{\thetamari}[1]{\theta^{\texttt{am}}_{#1}}
\newcommand{\thetada}[1]{\theta^{\texttt{ada}}_{#1}}
 \begin{align}
 \label{def:sgd:amari}
C_n^{-1} & = (1-a_n) C_{n-1}^{-1} + a_n \nabla \log f(Y_n; X_n, \thetamari{n-1})  \nabla \log f(Y_n; X_n, \thetamari{n-1})^\intercal \nn\\
\thetamari{n} & = \thetamari{n-1} + \gamma_n C_n \nabla \log f(Y_n; X_n, \thetamari{n-1}), 
\end{align}
  where $a_n = a_1/n$ is a learning rate sequence, separate from 
  $\gamma_n$.
  By standard stochastic approximation,  $C_n^{-1}$
  converges  to $\Fisher{\thetastar}$, and thus the procedure in Eq.~\eqref{def:sgd:amari} is asymptotically optimal. 
However, there are two important problems with this procedure. 
First, it is computationally costly because of matrix inversions.
A faster way is to apply quasi-Newton ideas. SGD-QN developed by \citet{bordes2009} is 
such a procedure where the first expensive matrix computations are substituted 
by the secant condition.
Second, the stochastic approximation of $\Fisher{\thetastar}$ is 
usually very noisy in high-dimensional problems and this affects the main 
approximation for $\thetastar$. Recently, more robust variants of SGD-QN have been proposed 
\citep{byrd2014stochastic}.

Another notable adaptive procedure is AdaGrad \citep{duchi2011}, 
which is defined as 
 \begin{align}
 \label{def:sgd:adagrad}
C_n^{-1} & = C_{n-1}^{-1} + \mathrm{diag}\left(
\nabla \log f(Y_n; X_n, \thetada{n-1})  \nabla \log f(Y_n; X_n, \thetada{n-1})^\intercal\right), \nn\\
\thetada{n} & = \thetada{n-1} + \gamma_1 C_n^{1/2} \nabla \log f(Y_n; X_n, \thetada{n-1}), 
\end{align}
where $\mathrm{diag}(\cdot)$ takes the diagonal matrix of its matrix 
argument, and the learning rate is set constant to $\gamma_n \equiv \gamma_1$.
AdaGrad can be  considered a second-order procedure because it tries to approximate the Fisher 
information matrix, however it only uses gradient information so technically 
it is first-order.
Under appropriate conditions, 
$C_n^{-1} \to \mathrm{diag}(\Fisher{\thetastar})$ and 
a simple modification in the proof of Theorem \ref{theorem:variance} 
can show that the 
asymptotic variance of the AdaGrad estimate is given by 
\begin{align}
\label{eq:variance_adagrad}
\sqrt{n} \Var{\thetada{n}} \to \frac{\gamma_1}{2} \mathrm{diag}(\Fisher{\thetastar})^{-1/2}. 
\end{align}
This result reveals an interesting trade-off achieved by AdaGrad 
and a subtle contrast to first-order SGD procedures.
The asymptotic variance of AdaGrad is $\bigO{1/\sqrt{n}}$, which indicates significant loss of information. 
However, this rate is attained {\em regardless} of the specification 
of the learning rate parameter $\gamma_1$.\footnote{
This follows from a property of recursions~\citep[Lemma 2.4]{toulis2016suppl}. 
On a high-level, the term $\gamma_{n-1}/\gamma_n$ is important for the variance rates of AdaGrad and SGD. When $\gamma_n\propto 1/n$, as in Theorem \ref{theorem:variance}, it holds that $\gamma_{n-1}/\gamma_n = 1 + \gamma_n/\gamma_1 + \bigO{\gamma_n^2}$, which explains the quantity $2\Fisher{\thetastar}-I/\gamma_1$ in 
first-order SGD. The rate $\bigO{1/n}$ is attained only if $2\Fisher{\thetastar}-I/\gamma_1\succ 0$. 
When $\gamma_n\propto 1/\sqrt{n}$, as in AdaGrad, it holds that $\gamma_{n-1}/\gamma_n = 1 + \littleO{\gamma_n}$ and the rate $\bigO{1/\sqrt{n}}$ is attained without any 
additional requirements.}
In contrast, as shown in Theorem \ref{theorem:variance}, first-order SGD procedures require
 $2\gamma_1\Fisher{\thetastar}-I\succ 0$ in order to achieve the $\bigO{1/n}$ rate, 
and the rate is significantly worse if this condition 
is not met. For instance, \citet{nemirovski2009robust} give 
an example of misspecification of $\gamma_1$ where the rate 
of first-order explicit SGD is $\bigO{n^{-\epsilon}}$, and 
$\epsilon$ can be arbitrarily small.
The variance result in Eq.~\eqref{eq:variance_adagrad} is illustrated in the numerical experiments of Section \ref{section:experiments_variance}.

\subsection{Optimality with averaging}
\label{section:averaging}
As shown in Section \ref{section:second_order}, Theorem \ref{theorem:variance} implies that first-order SGD procedures
can be statistically inefficient, especially in many dimensions. 
One surprisingly simple idea to achieve statistical efficiency
is to combine larger learning rates with averaging of the iterates.
In particular, we consider the procedure
\begin{align}
\label{def:aisgd}
\thetaim{n} & = \thetaim{n-1} +\gamma_n \nabla \log f(Y_n; X_n, \thetaim{n}),\nn\\
\overline{\thetaim{n}} & = \frac{1}{n} \sum_{i=1}^n \thetaim{i},
\end{align}
where $\thetaim{n}$ are the typical implicit SGD iterates 
in Eq.~\eqref{def:sgd:implicit}, and $\gamma_n = \gamma_1 n^{-\gamma}$, 
$\gamma \in [0.5, 1)$. Under suitable conditions, the iterates $\overline{\thetaim{n}}$ are asymptotically efficient.
This is formalized in the following theorem.

\begin{theorem}
\label{theorem:averaging}
\TheoremAveraging
\end{theorem}
{\em Remarks.} 
In the context of explicit stochastic approximations, 
averaging was first proposed and analyzed by \citet{ruppert1988efficient} 
and  \citet{bather1989stochastic}.  \citet{ruppert1988efficient} argued that larger learning rates in stochastic approximation 
uncorrelates the iterates allowing averaging to improve efficiency.
\citet{polyak1992} expanded the scope of averaging by proving asymptotic
optimality in more general explicit stochastic approximations that 
operate under suitable conditions similar to  Theorem \ref{theorem:averaging}. 
\citet{polyak1992} thus proved that slowly-converging stochastic approximations can be improved by using larger learning rates and averaging of the iterates. 
Recent work has analyzed explicit updates with averaging 
\citep{zhang2004solving, xu2011towards, bach2013non, shamir2012stochastic}, and has shown their superiority in numerous learning tasks.
More recently, \citet{toulis2016stability} derived the finite-sample error bounds of the averaged implicit SGD estimator.

\subsection{Asymptotic normality}
\label{section:normality}
Asymptotic distributions, or more generally invariance principles, are 
well-studied in classical stochastic approximation  \citep[Chapter II.8]{ljung1992stochastic}. 
In this section we leverage Fabian's theorem \citep{fabian1968} to show that iterates from implicit SGD are asymptotically normal. 

\begin{theorem}
\label{theorem:normality}
\TheoremNormality
\end{theorem}

{\em Remarks.} The combined results of Theorems \ref{theorem:mse2}, 
\ref{theorem:variance}, and \ref{theorem:normality} indicate that 
implicit SGD is numerically stable and has known asymptotic variance and distribution. 
Therefore, contrary to explicit SGD that has severe stability issues, 
implicit SGD emerges as a stable estimation procedure with known standard errors, 
which enables typical statistical tasks, such as confidence intervals, 
hypothesis testing, and model checking. We show empirical evidence supporting this claim 
in Section \ref{section:experiments_normality}.

\subsection{Stability}
\label{section:stability}

To illustrate the stability, or lack thereof, of both SGD estimators in small-to-moderate samples, 
we simplify the SGD procedures and inspect the size of the biases $\mathbb{E}(\thetasgd{n} - \vthetastar)$ and $\mathbb{E}(\thetaimpl{n} - \vthetastar)$.
In particular, based on Theorem \ref{theorem:mse2}, 
we simply assume the Taylor expansion
$\nabla \log f(Y_n; X_n, \theta_n) = - \Fisher{\thetastar}(\theta_n-\thetastar) + \bigO{\gamma_n}$; to simplify further we ignore the remainder term $\bigO{\gamma_n}$.

Under this simplification, the SGD procedures in Eq.~\eqref{def:sgd:explicit} and in Eq.~\eqref{def:sgd:implicit} 
can be written as follows:
\begin{align}
\label{eq:simple:recursions}
& 	\Ex{\thetasgd{n} - \vthetastar} =
\sgdM{n} 	\Ex{\thetasgd{n-1} - \vthetastar} = 
				\Pin{1}{n} b_0, \\
& 	\Ex{\thetaimpl{n} - \vthetastar} = \impM{n} 	\Ex{\thetaimpl{n-1} - \vthetastar} = 
				\Qin{1}{n} b_0,
\end{align}
where $\Pin{1}{n} = \prod_{i=1}^n \sgdM{i}$,
$\Qin{1}{n} = \prod_{i=1}^n \impM{i}$, and $b_0$ denotes 
the initial bias of the two procedures from a common starting point $\theta_0$.
Thus, the matrices $\Pin{1}{n}$ and $\Qin{1}{n}$ describe how fast the initial bias decays for the 
explicit and implicit SGD respectively. In the limit, $\Pin{1}{n} \to \m{0}$ and $\Qin{1}{n} \to \m{0}$~\citep[proof of Lemma~2.4]{toulis2016suppl}, and thus both methods are \emph{asymptotically stable}.

However, the explicit procedure has significant stability issues in small-to-moderate samples. 
By inspection of Eq. \eqref{eq:simple:recursions}, 
the magnitude of $\Pin{1}{n}$ is dominated by $\maxEigF$,
 the maximum eigenvalue of $ \Fisher{\vthetastar}$. 
 Furthermore, the rate of convergence is dominated by $\minEigF$, the minimum eigenvalue of $\Fisher{\vthetastar}$.\footnote{To see this, note that the eigenvalues of $\Pin{1}{n}$ are 
$\lambda_i' = \prod_j (1-\gamma_1 \lambda_i/j) = \bigO{n^{-\gamma_1 \lambda_i}}$
if $0 < \gamma_1 \lambda_i <1 $. See also proof of Lemma 
\ref{lemma:stability}. }
For stability, it is desirable $|1- \gamma_1 \lambda_i| < 1$,
for all eigenvalues $\lambda_i \in \eig{\Fisher{\thetastar}}$.
This implies the requirement $\gamma_1 < 2/\maxEigF$ for stability. 
Furthermore, Theorem~\ref{theorem:variance} implies the requirement $\gamma_1 > 1/2\minEigF$ for fast convergence.
This is problematic in high-dimensional settings because $\lmax$ is typically orders of magnitude larger than $\lmin$.
Thus, the requirements for stability and speed of convergence are in conflict in explicit procedures:
to ensure stability we need a small learning rate parameter $\gamma_1$, thus paying a high price in convergence which will be at the order of $\bigO{n^{-\gamma_1\lmin}}$, and vice versa.

In contrast, the implicit procedure is \emph{unconditionally stable}. 
The eigenvalues of $\Qin{1}{n}$ are $\lambda_i' = \prod_{j=1}^n 1 / (1+\gamma_1\lambda_i / j)
=\bigO{n^{-\gamma_1\lambda_i}}$. 
Critically, it is no longer required to have a small $\gamma_1$ for stability because the eigenvalues of $\Qin{1}{n}$ are always less than one.  
We summarize these findings in the following lemma.

\begin{lemma}
\label{lemma:stability}
\LemmaStability
\end{lemma}
\emph{Remarks}. Lemma \ref{lemma:stability} shows that in the explicit SGD procedure the effect from the initial bias can be amplified in an arbitrarily large way before fading out, if the learning rate is misspecified (i.e., if $\gamma_1>> 1/ \lmax$). 
This sensitivity of explicit 
SGD is well-known and requires problem-specific considerations to be avoided in practice, e.g., preprocessing, small-sample tests, projections, 
truncation \citep{chen1987convergence}. In fact, there exists 
voluminous work, which is still ongoing, 
in designing learning rates to stabilize explicit SGD; see, for example, a review 
by \citet{george2006adaptive}.
Implicit procedures render such ad-hoc designs obsolete 
because they remain stable regardless of learning rate design, and still maintain 
the asymptotic convergence and efficiency properties of explicit SGD.
%

\section{Applications}
\label{section:applications}
Here, we show how to apply 
implicit SGD in Eq.~\eqref{def:sgd:implicit} for estimation
in generalized linear models, Cox proportional hazards, 
and more general M-estimation problems.
We start by developing an algorithm that efficiently computes 
the implicit update in Eq.~\eqref{def:sgd:implicit}, and is applicable to all aforementioned models.

\subsection{Efficient computation of implicit updates}
The main difficulty in applying implicit SGD is the solution 
of the multidimensional fixed-point equation \eqref{def:sgd:implicit}.
In a large family of models where the likelihood depends on the 
parameter $\thetastar$ only through the natural parameter
$X_n^\intercal \thetastar$, the solution of the fixed-point equation 
is feasible and computationally efficient. 
We prove the general result in Theorem \ref{theorem:algo:implicit}.

For the rest of this section we will treat $\ell(X^\intercal \theta; Y)$ as a function of the natural parameter $X^\intercal \theta$ for a fixed outcome $Y$. Thus, $\ell'(X^\intercal \theta; Y)$ will refer to the first derivative of $\ell$ with respect to $X^\intercal \theta$ with fixed $Y$.


\begin{theorem}
\label{theorem:algo:implicit}
\LinearTheorem
\end{theorem}

{\em Remarks.} 
Theorem \ref{theorem:algo:implicit} implies that
 computing the implicit update in Eq.~\eqref{def:sgd:implicit} reduces to numerically solving the one-dimensional fixed-point equation for $\lambda_n$---this idea is implemented in Algorithm~\ref{algo:implicit}. As shown in the proof of Theorem~\ref{theorem:algo:implicit}, this implementation is fast because $\lambda_n$ lies on an interval $B_n$ of size $\bigO{\gamma_n}$.
\begin{algorithm}[hb!]
\caption{Efficient implementation of implicit SGD in Eq.~\eqref{def:sgd:implicit}}
\begin{algorithmic}[1]
	\normalsize
	\FORALL{$n \in \{1,2,\cdots\}$}
	\STATE \emph{\small \# compute search bounds} $B_n$
	\STATE $r_n \gets \gamma_n \ell'\left(X_n^\intercal \thetaim{n-1}; Y_n\right )$
	\STATE $B_n \gets [0, r_n]$
	\IF {$r_n \le 0$}
	\STATE $B_n \gets [r_n, 0]$
	\ENDIF
		\STATE  \emph{\small \# solve fixed-point equation by a 
		root-finding method}
	 \STATE $\xi = \gamma_n \ell'(X_n^\intercal \thetaim{n-1} + \xi X_n^\intercal C_n X_n; Y_n)$, $\xi \in B_n$
	 	\STATE $\lambda_n \gets \xi / r_n$
	 		\STATE \emph{\small \# following update is equivalent to update in Eq.~\eqref{def:sgd:implicit}}
 \STATE 	$\thetaim{n} \gets \thetaim{n-1} + \gamma_n \lambda_n \nabla \loglik{n}{\thetaim{n-1}}$
	\ENDFOR
\end{algorithmic}
\label{algo:implicit}
\end{algorithm}
We also note that Theorem~\ref{theorem:algo:implicit} can be readily extended to cases with 
linearly separable regularizers, for instance, regularizers using the $L_1$ norm $||\theta||=\sum_i |\theta_i|$. In such cases, there are additional fixed-point equations as in Step 9 of Algorithm~\ref{algo:implicit} that involve the components of the regularizer. More generally, for families of models that do not satisfy \assumeMain\assumeLinear\  there are methods to {\em approximately} perform the implicit update---we discuss one such method in Section~\ref{section:cox}.

\subsection{Generalized linear models}
\label{section:glm}

\newcommand{\xthetastar}{\mathbf{x}^\intercal \vthetastar}
\renewcommand{\xtheta}{\mathbf{x}^\intercal \vtheta}
In this section, we apply implicit SGD 
to estimate generalized linear models (GLMs).
In such models, $Y_n$ follows an exponential 
distribution conditional on $X_n$, 
and $\ExCond{Y_n}{X_n} = h(X_n^\intercal \thetastar)$,
where $h$ is the {\em transfer function} of the GLM model
\citep{nelder1972}.
%
Furthermore, the gradient of the GLM log-likelihood for parameter value $\theta$ at data point $(X_n, Y_n)$ is given by
\begin{align}
\label{glm:loglik}
\nabla \log f(Y_n; X_n, \theta) = [Y_n - h(X_n^\intercal \theta)] X_n.
\end{align}
The conditional variance of $Y_n$ is $\VarCond{Y_n}{X_n} = h'(X_n^\intercal\thetastar) X_n X_n^\intercal$, and thus the Fisher information matrix 
is $\Fisher{\theta} = \Ex{h'(X_n^\intercal \theta) X_n X_n^\intercal}$. Thus, the SGD procedures in Eq.~\eqref{def:sgd:explicit} and 
in Eq.~\eqref{def:sgd:implicit} can be written as
\begin{align}
\label{def:sgd:explicit_glm}
\thetasgd{n} & = \thetasgd{n-1} + \gamma_n C_n  [Y_n - h(X_n^\intercal \thetasgd{n-1})] X_n, \\
\label{def:sgd:implicit_glm}
\thetaim{n} & = \thetaim{n-1} + \gamma_n C_n  [Y_n - h(X_n^\intercal \thetaim{n})] X_n.
\end{align}
Implementation of explicit SGD is straightforward. 
Implicit SGD can be implemented through Algorithm \ref{algo:implicit}. In particular, $\log f(Y; X, \theta) \equiv \ell(X^\intercal \theta; Y)$
with $\ell(\eta; Y) = Y-h(\eta)$. In typical GLMs $h$ is twice-differentiable 
and also $h'(\eta) \ge 0$ because it is proportional to the 
conditional variance of $Y$ given $X$, thus fulfilling \assumeMain\assumeLinear.
In the simplified case where $C_n = I$, the identity matrix, for all $n$, 
Algorithm \ref{algo:implicit} simplifies to 
Algorithm \ref{algo:implicit_glm}, which was first derived by \citet{toulis2014statistical}.
\begin{algorithm}[b!]
\caption{Estimation of GLMs with implicit SGD}
\begin{algorithmic}[1]
	\normalsize
	\FORALL{$n \in \{1,2,\cdots\}$}
	\STATE $r_n \gets \gamma_n  \left [Y_n - h(X_n^\intercal \thetaim{n-1}) \right ]$
	\STATE $B_n \gets [0, r_n]$ \hspace{25px}  
	\IF {$r_n \le 0$}
	\STATE $B_n \gets [r_n, 0]$
	\ENDIF
	 \STATE $\xi = \gamma_n \left[ Y_n -  h \left(X_n^\intercal \thetaim{n-1} + \xi ||X_n||^2 \right) \right]$, $\xi \in B_n$
	\STATE $\thetaim{n} \gets \thetaim{n-1} + \xi X_n$ 
	\ENDFOR
\end{algorithmic}
\label{algo:implicit_glm}
\end{algorithm}
We make extensive experiments using Algorithm \ref{algo:implicit_glm} 
in Section \ref{section:experiments_performance}.

\subsection{Cox proportional hazards model}
\label{section:cox}
Here, we apply SGD to estimate a
Cox proportional hazards model, which is 
a popular model in survival analysis~\citep{cox1972regression, klein2003survival}.
Multiple variations of the model exist but for simplicity we will analyze 
one simple variation that is popular in practice \citep{davison2003statistical}.
Consider $N$ individuals, indexed by $i$, with observed survival times $Y_i$,
failure indicators $d_i$, and covariates $X_i$.
The survival times can be assumed ordered, $Y_1 <Y_2 \ldots < Y_N$,
whereas $d_i=1$ denotes failure (e.g., death) and $d_i=0$ indicates 
censoring (e.g., patient dropped out of study).
Given a failure for unit $i$ ($d_i=1$) 
at time $Y_i$, the {\em risk set} $\mathcal{R}_i$ 
is defined as the set of individuals that could possibly fail at $Y_i$, i.e., 
all individuals except those who failed or were censored before $Y_i$.
In our simplified model, $\mathcal{R}_i = \{i, i+1, \ldots, N\}$.
Define $\eta_i(\theta)  = \exp(X_i^\intercal \theta)$, 
then the log-likelihood $\ell$ for  $\theta$ 
is given by \citep[Chapter 10]{davison2003statistical}
\begin{align}
\label{cox:loglik}
\ell(\theta; X, Y) = \sum_{i=1}^N [d_i - H_i(\theta) \eta_i(\theta)] X_i,
\end{align}
where $H_i(\theta)  =\sum_{j : i \in \mathcal{R}_j} d_j (\sum_{k \in \mathcal{R}_j} \eta_k(\theta))^{-1}$.
In an online setting, where $N$ is infinite and data points $(X_i, Y_i)$ are observed one at a time, future observations affect the likelihood of previous ones, as can be seen by inspection of Eq. \eqref{cox:loglik}. 
Therefore, we apply SGD assuming fixed $N$ to estimate the MLE $\mle$. 
As mentioned in Section \ref{section:introduction}, our theory in Section \ref{section:theory} can be applied unchanged if we only substitute $\thetastar$, the true parameter, with the MLE $\mle$.

A straightforward implementation of explicit SGD in Eq.~\eqref{def:sgd:explicit}
 for the Cox model is 
shown in Algorithm \ref{algo:explicit_cox}. 
For implicit SGD in Eq.~\eqref{def:sgd:implicit} we have the update
\begin{align}
\label{eq:bad_implicit}
\thetaim{n} = \thetaim{n-1} + \gamma_n [d_i-H_i(\thetaim{n}) \eta_i(\thetaim{n})] X_i,
\end{align}
which is similar to the implicit procedure for GLMs in Eq.~\eqref{def:sgd:implicit_glm}.
However, the log-likelihood term $d_i-H_i(\thetaim{n}) \eta_i(\thetaim{n})$ 
does not satisfy the conditions of \assumeMain\assumeLinear\ because
$H_i(\theta)$ may be increasing or decreasing 
since it depends on terms $X_j^\intercal\theta, j \neq i$.
Thus, Theorem \ref{theorem:algo:implicit} cannot be applied.

One way to circumvent this problem is to simply compute 
$H_i(\cdot)$ on the previous update $\thetaim{n-1}$ instead 
of the current $\thetaim{n}$. Then, update \eqref{eq:bad_implicit} becomes
\begin{align}
\label{def:sgd:implicit_cox}
\thetaim{n} = \thetaim{n-1} + \gamma_n [d_i-H_i(\thetaim{n-1}) \eta_i(\thetaim{n})] X_i,
\end{align}
which now satisfies \assumeMain\assumeLinear\ since $H_i(\thetaim{n-1})$ 
is constant with respect to $\thetaim{n}$. In fact, this idea can be used to apply implicit SGD more generally beyond models that satisfy \assumeMain\assumeLinear;
see Section \ref{section:discussion} for a discussion.
%
\vspace{-10px}
\begin{figure}[h!]
\begin{center}
\begin{minipage}[t]{.5\textwidth}
 \begin{algorithm}[H]
 \label{algo:explicit_cox}
    \caption{\footnotesize Explicit SGD for Cox proportional hazards  model}
	\For{$n=1,2,\ldots$}{
		$i \leftarrow \text{sample}(1, N)$\\
		$\widehat{H}_i \leftarrow \sum_{j : i \in \mathcal{R}_j}
				 \frac{d_j}{\sum_{k \in \mathcal{R}_j} \eta_k(\thetasgd{n-1})}$\\
		$w_{n-1} \leftarrow  \left[d_i - \widehat{H}_i \eta_i(\thetasgd{n-1})\right] $\\
		$ \thetasgd{n} = \thetasgd{n-1} + \gamma_n w_{n-1} C_n X_i$ \\
		\vspace{20.2pt}
	}
  \end{algorithm}
\end{minipage}%
\begin{minipage}[t]{.5\textwidth}
 \begin{algorithm}[H]
    \caption{\footnotesize Implicit SGD for Cox proportional hazards model}
     \label{algo:implicit_cox}
    \For{$n=1,2,\ldots$}{
		$i \leftarrow \text{sample}(1, N)$\\
		$\widehat{H}_i \leftarrow \sum_{j : i \in \mathcal{R}_j}
				 \frac{d_j}{\sum_{k \in \mathcal{R}_j} \eta_k(\thetaim{n-1})}$\\
		$w(\theta) = d_i - \widehat{H}_i \eta_i(\theta)$ \\
		$W_n \leftarrow w(\thetaim{n-1}) C_n X_i$ \\
		$ \lambda_n w(\thetaim{n-1}) = 
			w\left(\thetaim{n-1} + \gamma_n \lambda_n W_n\right) $\\
		$ \thetaim{n} = \thetaim{n-1} + \gamma_n \lambda_n W_n$
	}
  \end{algorithm}
 \end{minipage}
\end{center}
\end{figure}


\subsection{M-Estimation}
\label{section:m}
Given $N$ observed data points $(X_i, Y_i)$ 
and a convex function $\rho : \Reals{} \to \Reals{+}$, 
the M-estimator is defined as 
\begin{align}
\label{def:m}
\hat{\theta}^m = \arg \min_\theta \sum_{i=1}^N \rho(Y_i - X_i^\intercal \theta),
\end{align}
where it is assumed $Y_i = X_i^\intercal \thetastar + \epsilon_i$, 
and $\epsilon_i$ are i.i.d. zero mean-valued noise.
M-estimators are especially useful in robust statistics 
\citep{huber1964robust} because 
appropriate choice of $\rho$ can 
reduce the influence of outliers in data.
Typically, $\rho$ is twice-differentiable around zero. In this case,
\begin{align}
\Ex{\rho'(Y - X^\intercal \hat{\theta}^m) X} = 0,
\end{align}
where the expectation is over the empirical data distribution.
Thus, according to Section \ref{section:introduction},
SGD procedures can be applied to approximate 
the M-estimator $\hat{\theta}^m$.
There has been increased interest in the literature  for fast approximation of M-estimators due  to their robustness~\citep{donoho2013high, jain2014iterative}.
The implicit SGD procedure for approximating M-estimators is
defined in Algorithm \ref{algo:implicit_m}, 
and is a simple adaptation of Algorithm \ref{algo:implicit}.

Importantly, the conditions of \assumeMain\assumeLinear\ are met because $\rho$ is convex and thus $\rho'' \ge 0$. Thus, Step 4  of Algorithm \ref{algo:implicit_m} 
is a straightforward application of Algorithm \ref{algo:implicit} 
by simply setting $\ell'(X_n'\thetaim{n-1}; Y_n) \equiv 
\rho'(Y_n-X_n^\intercal \thetaim{n})$. 
The asymptotic variance of $\thetaim{n}$ is also easy to derive.
If $S = \Ex{X_n X_n^\intercal}$, $C_n \to C>$ 
such that $S$ and $C$ commute, $\psi^2 = \Ex{\rho'(\epsilon_i)^2}$, 
and $v(z) = \Ex{\rho'(\epsilon_i+z)}$, Theorem \ref{theorem:variance} 
can be leveraged to show that 
\begin{align}
\label{var:m}
n \Var{\thetaim{n}} \to \psi^2 (2v'(0) C S - I)^{-1} C S C.
\end{align}
Historically, one of the first applications of explicit stochastic approximation 
procedures in robust estimation was due to \citet{martin1975robust}.
The asymptotic variance  \eqref{var:m} was first 
derived, only for the explicit SGD case, by \citet{poljak1980robust} using stochastic approximation
 theory from \citet{nevelson1973stochastic}.

\begin{algorithm}[t!]
\normalsize
    \caption{Implicit SGD for M-estimation}
     \label{algo:implicit_m}
    \For{$n=1,2,\ldots$}{
		$i \leftarrow \text{sample}(1, N)$\\
		$w(\theta) = \rho'(Y_i-X_i^\intercal \theta)$ \\
		$ \lambda_n w(\thetaim{n-1}) \leftarrow 
			w\left(\thetaim{n-1} + \gamma_n \lambda_n w(\thetaim{n-1}) C_n X_i\right)$ \quad {\em\#  implicit update} \\
$\thetaim{n} \leftarrow \thetaim{n-1} + \gamma_n \lambda_n  w(\thetaim{n-1}) C_n X_i$}
  \end{algorithm}


\section{Simulation and data analysis}
\label{section:experiments}
\renewcommand{\Vx}{\m{V}_{\texttt{x}}}
\newcommand{\glm}{\texttt{glm()}}
In this section, we demonstrate the computational and statistical 
advantages of SGD estimation procedures in Eq.~\eqref{def:sgd:explicit} and in Eq.~\eqref{def:sgd:implicit}.
For our experiments we developed a new \texttt{R} package, namely \texttt{sgd},
which has been published on CRAN. All experiments were conducted on a single laptop  running Linux Ubuntu 13.x 
with 8 cores{@}2.4GHz, 16Gb of RAM memory 
and 256Gb of physical storage with SSD technology.
A separate set of experiments, which is presented in the supplemental article~\citep[Section 3]{toulis2016suppl},
focuses on comparisons of implicit SGD with popular machine learning methods on typical estimation tasks.

\subsection{Numerical results}
\label{section:experiments_theory}

In this section we aim to illustrate the theoretical results of Section \ref{section:theory}, 
namely the result on asymptotic variance 
(Theorem \ref{theorem:variance}) and asymptotic normality (Theorem \ref{theorem:normality}) of SGD procedures.
%


\subsubsection{Asymptotic variance}
\label{section:experiments_variance}
In this experiment we use a normal linear model following~\citet{xu2011towards}.
The procedures we test are explicit SGD in Eq.~\eqref{def:sgd:explicit}, 
implicit SGD in Eq.~\eqref{def:sgd:implicit}, and AdaGrad in Eq.~\eqref{def:sgd:adagrad}.
For simplicity we use first-order SGD, i.e., $C_n=I$.
In the experiment we calculate the empirical variance of said procedures 
for 25 values of their common learning rate parameter $\gamma_1$ 
in the interval $[1.2, 10]$. 
For every value of $\gamma_1$ we calculate the 
empirical variances through the following process, repeated for 150 times.
First, we set $\thetastar = (1, 1, \cdots, 1)^\intercal \in \Reals{20}$
 as the true parameter value. 
For iterations $n=1, 2,\ldots, 1500$, we sample 
covariates as $X_n \sim \mathcal{N}_p(\m{0}, S)$, 
where $S$ is diagonal with elements uniformly on $[0.5, 5]$.
The outcome $Y_n$ is then sampled as 
$Y_n| X_n \sim \mathcal{N}(X_n^\intercal \thetastar, 1)$.
In every repetition we store the iterate $\theta_{1500}$ for every 
tested procedure and then calculate the  empirical variance 
of stored iterates over all 150 repetitions. 

For any fixed learning rate parameter $\gamma_1$ we set $\gamma_n=\gamma_1/n$ for implicit SGD and $\gamma_n=\gamma_1$ for AdaGrad.
For explicit SGD we set $\gamma_n  =\min(0.3, \gamma_1 / (n + ||X_n||^2)$ in order to stabilize its updates. This trick is necessary by the analysis of Section \ref{section:stability}. In particular, the Fisher information matrix here
is $\Fisher{\thetastar} = \Ex{X_nX_n^\intercal} = S$, and thus 
the minimum eigenvalue is $\lmin=0.5$ and the maximum is $\lmax=5$.
Therefore, for stability we require $\gamma_1  < 2/\lmax=0.4$ and for 
fast convergence we require $\gamma_1 > 1/(2 \lmin) = 1$. 
The two requirements are incompatible, which indicates that explicit SGD 
can have serious stability issues.

\begin{figure}[t]
 \centering
 \includegraphics[width=\textwidth]{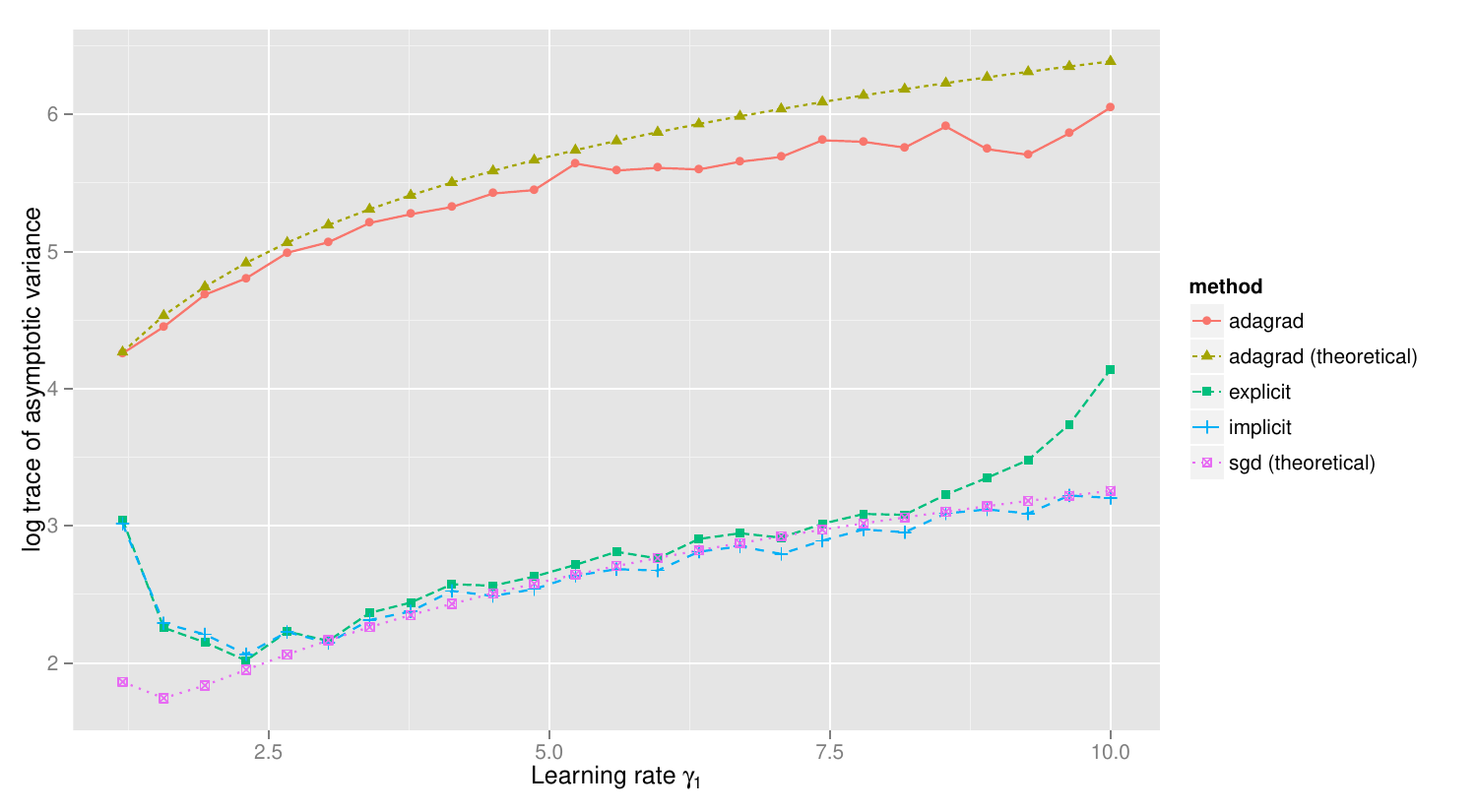}
 \caption{Simulation with normal model. The x-axis corresponds to 
learning rate parameter $\gamma_1$; the y-axis curves corresponds to 
 log trace of the empirical variance of tested procedures (explicit/implicit SGD, AdaGrad).
  Theoretical asymptotic variances of SGD and AdaGrad are plotted as well.
 Implicit SGD is stable and its empirical variance is very close
  to its asymptotic value. Explicit SGD becomes unstable at large $\gamma_1$.
AdaGrad is statistically inefficient but remains stable to large learning rates. \label{figure:normal}}
 \end{figure}

For given $\gamma_1 > 1$, the asymptotic variance of SGD procedures after $n$ iterations is $(1/n) \gamma_1^2 (2\gamma_1 S - I)^{-1} S$, by Theorem \ref{theorem:variance}.
The asymptotic variance of AdaGrad after $n$ iterations is equal to
$(\gamma_1/2\sqrt{n}) S^{-1/2}$ by Eq.~\eqref{eq:variance_adagrad}.
 The log traces of the empirical variance of the SGD procedures 
 and AdaGrad in this experiment are shown in Figure \ref{figure:normal}.
The x-axis corresponds to different values of 
the learning rate parameter $\gamma_1$, 
and the y-axis corresponds to the log 
trace of the empirical variance of the iterates for all three different procedures. 
We also include curves for the theoretical  values
of the empirical variances.

We see that our theory predicts well the empirical variances of all 
methods. Explicit SGD performs on par with implicit SGD 
for moderate values of $\gamma_1$, however, it required 
a modification in its learning rate to make it work. 
Furthermore, explicit SGD quickly becomes unstable at larger values of $\gamma_1$
(see, for example, its empirical variance for $\gamma_1=10$), 
and in several instances, not considered in Figure \ref{figure:normal}, it numerically diverged.
On the other hand, AdaGrad is stable to the specification 
of $\gamma_1$ and tracks its theoretical variance well. 
However, it gives inefficient estimators because their variance 
has order $\bigO{1/\sqrt{n}}$.
Implicit SGD effectively combines stability and good statistical efficiency.
First, it remains very stable to the entire range of the learning 
rate parameter $\gamma_1$. Second, its empirical variance is $\bigO{1/n}$ and is 
tracks closely the theoretical value predicted by Theorem \ref{theorem:variance} for 
all $\gamma_1$.

\subsubsection{Asymptotic normality}
\label{section:experiments_normality}

In this experiment we use the normal linear model 
in the setup of Section \ref{section:experiments_variance} 
to check the asymptotic normality result of Theorem \ref{theorem:normality}.
For simplicity, we only test first-order implicit SGD in Eq.~\eqref{def:sgd:implicit} 
and first-order explicit SGD.
%

In the experiment we define a set of learning rates $(0.5, 1, 3, 5, 6, 7)$.
For every learning rate 
we take 400 samples of $N (\theta_N-\thetastar)^\intercal \Sigma^{-1} (\theta_N-\thetastar)$, where $N=1200$ and $\theta_N$ denotes either $\thetasgd{N}$ or $\thetaim{N}$.
The matrix $\Sigma$ is the asymptotic variance matrix in Theorem~\ref{theorem:normality}, and $\thetastar = 
10 \exp\left(-2 \cdot(1, 2, \ldots, p)\right)$, is the true parameter value.
We use the ground-truth values both for $\Sigma$ and $\thetastar$, 
as we are only interested to test normality of the iterates in this experiment.
We also tried $p=5, 10, 100$ as the parameter dimension. 
Because the explicit SGD procedure was very unstable across experiments 
we only report results for $p=5$. Results on the implicit procedure 
for larger $p$ are given in the supplemental article~\citep{toulis2016suppl}, 
where we also include results for a logistic regression model.

By Theorem \ref{theorem:normality} for implicit SGD, and by classical normality results for explicit SGD~\citep{fabian1968, ljung1992stochastic}, the quadratic form $N (\theta_{N}-\thetastar)^\intercal \Sigma^{-1} (\theta_{N}-\thetastar)$ is a chi-squared random variable 
with $p$ degrees of freedom. Thus, for every procedure 
we plot this quantity against independent samples from a $\chi^2_p$ distribution and visually check for deviations.
As before, we tried to stabilize explicit SGD as much as possible 
by setting $\gamma_n  =\min(0.3, \gamma_1 / (n + ||X_n||^2))$. This worked in many iterations, but not for all. 
Iterations for which explicit SGD diverged were not considered.
For implicit SGD we simply set $\gamma_n=\gamma_1/n$ without 
additional tuning.

\begin{figure}[t]
 \centering
 \includegraphics[width=\textwidth]{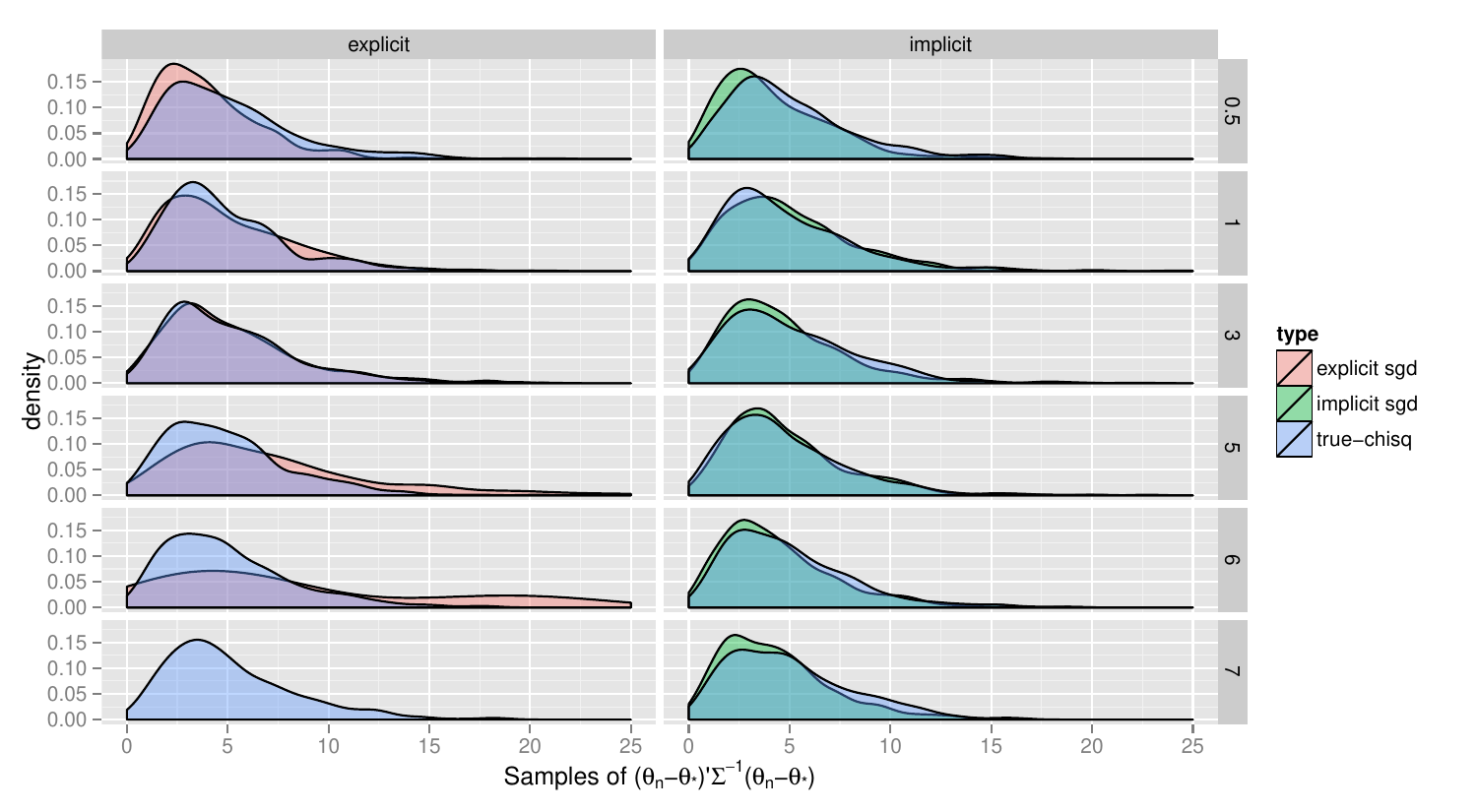}
 \caption{Simulation with normal model. The x-axis corresponds to 
the SGD procedure (explicit or implicit) 
for various values of the learning rate parameter, $\gamma_1 \in \{0.5, 1, 3, 5, 7\}$. 
The histograms (x-axis) for the SGD procedures are 
500 replications of SGD where at each replication 
we only store the quantity $N (\theta_{N}-\thetastar)^\intercal \Sigma^{-1} (\theta_{N}-\thetastar)$, for every method ($N=1200$);
the theoretical covariance matrix $\Sigma$ is different for every learning rate and 
is given in Theorem \ref{theorem:variance}.  
The data generative model is the same as in Section \ref{section:experiments_variance}.
We observe that implicit SGD is stable and follows the nominal chi-squared distribution.
Explicit SGD becomes unstable at larger $\gamma_1$ and its distribution does 
not follow the nominal one well. In particular, 
the distribution of $N (\thetasgd{N}-\thetastar)^\intercal \Sigma^{-1} (\thetasgd{N}-\thetastar)$  becomes increasingly heavy-tailed as 
the learning rate parameter gets larger, and eventually diverges for $\gamma_1\ge 7$.}
\label{figure:normality}
 \end{figure}

 The results of this experiment are shown in Figure \ref{figure:normality}.
The vertical axis on the grid corresponds to 
different values of the learning rate parameter $\gamma_1$,
and the horizontal axis has histograms 
of $N (\theta_{N}-\thetastar)^\intercal \Sigma^{-1} (\theta_{N}-\thetastar)$, and 
also includes samples from a $\chi^2_5$ distribution for visual comparison.

We see that the distribution $N (\thetaim{N}-\thetastar)^\intercal \Sigma^{-1} (\thetaim{N}-\thetastar)$ of the implicit iterates follows the nominal chi-squared distribution. This also seems to be 
unaffected by the learning rate parameter.
However, the distribution of $N (\thetasgd{N}-\thetastar)^\intercal \Sigma^{-1} (\thetasgd{N}-\thetastar)$ does not follow a chi-squared distribution, except for small 
learning rate parameter values. 
For example, as the learning rate parameter increases, 
the distribution becomes more heavy-tailed (e.g., for $\gamma_1=6$), 
indicating that explicit SGD becomes unstable.
Particularly for $\gamma_1=7$ explicit SGD diverged in almost all replications, 
and thus a histogram could not be constructed.
 
\subsection{Comparative performance results}
\label{section:experiments_performance}
In this section we aim to illustrate the performance of
implicit SGD estimation against deterministic estimation procedures 
that are optimal. 
%
%
The goal is to investigate the extent to which implicit SGD can be as fast as deterministic methods, and to quantify how much statistical efficiency needs be sacrificed to accomplish that. 

\subsubsection{Experiments with \glm\ function}
\label{section:experiment:glm}
The built-in function \glm\ in \texttt{R}
performs deterministic maximum-likelihood estimation through iterative reweighted 
least squares. 
In this experiment, we wish to compare computing time and MSE 
between first-order implicit SGD and \glm.
Our simulated data set is a simple normal linear model constructed as follows. 
First, we sample a binary $p \times p$ design matrix $\m{X} = (x_{ij})$ such that $x_{i1} = 1$ (intercept) and $P(x_{ij}=1) = s$ i.i.d, where $s \in (0, 1)$ determines the sparsity of $\m{X}$. 
We set $s=0.08$ indicating that roughly 8\% of the $\m{X}$ matrix will be nonzero. %
We generate $\thetastar$ by sampling  $p$ elements from $(-1, -0.35, 0, 0.35, 1)$ with replacement.
The outcomes are $Y_i = X_i^\intercal \thetastar + \epsilon_i$, where 
$\epsilon_i \sim \mathcal{N}(0, 1)$ i.i.d., and $X_i = (x_{ij})$ is the $p\times1$ vector of $i$'s covariates. 
By GLM properties, 
\[ \Fisher{\vthetastar} = \Ex{h'(X_i^\intercal \thetastar) X_i X_i^\intercal} = \left( \begin{array}{ccccc}
1 & s & s &  \cdots & s \\
s & s & s^2 & \cdots & s^2\\
s & s^2 & s & s^2 & \cdots \\
\cdots & s^2 & \cdots & s & \cdots \\
s & s^2 & \cdots & \cdots  & s \\
 \end{array} \right).\]

Slightly tedious algebra can show that the eigenvalues of $\Fisher{\vthetastar}$ are 
$s (1-s)$ with multiplicity $(p-2)$ and the two solutions of $x^2 - A(s) x + B(s)=0$, where 
$A(s)=1+s + s^2(p-2)$ and $B(s) = s(1-s)$.
It is thus possible to use the analysis of Section \ref{section:asymptotics} and Eq. \eqref{eq:optimal:rates} to derive a theoretically optimal learning rate.
We sample 200 pairs $(p,N)$ for the problem size, 
uniformly in the ranges $p \sim [10, 500]$ and $N \sim [500, 50000]$, and obtain running times and MSE of the estimates 
from implicit SGD and \texttt{glm()}. 
Finally, we then run a regression of computing time and MSE 
against the problem size $(N, p)$. 

 The results are shown in Table 
\ref{table:experiment:glm}. We observe that implicit SGD scales better in both
sample size $N$, and especially in the model size $p$. 
We also observe that this significant computational gain does not come with much efficiency loss. 
In fact, averaged over all samples, the MSE of the implicit SGD is 10\% higher than the MSE of 
\texttt{glm()}, with a standard error of $\pm 0.005$.
Furthermore, the memory requirements (not reported in Table \ref{table:experiment:glm}) are roughly $\bigO{N p^2}$ for \texttt{glm()} and only $\bigO{p}$ for implicit SGD. 

\renewcommand*{\arraystretch}{1.3}
\begin{table}[t]
\caption{Parameters from regressing computation 
time and MSE against $(N, p)$ in $\log$-scale for \texttt{glm()} and implicit GLM. Computation time for \texttt{glm()} 
is roughly $\bigO{p^{1.47} N}$ and for implicit SGD, it is $\bigO{p^{0.2} N^{0.9}}$.
Implicit SGD scales better in parameter dimension $p$, whereas MSE for both methods are comparable, at the order of $\bigO{\sqrt{p/N}}$.
}
\label{table:experiment:glm}
\begin{center}
\begin{small}
\begin{sc}
\begin{tabular}{l | cc | cc}
method   & \multicolumn{2}{c}{Time(sec)} & \multicolumn{2}{c}{MSE} \\
& $\log p$ (se) & $\log N$ (se) & $\log p$  (se) & $\log N$ (se)  \\
\hline
\texttt{glm()} function & 1.46 (0.019) & 1.03 (0.02)  & 0.52 (0.007) & -0.52 (0.006) \\
implicit SGD & 0.19 (0.012) & 0.9 (0.01) & 0.58 (0.007) & -0.53 (0.006) \\
\end{tabular}
\end{sc}
\end{small}
\end{center}
\vskip 0.0in
\end{table}

\subsubsection{Experiments with \biglm}
\label{section:experiment:biglm}
The package \biglm\ is a popular choice for fitting GLMs with data sets where $N$ is large but $p$ is small.\footnote{See \url{http://cran.r-project.org/web/packages/biglm/index.html} for the \biglm\ package. \biglm\ is part of the High-Performance Computing (HPC) task view of the CRAN project here~\url{http://cran.r-project.org/web/views/HighPerformanceComputing.html}.}
It works in an iterative way by splitting the data set in many parts, 
and by updating the model parameters using incremental QR decomposition \citep{miller1992algorithm}, which results in only $\bigO{p^2}$ memory requirement.
In this experiment, we compare implicit SGD with \biglm\ on larger 
data sets of Section \ref{section:experiment:glm}.
  with small $p$ and large $N$ such that $Np$ remains roughly constant. 
 
 The results are shown in Table \ref{table:experiment:biglm}. 
 We observe that implicit SGD is significantly faster at a very small
efficiency loss. The difference is more dramatic at large $p$; for example, when $p=10^3$ or $p=10^4$,
\biglm\ quickly runs out of memory, whereas  implicit SGD works without problems.

\renewcommand*{\arraystretch}{1.25}
\begin{table}[t]
\caption{Comparison of implicit SGD with \biglm. 
MSE is defined as $||\theta_N-\thetastar||/\theta_0-\thetastar||$.
Values ``*" indicate out-of-memory errors.
\biglm\ was run in combination with the \texttt{ffdf} package to map big data files to memory. 
Implicit SGD used a similar but slower ad-hoc method. The table reports computation times excluding file access.
} 
\label{table:experiment:biglm}
\begin{center}
\begin{small}
\begin{sc}
\begin{tabular}{ccc | cc | cc}
 & & & \multicolumn{4}{c}{Procedure} \\
 & & &  \multicolumn{2}{c}{\biglm} & \multicolumn{2}{c}{Implicit SGD} \\
$p$ & $N$ & size (GB) & time(secs) & 
MSE & time(secs) & MSE \\
\hline
1e2 & 1e5 & 0.021  &  2.32 & 0.028  & 2.4 & 0.028 \\ 
1e2 & 5e5 & 0.103 &  8.32 &  0.012 & 7.1  & 0.012 \\ 
1e2 & 1e6 & 0.206 &   16 & 0.008 &  14.7& 0.009 \\ 
1e2 & 1e7 & 2.1 & 232 & 0.002 & 127.9 & 0.002 \\
1e2 & 1e8 & 20.6  &  * & *  & 1397 & 0.00 \\ 
1e3 & 1e6 & 2.0 &  * & * &  31.38 & 0.153 \\ 
1e4 & 1e5 & 2.0 &  * & * & 25.05 &  0.160 \\ 
\end{tabular}
\end{sc}
\end{small}
\end{center}
\vskip 0.0in
\end{table}

\subsubsection{Experiments with \texttt{glmnet}}
\label{section:experiment:glmnet}
The \texttt{glmnet} package in R \citep{friedman2010regularization}
is a deterministic optimization algorithm for generalized linear models that uses the elastic net.
It performs a component-wise update of the parameter vector, utilizing thresholding from the regularization penalties for more computationally
efficient updates.
One update over all parameters costs roughly $\bigO{N p}$ operations. Additional computational gains are achieved when the design matrix is sparse because fewer components are updated per each iteration.

In this experiment, we compare  implicit SGD with \texttt{glmnet} on a subset of experiments in the original package release \citep{friedman2010regularization}.
In particular, we implement the experiment of Subsection 5.1 in that paper, as follows.
First, we sample the design matrix $\m{X} \sim \mathcal{N}_p(0, \m{\Sigma})$, where $\m{\Sigma} = b^2 \m{U} + \m{I}$ 
and $U$ is the $p \times p$ matrix of ones.
The parameter $b = \sqrt{\rho / (1-\rho)}$, where $\rho$ is the target correlation of columns of $\m{X}$,
is controlled in the experiments.
The outcomes are $Y = \m{X} \thetastar + \sigma^2 \epsilon$,
where $\theta^*_j = (-1)^j \exp(-2(j-1) /20)$,
 and $\epsilon$ is a standard $p$-variate normal.
The parameter $\sigma$ is tuned to achieve a pre-defined signal-noise ratio. We report average computation times in Table \ref{table:experiment:glmnet:gaussian} over 10 replications, 
which expands Table 1 of \citet{friedman2010regularization}.

First, we observe that implicit SGD is consistently faster than the \texttt{glmnet} method.
In particular, the SGD method scales better at larger $p$ following a
sublinear growth as noted in Section \ref{section:experiment:glm}. 
Interestingly, it is also not affected by covariate correlation, whereas
\texttt{glmnet} gets slower as more components need to be updated at every iteration. For example, with correlation $\rho=0.9$ and $N=1e5$, $p=200$, the SGD method is almost 10x faster. 

Second, to compare \texttt{glmnet} with implicit SGD in terms of MSE 
we picked the median MSE produced by the grid of regularization parameters 
computed by \texttt{glmnet}. We picked the median because \texttt{glmnet} is a deterministic method and so at the best regularization value its MSE will be lower than the MSE of implicit SGD. 
However, implicit SGD seems to perform better against the median performance of \texttt{glmnet}.
Furthermore, Table \ref{table:experiment:glmnet:gaussian} indicates 
a clear trend where, for bigger dimensions $p$ and higher correlation $\rho$,
implicit SGD is performing better than \texttt{glmnet} in terms of efficiency as well.
We obtain similar results in a comparison on a logistic regression model, 
which we present in Section 3 of the supplemental article~\citep{toulis2016suppl}.

\renewcommand*{\arraystretch}{0.75}
\begin{table}[t]
\caption{Comparing implicit SGD with \texttt{glmnet}. 
Table reports running times (in secs.) and MSE for both procedures. 
The MSE of \texttt{glmnet} is calculated as the median MSE over the 100 grid values 
of regularization parameter computed by default \citep{friedman2010regularization}.}
\label{table:experiment:glmnet:gaussian}
\begin{center}
\begin{small}
\begin{sc}
 \begin{tabular}{l ccccc }
 method & metric & \multicolumn{ 4 }{c}{correlation ($\rho$)} \\
 & & 0 & 0.2 & 0.6 & 0.9 \\
 & &  \cline{1-4} \\
 & &  \multicolumn{4}{c}{$N=1000, p=10$}  \\
 & & \cline{1-4} \\
 \multirow{2}{*}{\texttt{glmnet}}  & time(sec) & 0.005 & 0.005 & 0.008 & 0.022 \\
                 & mse & 0.083 & 0.085 & 0.099 & 0.163 \\
  &  &  &  &  \\
 \multirow{2}{*}{\texttt{sgd}}  & time(sec) &  0.011 & 0.011 & 0.011 & 0.011 \\
                 & mse & 0.042 & 0.042 & 0.049 & 0.053 \\
 & &  \cline{1-4} \\
 & & \multicolumn{4}{c}{$N=5000, p=50$}  \\
 &&  \cline{1-4} \\
 \multirow{2}{*}{\texttt{glmnet}}  &&  0.058 & 0.067 & 0.119 & 0.273 \\
                 & & 0.044 & 0.046 & 0.057 & 0.09 \\
 &  &  &  &  &  \\
 \multirow{2}{*}{\texttt{sgd}}  &&  0.059 & 0.056 & 0.057 & 0.057 \\
                 &&  0.019 & 0.02 & 0.023 & 0.031 \\
 &&  \cline{1-4} \\
 &&  \multicolumn{4}{c}{$N=100000, p=200$}  \\
 &&  \cline{1-4} \\
 \multirow{2}{*}{\texttt{glmnet}}  &&  2.775 & 3.017 & 4.009 & 10.827 \\
                 &&  0.017 & 0.017 & 0.021 & 0.033 \\
  & &  &  &  &  \\
 \multirow{2}{*}{\texttt{sgd}}  & & 1.475 & 1.464 & 1.474 & 1.446 \\
                 &&  0.004 & 0.004 & 0.004 & 0.006 \\
 \end{tabular}
\end{sc}
\end{small}
\end{center}
\vskip -0.2in
\end{table}

\subsubsection{Cox proportional hazards}
\label{section:experiments_cox}
In this experiment we test the performance of
implicit SGD on estimating the parameters of a Cox proportional hazards model 
in a setup that is similar to the numerical example of \citet[Section 3]{simon2011regularization}.

We consider $N=1000$ units 
with covariates $X \sim \mathcal{N}(0, \Sigma)$, 
where $\Sigma = 0.2 U + I$, and $U$ is the matrix of ones.
We sample times as $Y_i \sim \mathrm{Expo}\left(\eta_i(\thetastar)\right)$,
where $\eta_i(\theta) = \exp(X_i^\intercal \theta)$, 
and $\thetastar = (\theta_{\star,k})$ is a vector with $p=20$ elements 
defined as $\theta_{\star, k}= 2 (-1)^{-k} \exp(-0.1k)$.
Time $Y_i$ is censored, and thus $d_i=0$, 
according to probability $\left(1+\exp(-a (Y_i-q)\right)^{-1}$,
where $q$ is a quantile of choice (set here as $q=0.8$), 
and $a$ is set such that $\min\{Y_i\}$ is censored with 
a prespecified probability (set here as 0.1\%).
We replicate 50 times the following process.
First, we run implicit SGD for $2 N$ iterations, and then measure MSE $||\thetaim{n}-\thetastar||^2$, for all $n =1, 2, \ldots 2N$. To set the learning rates we use Eq. \eqref{eq:optimal:rates}, 
where the Fisher matrix is diagonally approximated, through 
the AdaGrad procedure \eqref{def:sgd:adagrad}.
We then take the 5\%, 50\% and 95\% quantiles of MSE across all repetitions and plot them against iteration number $n$.

\begin{figure}[t]
\centering
\includegraphics[width=0.51\textwidth]{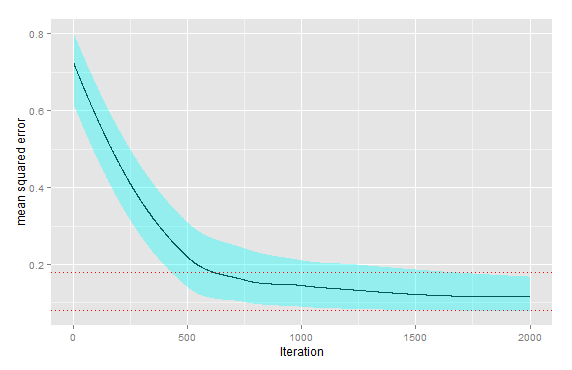}
\includegraphics[width=0.48\textwidth]{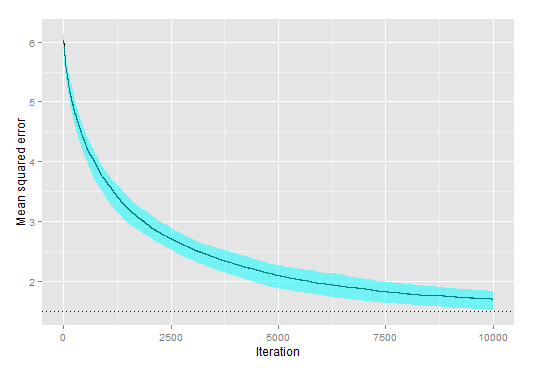}
\vspace{-10px}
\caption{{Left panel:} 5\%-95\% quantile band of implicit SGD estimates (in cyan) against 5\%-95\% band of the MLE (dashed lines) for a Cox proportional hazards model (50 replications); {Right panel:} 5\%-95\% quantile band of implicit SGD estimates (in cyan) against median MLE (dashed line) on an M-estimation task (100 replications).}
\label{figure:example_cox}
\end{figure}

The results are shown in Figure \ref{figure:example_cox} (left panel).
In the figure we also plot (horizontal dashed lines) the 
5\% and 95\% quantiles of the MSE of the MLE, assumed 
to be the best MSE achievable for SGD.
We observe that implicit SGD performs well compared to MLE
in this small-sized problem. In particular, 
implicit SGD, under the aforementioned generic 
tuning of learning rates, converges to the region of optimal MLE in a few thousands of iterations. In experiments with explicit SGD we were not able to replicate this performance because of numerical 
instability. We note that there are no standard implementations of explicit SGD for estimating Cox proportional hazards models, to our best knowledge.

\subsubsection{M-estimation}
\label{section:experiments_m}
In this experiment we test the performance of
implicit SGD, in particular Algorithm \ref{algo:implicit_m}, 
on a M-estimation problem 
in a setup that is similar to the simulation example 
of \citet[Example 2.4]{donoho2013high}. 

We set $N = 1000$ data points and $p = 200$ as the parameter dimension. We sample $\thetastar$  as a random 
 vector with norm $||\thetastar|| = 6\sqrt{p}$, and sample 
 the design matrix as $X \sim \mathcal{N}(0, (1/N) I)$.
 The outcomes are sampled i.i.d. from a contaminated normal distribution,  i.e., with probability 95\%, $Y_n  \sim \mathcal{N}(X_n^\intercal \thetastar, 1)$, and $Y_n=10$ with probability 5\%.
 
 The results over 2000 iterations of implicit SGD are shown in Figure \ref{figure:example_cox} (right panel). In the figure we plot the 5\% and 95\% 
 quantiles of MSE of implicit SGD over 100 replications of the 
 experiment. 
 We also plot (horizontal dashed line) the median MSE of the MLE estimator, computed using the \texttt{coxph} built-in command of \texttt{R}.
We observe that SGD converges steadily to the best possible MSE.
Similar behavior was observed under various modifications of the 
simulation parameters.




\subsection{National Morbidity-Mortality Air Pollution (NMMAPS) study}
\label{section:experiments:nmmaps}
The NMMAPS study~\citep{samet2000national,
dominici2002air} analyzed the risks of air pollution to public health. Several cities 
(108 in the US) are included in the study with daily measurements covering more than 13 years 
(roughly 5,000 days) including air pollution data (e.g. concentration of CO in the atmosphere) together with health outcome variables such as number of respiratory-related deaths.

The original study fitted a Poisson generalized additive model (GAM), separately for each city due to data set size. 
Recent research \citep{wood2014generalized} has developed procedures similar to \biglm's iterative QR decomposition to fit all cities simultaneously on the full data set with approximately $N=1.2$ million observations and $p=802$ covariates (7 Gb in size).
In this experiment, we construct a GAM model using data from all cities in the NMMAPS study in 
a process that is very similar (but not identical) to the data set of \citet{wood2014generalized}.

\begin{figure}[ht!]
\centering
  \includegraphics[width=0.8\textwidth]{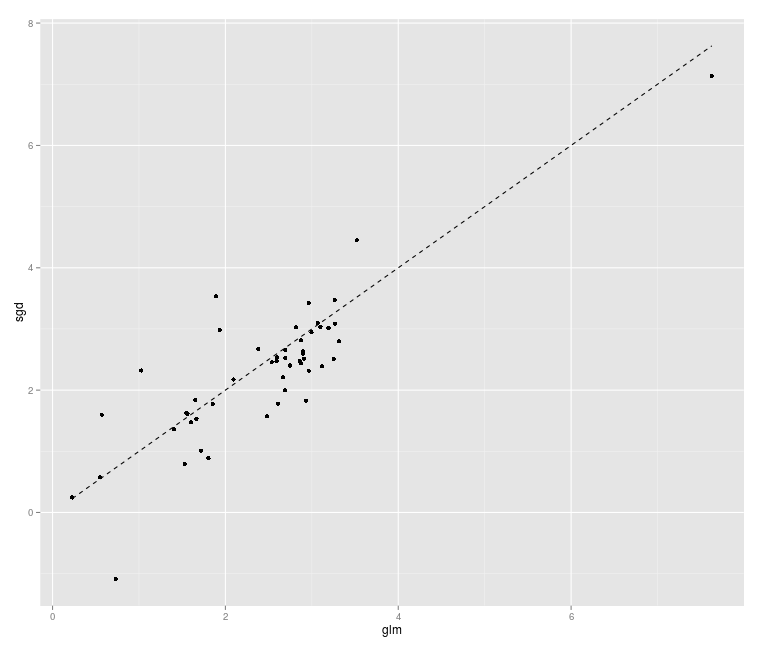}
  \caption{Estimates of implicit SGD (y-axis) and \glm\ (x-axis) on a subset of the NMMAPS data set 
	with $N=50,000$ observations and $p=50$ covariates, which is roughly 5\% of the entire data set.
		  }
	\label{figure:nmmaps}
\end{figure}

Our final data set has $N=1,426,806$ observations and $p=794$ covariates including all cities 
in the NMMAPS study (8.6GB in size), and is fit using 
a simple first-order implicit SGD procedure with 
$C_n=I$ and $\gamma_1=1$.
The runtime for implicit SGD was roughly 120 seconds, which is 
 6x faster than the 12 minutes reported by~\citet{wood2014generalized} on a similar  computer. 
We cannot directly compare the estimates from the two procedures 
because the datasets used were different.
However, we can compare the estimates of our model with the estimates of \glm\ on a random small 
subset of the data. For that purpose, we subsampled $N=50,000$ observations and $p=50$ covariates 
(19.5MB in size) and fit the smaller data set using implicit SGD and \glm. 
A scatter plot of the estimates is shown in Figure \ref{figure:nmmaps}.
The estimates of the implicit of the SGD procedure are very close to MLE, while further replications of the aforementioned testing process revealed the same pattern indicating that implicit SGD converged on all replications.

\section{Discussion}
\label{section:discussion}
The theory in Section \ref{section:theory} suggests that implicit SGD is numerically stable and has known asymptotic variance and asymptotic distribution. 
The experiments in Section \ref{section:experiments} show that 
the empirical properties of SGD are well predicted by theory. 
In contrast, explicit SGD is unstable and cannot work well 
without problem-specific tuning.
Thus, we conclude that implicit SGD is a principled estimation procedure and is superior to widely-used explicit SGD procedures.

Intuitively, implicit SGD leverages second-order
information at every iteration, although second-order quantities do not need to be computed in Eq. \eqref{def:sgd:implicit}. To demonstrate this, we build upon the argument that was first introduced in Section~\ref{section:introduction}. Assume both explicit and implicit
SGD are at the same estimate $\theta_0$. 
Then, using definitions in Eq.~\eqref{def:sgd:explicit} and
in Eq.~\eqref{def:sgd:implicit}, a Taylor approximation of $\nabla \log f(Y_n; X_n, \thetaim{n})$ 
yields
\begin{align}
\label{eq:approx:implicit}
\Delta \thetaim{n} \approx [I + \gamma_n \hat{\mathcal{I}}(\theta_0; X_n, Y_n)]^{-1} \Delta \thetasgd{n},
\end{align}
where $\Delta \thetaim{n} = \thetaim{n}-\theta_0$ 
and $\Delta \thetasgd{n} = \thetasgd{n}-\theta_0$, and the matrix $\hat{\mathcal{I}}(\theta_0; X_n, Y_n)= -\nabla^2 \log f(Y_n; X_n, \theta)|_{\theta=\theta_0}$ is the observed Fisher information at $\theta_0$.
In other words, the implicit procedure is a \emph{shrinked} version of the explicit one, where the shrinkage factor depends on the observed information. 

Naturally, the implicit SGD iterate $\thetaim{n}$ has also a Bayesian interpretation.  In particular, $\thetaim{n}$ is the posterior mode of a Bayesian model defined as
\begin{align}
\label{eq:implicit_bayes}
\theta | \thetaim{n-1}& \sim \mathcal{N}(\thetaim{n-1}, \gamma_n C_n) \nn\\
Y_n | X_n, \theta & \sim f(.; X_n, \theta).
\end{align}
The explicit SGD update $\thetasgd{n}$ can be written as in Eq.~\eqref{eq:implicit_bayes}, however 
$f$ needs to be substituted with its linear approximation around $\thetasgd{n-1}$.
Thus, Eq.~\eqref{eq:implicit_bayes} provides an alternative 
explanation why implicit SGD is more principled than explicit SGD. Furthermore, it indicates possible improvements for implicit SGD.
For example, the prior in Eq. \eqref{eq:implicit_bayes} could
be chosen to fit better the parameter space (e.g., $\thetastar$ being on the simplex).
\citet{krakowskigeometric} and \citet{nemirovski2009robust}  have argued that appropriate 
implicit updates can fit better in the geometry of the parameter space, and thus converge faster. Setting up the parameters of the prior is also crucial.
Whereas in explicit SGD there is no statistical intuition behind learning rates $\gamma_n$, 
Eq. \eqref{eq:implicit_bayes} reveals that in implicit SGD the terms $(\gamma_n C_n)^{-1}$ encode 
the statistical information up to iteration $n$. It follows immediately 
that it is optimal, in general, to set $\gamma_n C_n = \Fisher{\thetastar}^{-1}/n$, which is a special case of Theorem \ref{theorem:variance}. 
%

The Bayesian formulation of Eq.~\eqref{eq:implicit_bayes} also explains the stability of implicit SGD.
In Theorem \ref{theorem:mse2} we showed that the initial conditions are discounted at an exponential rate, regardless of misspecification of the learning rates.
This stability of implicit SGD allows several ideas for improvements. 
For example, constant learning rates could be used in implicit SGD to speed up convergence towards a region around $\thetastar$. 
A sequential hypothesis test could decide on whether $\thetaim{n}$ has reached that region or not, and switch to the theoretically optimal $1/n$ rate accordingly.
Alternatively, we could run implicit SGD with AdaGrad learning rates
and switch to $1/n$ rates when the theoretical $\bigO{1/\sqrt{n}}$ variance of AdaGrad 
becomes larger than the $\bigO{1/n}$ variance of implicit SGD.
Such schemes using constant rates with explicit SGD are very hard to do in practice 
because of instability.

Regarding statistical efficiency, a key technical result in this paper is that 
the asymptotic variance of implicit SGD can be calculated exactly using Theorem \ref{theorem:variance}.
Optimal learning rates were suggested in Eq. \eqref{eq:optimal:rates} that depend 
on the eigenvalues of the unknown Fisher matrix $\Fisher{\thetastar}$. 
In this paper, we used second-order procedures of Section \ref{section:second_order} to iteratively estimate the eigenvalues,
however better methods are certainly possible and could 
improve the performance of implicit SGD.
For example, it is known that typical iterative methods usually overestimate the largest eigenvalue and underestimate the 
smallest eigenvalue, in small-to-moderate samples. This  crucially 
affects the behavior of stochastic approximations with learning 
rates that depend on sample eigenvalues.
Empirical Bayes methods have been shown to be superior 
in iterative estimation of eigenvalues of large matrices~\citep{mestre2008improved},
and it would be interesting to apply such methods to 
design the learning rates of implicit SGD procedures.

Regarding computational efficiency, we developed Algorithm \ref{algo:implicit} which implements implicit SGD on a large family of statistical models. However, the trick used in fitting the Cox proportional hazards model in Section \ref{section:cox} can be more generally applied to models outside this family. 
For example, assume a log-likelihood gradient of the form $s(X^\intercal \theta; Y) G(\theta; X, Y)$, where both its scale $s(\cdot)$ and direction $G(\cdot)$ depend on model parameters $\theta$; 
this violates conditions of \assumeMain\assumeLinear. The implicit update in Eq.~\eqref{def:sgd:implicit}---where $C_n=I$ for simplicity---would be $\thetaim{n} = \thetaim{n-1} + \gamma_n s(X_n^\intercal \thetaim{n}; Y_n) G(\thetaim{n}; X_n, Y_n)$, 
which cannot be computed by Algorithm \ref{algo:implicit}. 
One way to circumvent this problem is to use 
an implicit update only on the scale and use an explicit 
update on the direction, i.e., 
 $\thetaim{n} = \thetaim{n-1} + 
 \gamma_n s(X_n^\intercal \thetaim{n}; Y_n) G(\thetaim{n-1}; X_n, Y_n)$. This form of updates expands the applicability 
 of implicit SGD.

Finally, hypothesis testing and construction of confidence intervals
 using SGD estimates is an important issue that 
 has  remained unexplored.
In experiments of Section \ref{section:experiments_normality} we 
showed that implicit SGD is indeed asymptotically normal in several simulation scenarios.
However, as SGD procedures are iterative, 
 there needs to be a rigorous and general method to decide whether 
 SGD iterates have converged to the asymptotic regime. 
Several methods, such as bootstrapping the data set, could be used 
for that. Furthermore, conservative confidence intervals 
could be constructed through multivariate Chebyshev inequalities or other strategies \citep{marshall1960multivariate}. 

\subsection{Concluding remarks}
\label{section:conclusion}
In this paper, we introduced a new stochastic gradient descent procedure that uses implicit updates at every iteration, which we termed implicit SGD. Equation \eqref{eq:approx:implicit} shows, intuitively, that the iterates of implicit SGD are a shrinked version of the standard iterates, where the shrinkage factor depends on the observed Fisher information matrix.
Thus, implicit SGD combines the computational efficiency of first-order methods with the numerical stability of second-order methods.

In a theoretical analysis, 
we derived non-asymptotic upper bounds for the mean-squared errors 
of implicit SGD iterates, and the asymptotic variance of both explicit and implicit SGD iterates.
Our analysis quantifies the efficiency loss of SGD procedures, and suggests principled strategies to calibrate a hyperparameter that is common to both explicit and implicit SGD procedures, 
known as the learning rate.
We illustrated the use of implicit SGD for statistical estimation 
in generalized linear models, Cox proportional hazards model, and general M-estimation problems.

Viewed as statistical estimation procedures, 
our results suggest that implicit SGD has the same asymptotic 
efficiency to explicit SGD. However, the implicit procedure 
is significantly more stable than the explicit one with respect 
to misspecification of the learning rate.
In general, explicit SGD procedures are sensitive to outliers and to misspecification of the learning rates, making it impossible 
to apply without problem-specific tuning.
In theory and in extensive experiments, implicit procedures 
emerge as principled iterative estimation methods because 
they are numerically stable, they are robust to tuning of hyper-parameters, and their standard errors are well-predicted by theory.
Thus, implicit stochastic gradient descent is poised to become a workhorse of estimation from large data sets in statistical practice.
%

%

\bibliographystyle{plainnat}
\bibliography{sgd-aos}

\newpage
\appendix
\newcommand{\EqSGD}{Eq.~\eqref{def:sgd:explicit}}
\newcommand{\EqIM}{Eq.~\eqref{def:sgd:implicit}}
\newcommand{\theoremVar}{\ref{theorem:variance}}

\section{Appendix: R code}
All experiments were run using the \texttt{R} package \texttt{sgd}, which implements explicit SGD and implicit SGD defined in Eqs.~(1) and (4) of the main paper.
The package is published at CRAN here \url{http://cran.r-project.org/web/packages/sgd/index.html}.
%
%

\section{Appendix: Useful lemmas}
Next, we prove lemmas on recursions
that will be useful for subsequent analysis.
All results are stated under a combination of \assumeMains.

\begin{lemma}
\label{lemma:decay_factor}
Consider a sequence $b_n$ such that $b_n \downarrow 0$
and $\sum_{i=1}^\infty b_i = \infty$. Then,
there exists a positive constant $K>0$, such that
\begin{align}
\label{eq:decay_factor}
\prod_{i=1}^n \frac{1}{1+b_i} \le \exp(-K  \sum_{i=1}^n b_i).
\end{align}
\end{lemma}
\begin{proof}
The function $x \log(1+1/x)$ is increasing-concave in $(0, \infty)$. 
From $b_n \downarrow 0$ it follows that $\log(1+b_n) / b_n$ is non-increasing.
Consider the value $K  =  \log(1 + b_1) / b_1$.
Then, $\log(1+b_1)/b_1 \ge \log(1+b_n)/b_n$ implies that $(1+b_n)^{-1} \le \exp(-K b_n)$.
Successive applications of this inequality
yields Ineq.~\eqref{eq:decay_factor}.
\end{proof}
%

\begin{lemma}
\label{lemma:implicit_recursion}
Consider sequences $a_n \downarrow 0, b_n \downarrow 0$, and $c_n \downarrow 0$ such that, $a_n = \littleO{b_n}$,  $\sum_{i=1}^\infty a_i  =  A < \infty$, and there is $n'$ such that $c_n/b_n < 1$ for all $n>n'$.
Define, 
\begin{align}
\label{new:defs}
\delta_n  =  \frac{1}{a_n} (a_{n-1}/b_{n-1} -a_n/{b_n}) \text{ and }
\zeta_n  =  \frac{c_n}{b_{n-1}} \frac{a_{n-1}}{a_n},
\end{align}
and suppose that $\delta_n \downarrow 0$ and $\zeta_n  \downarrow 0$.

Consider a positive sequence $y_n>0$
that satisfies the following recursive inequality,
\begin{align}
\label{ineq:implicit_recursion}
y_n \le \frac{1+c_n}{1 + b_n} y_{n-1}  + a_n.
\end{align}
Then, for every $n>0$, there exist constants 
$K_0, n_0$ such that
\begin{align}
\label{eq:result:implicit_recursion}
y_n \le K_0 \frac{a_n}{b_n} + Q_{1}^n y_0 + Q_{n_0+1}^n (1+c_1)^{n_0} A,
 \end{align}
 where $Q_i^n = \prod_{j=i}^n (1+c_i)/(1+b_i)$,
with $Q_i^n = 1$ if $n < i$, by definition.
\end{lemma}
\begin{proof}
Pick a positive $n_0$ such that $\delta_n + \zeta_n < 1$
and $(1+c_n)/(1+b_n) < 1$, for 
all $n \ge n_0$. Also, define 
$K_0  =  (1+b_1) (1-\delta_{n_0}-\zeta_{n_0})^{-1}$.
We consider two separate cases, namely, $n<n_0$ and $n \ge n_0$, 
and then we will combine the respective bounds.

{\bf Analysis for $n < n_0$}.
We first find a crude bound for $Q_{i+1}^n$. 
It holds,
\begin{align}
\label{lemma1:q_bound}
Q_{i+1}^n \le (1+c_{i+1}) (1+c_{i+2})\cdots(1+c_n) \le (1+c_1)^{n_0},
\end{align}
since $c_1 \ge c_n$ ($c_n \downarrow 0$ by definition) and there are no more than $n_0$ terms 
in the product.
From Ineq. \eqref{ineq:implicit_recursion} we get
\begin{align}
\label{ineq0}
y_n & \le Q_1^n y_0 + \sum_{i=1}^n Q_{i+1}^n a_i
	\commentEq{by expanding recursive Ineq. \eqref{ineq:implicit_recursion}}\nn \\
	& \le  Q_1^n y_0 + (1+c_1)^{n_0} \sum_{i=1}^n  a_i  \nn
	\commentEq{using Ineq. \eqref{lemma1:q_bound}}\\
	& \le  Q_1^n y_0 + (1+c_1)^{n_0} A.
\end{align}
This inequality also holds for $n=n_0$.

{\bf Analysis for $n \ge n_0$}.
In this case, we have for all $n \ge n_0$,
\begin{align}
\label{ineq:an}
 (1+b_1) \left(
 1-\delta_n - \zeta_n\right)^{-1} \le K_0 &
 	\commentEq{by definition of $n_0, K_0$}\nn \\
 	K_0 (\delta_n+ \zeta_n) + 1+ b_1  \le K_0 & \nn \\
	K_0 (\delta_n+ \zeta_n) + 1+ b_n  \le K_0 &
	\commentEq{because $b_n \le b_1$, since $b_n \downarrow 0$} \nn \\
	\frac{1}{a_n} K_0 (\frac{a_{n-1}}{b_{n-1}} - \frac{a_n}{b_n}) +
	\frac{1}{a_n} K_0 \frac{c_n a_{n-1}}{b_{n-1} } + 1+ b_n  \le K_0 &
	\commentEq{by definition of $\delta_n, \zeta_n$} \nn  \\
	a_n (1+b_n)  \le K_0 & a_n - K_0 \left(\frac{(1+c_n)a_{n-1}}{b_{n-1}} - \frac{a_n}{b_n}\right)  \nn \\
	a_n  \le K_0 & (\frac{a_n}{b_n} - \frac{1+c_n}{1+b_n}
		\frac{a_{n-1}}{b_{n-1}}).&
\end{align}
Now combine Ineq. \eqref{ineq:an} and  Ineq. \eqref{ineq:implicit_recursion} 
to obtain
\begin{align}
\label{ineq1}
(y_n - K_0 \frac{a_n}{b_n}) \le
\frac{1+c_n}{1+b_n}
(y_{n-1} - K_0 \frac{a_{n-1}}{b_{n-1}}).
\end{align}
Define $s_n  =  y_n - K_0 a_n/b_n$.
Then, from Ineq. \eqref{ineq1}, 
$s_n \le \frac{1+c_n}{1+b_n} s_{n-1}$,
where $\frac{1+c_n}{1+b_n} < 1$ since $n\ge n_0$.
Let $n_1$ be the smallest integer such that
$n_1 \ge n_0$ and $s_{n_1} \le 0$.
If $n_1$ does not exist then $s_n$ are all positive, and thus 
$y_n \le K_0 a_n/b_n$, which satisfies Ineq.~\eqref{ineq:implicit_recursion}, for all $n\ge n_0$.
If $n_1$ exists then for all $n \ge n_1$, it follows $s_n \le 0$, 
and thus $y_n \le K_0 a_n/b_n$ for all $n \ge n_1$.
For $n_0 \le n < n_1$ all $s_n$ are positive.
Using Ineq. \eqref{ineq1}, we have
$ s_n \le (\prod_{i=n_0+1}^n \frac{1+c_i}{1+b_i}) s_{n_0}
 =  Q_{n_0+1}^n s_{n_0}$, and thus
\begin{align}
\label{ineq2}
y_n - K_0 \frac{a_n}{b_n} & \le  Q_{n_0+1}^n s_{n_0}
\commentEq{by definition of $s_n$} \nn \\
y_n & \le K_0 \frac{a_n}{b_n} + Q_{n_0+1}^n y_{n_0} 
\commentEq{because $s_n \le y_n$} \nn \\
y_n & \le K_0 \frac{a_n}{b_n} + Q_{1}^n y_0 + Q_{n_0+1}^n (1+c_1)^{n_0} A.
\commentEq{by Ineq. \eqref{ineq0} on $y_{n_0}$} 
\end{align}
Combining this result with Ineq. \eqref{ineq0} and Ineq. \eqref{ineq2}, we obtain
\begin{align}
y_n \le K_0 \frac{a_n}{b_n} + Q_{1}^n y_0 + Q_{n_0+1}^n (1+c_1)^{n_0} A,
\end{align}
since $Q_{i}^n=1$ for $n < i$, by definition.
\end{proof}

\begin{corollary}
\label{corollary:implicit_recursion}
In Lemma \ref{lemma:implicit_recursion}
assume $a_n = a_1 n^{-\alpha}$ and $b_n = b_1 n^{-\beta}$, and $c_n=0$, 
where $\alpha > \beta$, and $a_1, b_1, \beta>0$ and $\alpha > 1$. 
Then, there exists $n_0>0$ such that for all $n \ge n_0$,
\begin{align}
\label{thm1:eq4}
y_n \le 2\frac{a_1 (1+b_1)}{b_1} n^{-\alpha + \beta} + \exp(-\log(1+b_1) \phi_\beta(n)) [y_0 +(1+b_1)^{n_0}A],
 \end{align}
 where $A=\sum_i a_i < \infty$, and $\phi_\beta$ is defined as in Theorem~(2.1) of the main paper; i.e., $\phi_\beta(n) = n^{1-\beta}$ if $\beta \in(0.5, 1)$, and $\phi_\beta(n) = \log n$ if $\beta=1$.
\end{corollary}
\begin{proof}
For every $n>2$ and $\gamma\in (0.5, 1]$ it is easy to show through induction that 
\begin{align}
\label{ineq:recur1}
(n-1)^{-\gamma} - n^{-\gamma} & \le 2 n^{-1-\gamma},\\
\label{ineq:recur2}
\sum_{i=1}^n i^{-\gamma} & \ge \phi_\gamma(n).
\end{align}
By definition of $\delta_n$ and Ineq.~\eqref{ineq:recur1},
\begin{align}
\label{ineq:recur3}
\delta_n = \frac{1}{a_n} (\frac{a_{n-1}}{b_{n-1}}-\frac{a_n}{b_n}) = 
\frac{1}{a_1 n^{-\alpha}} \frac{a_1}{b_1} ((n-1)^{-\alpha+\beta} - n^{-\alpha+\beta}) \le
\frac{2}{b_1} n^{-1+\beta}.
\end{align}
Also, $\zeta_n = 0$ since $c_n=0$.
For the rest of the proof we will suppose that Ineq.~\eqref{ineq:recur3} holds for every $n$ since for $n=1$ we can simply define $\delta_1 \le 1/2$.

Next, we take $n_0 = \lceil (4/b_1)^{1/(1-\beta)} \rceil$ so that 
$\delta_n < 1/2$ and $\delta_n + \zeta_n < 1$ for all $n \ge n_0$.
Therefore, 
$K_0 = (1+b_1)(1-\delta_{n_0})^{-1} \le 2 (1+b_1)$;
define $K_0 =2(1+b_1)$.
Since $c_n =0$, it follows $Q_{i}^n = \prod_{j=i}^n (1+b_i)^{-1}$.
Thus, for a lower bound,
\begin{align}
\label{cor:ineqs1}
Q_1^n & \ge (1+b_1)^{-n},
\end{align}
and for an upper bound,
\begin{align}
\label{cor:ineqs2}
Q_1^n & \le \exp(-\log(1+b_1)/b_1 \sum_{i=1}^n b_i),
 \commentEq{by Lemma~\ref{lemma:decay_factor}} \nonumber \\
 Q_1^n & \le  \exp(-\log(1+b_1) \phi_\beta(n)).
 \commentEq{by Ineq.~\eqref{ineq:recur2}}
\end{align}
 Lemma~\ref{lemma:implicit_recursion}, Ineq.~\eqref{cor:ineqs1} and Ineq.~\eqref{cor:ineqs2} imply that
\begin{align}
y_n & \le K_0 \frac{a_n}{b_n} + Q_{1}^n y_0 + Q_{n_0+1}^n (1+c_1)^{n_0} A 
\commentEq{by Lemma ~\ref{lemma:implicit_recursion}}
\nn \\
 & \le 2\frac{a_1 (1+b_1)}{b_1} n^{-\alpha + \beta} + Q_1^{n} [y_0 + (1+b_1)^{n_0}A] 
 \commentEq{by Ineq.~\eqref{cor:ineqs1}, $c_1=0$}
 \nn \\
&  \le  2\frac{a_1 (1+b_1)}{b_1} n^{-\alpha + \beta} + \exp(-\log(1+b_1) \phi_\beta(n)) [y_0 +(1+b_1)^{n_0}A],
\end{align}
where the last inequality follows from Ineq.~\eqref{cor:ineqs2}.
\end{proof}

\begin{lemma}
\label{lemma:useful}
Suppose \assumeMains\assumeLinear, \assumeLip, and \assumeStrong\ hold.
Then, almost surely it holds
\begin{align}
\label{eq:useful:lam}
\lambda_n  & \le \frac{1}{1 + \gamma_n \minEigC\FisherMin},\\
\label{eq:useful:mse2}
||\thetaim{n}-\thetaim{n-1}||^2 & \le 4 L_0^2 \gamma_n^2,
\end{align}
where $\lambda_n$ is defined in Theorem~(3.1), 
and $\thetaim{n}$ is the $n$-th iterate of implicit SGD, defined by {\EqIM}  in the main paper.
\end{lemma}
\begin{proof}
For the first part, from Theorem~(3.1) we have
\begin{align}
\label{useful:eq1}
\ell'(X_n^\intercal \thetaim{n}; Y_n)  = 
\lambda_n \ell'(X_n^\intercal \thetaim{n-1}; Y_n),
\end{align}
where the derivative of the log-likelihood $\ell$ is with respect to the natural parameter $X^\intercal \theta$.
Using definition in \EqIM, 
\begin{align}
\thetaim{n} = \thetaim{n-1} + \gamma_n \lambda_n \ell'(X_n^\intercal \thetaim{n-1}; Y_n) C_n X_n.
\end{align}
We use this definition of $\thetaim{n}$ into Eq.\eqref{useful:eq1} and
perform a Taylor approximation on $\ell'$ to obtain
\begin{align}
\label{useful:eq2}
\ell'(X_n^\intercal \thetaim{n}; Y_n)  = \ell'(X_n^\intercal \thetaim{n-1}; Y_n) + 
 \tilde{\ell}''\gamma_n \lambda_n \ell'(X_n^\intercal\thetaim{n-1}; Y_n) X_n^\intercal C_n X_n,
\end{align}
where $\tilde{\ell}'' = \ell''(\delta X_n^\intercal\thetaim{n-1} + (1-\delta) 
X_n^\intercal\thetaim{n}; Y_n)
\equiv \ell''(X_n^\intercal \tilde{\theta}; Y_n)$, and $\delta \in [0, 1]$. 
By combining Eq.~\eqref{useful:eq1} with Eq.~\eqref{useful:eq2} and cancelling out the first derivative term we get
\begin{align}
\lambda_n & = 1 + \tilde{\ell}''\gamma_n \lambda_n X_n^\intercal C_n X_n\nonumber\\
\lambda_n & \le 1 + \tilde{\ell}''\gamma_n \lambda_n \minEigC ||X_n||^2
\commentEq{by \assumeMain\assumeCn\ and 
$\ell'' <0$}
\nonumber\\
\lambda_n  (1 - \gamma_n \minEigC \tilde{\ell}'' ||X_n||^2)    & \le 1 
 \nn \\
\left(1 + \gamma_n \minEigC \mathrm{trace}(\hat{\mathcal{I}}(\tilde{\theta}))\right) \lambda_n & \le 1 
\commentEq{where $\hat{\mathcal{I}}$ is the observed Fisher information}\nn \\
(1 + \gamma_n \minEigC\FisherMin) \lambda_n & \le 1 
\commentEq{by \assumeMain\assumeStrong}.
\end{align}
For the second part, since the log-likelihood is differentiable (\assumeMain\assumeLinear) we can rewrite the definition of implicit SGD in \EqIM\ (in the main paper) as
\begin{align}
\thetaim{n} = \arg\max \{-\frac{1}{2\gamma_n} ||\theta-\thetaim{n-1}||^2 + \ell(X_n^\intercal \theta; Y_n) \}.\nn
\end{align}
Therefore, setting $\theta= \thetaim{n-1}$ in the above equation yields
\begin{align}
-\frac{1}{2\gamma_n} ||\thetaim{n}-\thetaim{n-1}||^2 + \ell(X_n^\intercal \thetaim{n}; Y_n)  & \ge \ell(X_n^\intercal \thetaim{n-1}; Y_n) \nn\\
||\thetaim{n}-\thetaim{n-1}||^2 & 
\le 2\gamma_n \left(\ell(X_n^\intercal \thetaim{n}; Y_n)  - \ell(X_n^\intercal \thetaim{n-1}; Y_n) \right) \nn\\
||\thetaim{n}-\thetaim{n-1}||^2 & 
\le 2\gamma_n L_0 ||\thetaim{n} - \thetaim{n-1}||
\commentEq{By \assumeMain\assumeLip}
\nn\\
||\thetaim{n}-\thetaim{n-1}|| 
& \le 2 L_0 \gamma_n\nn\\
||\thetaim{n}-\thetaim{n-1}||^2
& \le 4 L_0^2 \gamma_n^2.\nn
\end{align}
\end{proof}

\section{Appendix: Theoretical analysis}
\subsection{Finite-sample analysis}
\begin{customthm}{2.1}
\TheoremMSE
\end{customthm}
\begin{proof}
Starting from the procedure defined by \EqIM\ in the main paper, we have 
\begin{align}
 \label{mse2:eq1}
\thetaim{n} - \thetastar  = & \thetaim{n-1}-\thetastar + \gamma_n C_n \nabla \log f(Y_n; X_n, \thetaim{n})\nn\\
\thetaim{n} - \thetastar   = &  \thetaim{n-1}-\thetastar +  \gamma_n \lambda_n C_n \nabla\log f(Y_n; X_n, \thetaim{n-1})
\commentEq{By Theorem~(3.1)}\nn\\
||\thetaim{n}-\thetastar||^2  = & 
	||\thetaim{n-1}-\thetastar||^2  \nn\\
	& + 2 \gamma_n \lambda_n (\thetaim{n-1}-\thetastar)^\intercal C_n \nabla\log f(Y_n; X_n, \thetaim{n-1}) \nn\\
	& + \gamma_n^2 ||C_n \nabla\log f(Y_n; X_n, \thetaim{n})||^2.
\end{align}
The last term can be simply bounded since $\nabla\log f(Y_n; X_n, \thetaim{n}) = 
\thetaim{n} - \thetaim{n-1}$ by definition; thus,
\begin{align}
\label{mse2:eq2}
 || C_n \nabla\log f(Y_n; X_n, \thetaim{n})||^2 \le \maxEigC^2 ||\thetaim{n}-\thetaim{n-1}||^2 \le 4 L_0^2 \maxEigC^2 \gamma_n^2,
\end{align}
which holds almost surely by Lemma \ref{lemma:useful}-Eq.\eqref{eq:useful:mse2}. 
For the second term we can bound its expectation as 
\begin{align}
\label{mse2:eq3}
\mathbb{E}(2 \gamma_n & \lambda_n (\thetaim{n-1}-\thetastar)^\intercal  C_n \nabla\log f(Y_n; X_n, \thetaim{n-1}))  \nn\\
& \le \frac{2\gamma_n}{1 + \gamma_n \minEigC\FisherMin} \Ex{(\thetaim{n-1}-\thetastar)^\intercal C_n \nabla\log f(Y_n; X_n, \thetaim{n-1})}\nn
\commentEq{by Lemma~\ref{lemma:useful}}\\
& \le \frac{2\gamma_n}{1 + \gamma_n \minEigC\FisherMin} \Ex{(\thetaim{n-1}-\thetastar)^\intercal C_n \nabla h(\thetaim{n-1})}\nn
\commentEq{where $\nabla h(\thetaim{n-1})  = \mathbb{E}(\nabla\log f(Y_n; X_n, \thetaim{n-1}) | \mF{n-1})$} \nn\\
& \le -\frac{2\gamma_n \minEigF\minEigC}{1 + \gamma_n \minEigC\FisherMin} 
||\thetaim{n-1}-\thetastar||^2
\commentEq{by strong convexity, \assumeMain\assumeStrong.}
\end{align}

Taking expectations in Eq.~\eqref{mse2:eq1}
and substituting Ineqs.~\eqref{mse2:eq2} and \eqref{mse2:eq3} into Eq.~\eqref{mse2:eq1} yields the recursion,
\begin{align}
\label{mse2:eq4}
\Ex{||\thetaim{n}-\thetastar||^2} \le (1-\frac{2\gamma_n \minEigF\minEigC}{1 + \gamma_n \minEigC\FisherMin}) \Ex{||\thetaim{n-1}-\thetastar||^2}
+ 4L_0^2 \maxEigC^2 \gamma_n^2.
\end{align}
Define $\mu_1 = 2\minEigF$, $\mu_2 = \max\{\gamma_1 \minEigC\ \mu_1 (b-\mu_1), 0\}$ and $\mu = \mu_1 / (\mu_1 + \mu_2)$; note that 
$\mu \in (0, 1]$, and $\mu=1$ only when $\mu_2=0$, i.e., $2\minEigF \ge b$. Through simple algebra we obtain
\begin{align}
\label{mse2:eq5}
(1-\frac{2\gamma_n \minEigF\minEigC}{1 + \gamma_n \minEigC\FisherMin})
\le \frac{1}{1 + 2\gamma_n \mu \minEigF\minEigC },
\end{align}
for all $n >0$. Therefore we can write recursion \eqref{mse2:eq4} as
\begin{align}
\label{mse2:eq6}
\Ex{||\thetaim{n}-\thetastar||^2} \le \frac{1}{1 + 2\gamma_n \mu\minEigF\minEigC}  \Ex{||\thetaim{n-1}-\thetastar||^2} + 4L_0^2 \maxEigC^2 \gamma_n^2.
\end{align}
We can now apply Corollary \ref{corollary:implicit_recursion} 
with $a_n= 4L_0^2\maxEigC^2 \gamma_n^2$ and $b_n = 2\gamma_n \mu \minEigF\minEigC$.
 \end{proof}

\textbf{Note.} 
Assuming Lipschitz continuity of the gradient $\nabla \ell$ instead 
of function $\ell$ would not critically alter the main result 
of Theorem~(2.1). In fact, assuming 
Lipschitz continuity with constant $L$ of $\nabla \ell$ and boundedness of 
$\Ex{||\nabla \log f(Y_n; X_n, \thetastar)||^2} \le \sigma^2$, 
as it is typical in the literature, 
would simply add a term $\gamma_n^2 L^2 \Ex{||\thetaim{n}-\thetastar||^2} + \gamma_n^2 \sigma^2$ in the right-hand side of Eq.\eqref{mse2:eq1}.
In this case the upper-bound is always satisfied for $n$ such that 
$\gamma_n^2 L^2  > 1$, which also highlights a difference of implicit 
SGD with explicit SGD, as in explicit SGD the term 
$\gamma_n^2 L^2 ||\thetasgd{n-1}-\thetastar||^2$ increases the upper bound 
and can make $||\thetasgd{n}-\thetastar||^2$ diverge.
For, $\gamma_n^2 L^2 < 1$, the 
discount factor for implicit SGD would be $(1-\gamma_n^2 L^2)^{-1}(1+2\gamma_n \mu\minEigF\minEigC)^{-1}$, 
which could then be bounded by a quantity 
$(1+\gamma_n d)^{-1}$ for some constant $d$. 
This would lead to a solution that is similar to Theorem~(2.1).

\subsection{Asymptotic analysis}
\label{appendix:asymptotic}
Here, we prove the main result on the 
asymptotic variance of implicit SGD. 
First, we introduce linear maps $\linearMap{B}{\cdot}$ defined as
$\linearMap{B}{X} = \frac{1}{2}(\m{B X} + \m{X B})$, where 
$\m{B}$ is symmetric positive definite matrix and $\m{X}$ is bounded.
The identity map is denoted as $\linearMapIdentityOne$ and 
it holds $\linearMapIdentity{\m{X}} = \m{X}$, for all $\m{X}$. Also, 
$\linearMapNullOne$ is the null operator for which $\linearMapNull{\m{X}} = \m{0}$, for all $\m{X}$.
By the Lyapunov theorem \citep{lyapunov1992general} the map
$\linearMapOne{B}$ is one-to-one and thus the inverse operator $\invLinearMap{B}{\cdot}$ is well-defined. Furthermore, we define 
the norm of a linear map as $||\linearMapOne{B}|| = \max_{||\m{X}||=1}
|| \linearMap{B}{\m{X}}||$. For bounded inputs $\m{X}$, it holds $||\linearMapOne{B}|| = \bigO{||\m{B}||}$.

%
\begin{lemma}
\label{lemma:recursions}
Suppose that the sequence $\{\gamma_n\}$ 
satisfies \assumeMain\assumeGn.
Consider the matrix recursions
\begin{align}
	\label{matrix:recursion:sgd}
	\m{X}_n & = \linearMap{I-\gamma_n B_n}{\m{X}_{n-1}} + \gamma_n (\m{C} + \Mn{D}{n}), \\ 
	\label{matrix:recursion:implicit}
	\m{Y}_n & = \invLinearMap{I + \gamma_n B_n}{\m{X}_{n-1} + \gamma_n (\m{C} + \m{D}_n)},
\end{align}
such that
\begin{enumerate}[(a)]
\item All matrices $\m{X}_n, \m{Y}_n, \m{B}_n, \m{D}_n$ and $\m{C}$ are bounded,
\item $\Mn{B}{n} \to \m{B}$ is positive definite and $|| \Mn{B}{n} - \Mn{B}{n-1} || = \bigO{\gamma_n^2}$, 
\item $\m{C}$ is a fixed matrix and $\m{D}_n \to \m{0}$.
\end{enumerate}
Then, both recursions approximate the matrix $\invLinearMap{B}{\m{C}}$ i.e.,
\begin{equation}
|| \m{X}_n \m{B} + \m{B} \m{X}_n - 2 \m{C} ||  \to 0 \text{ and } 
	| \m{Y}_n \m{B} + \m{B} \m{Y}_n - 2 \m{C}|| \to  0.
\end{equation}
If, in addition, $\m{B}$ and $\m{C}$ commute then 
$\m{X}_n \to \m{B}^{-1} \m{C}$ and $\m{Y}_n \to \m{B}^{-1} \m{C}$.
\end{lemma}
\newcommand{\G}[1]{\m{\Gamma}_{#1}}
\renewcommand{\P}[2]{\mt{P}{#1}^{#2}}
\newcommand{\Q}[2]{\mt{Q}{#1}^{#2}}
\newcommand{\C}[1]{\mt{C}{#1}}
\newcommand{\convSum}[1]{\boldsymbol{S}_n^{#1}}
\newcommand{\deltaB}[1]{\invLinearMapOne{B_{#1-1}}  - \invLinearMapOne{B_{#1}}}
\begin{proof}
We make the following definitions.
\begin{align}
\label{eq:gamma}
& \m{\Gamma}_n   =  \I - \gamma_n \m{B}_n, \\
\label{eq:partial}
& \P{i}{n}  =  \linearMapOne{\Gamma_n} \circ 
					\linearMapOne{\Gamma_{n-1}} \circ 
					\cdots \linearMapOne{\Gamma_{i}},
\end{align}
where the symbol $\circ$ denotes successive application of the linear maps, and $\P{i}{n} = \linearMapIdentityOne$ if 
$n<i$, by definition. It follows,
\begin{align}
\label{eq:bounded}
||\P{i}{n}||  = \bigO{\prod_{j=i}^n ||\I - \gamma_i \m{B}_i||} \le K_0 e^{-K_1 \sum_{j=i}^n \gamma_j},
\end{align}
for suitable constants $K_0, K_1$ \citep[see][Appendix, Part 3]{polyak1992}. 
Let $\Gamma(n) = K_1 \sum_{i=1}^n \gamma_i$.
By \assumeMain\assumeGn, 
$\Gamma(n) \to \infty$ and thus $\P{i}{n} \to \linearMapNullOne$ as $n \to \infty$ and $i$ is fixed.
The matrix recursion in Lemma \ref{lemma:recursions} can be rewritten as $\mt{X}{n} = \linearMap{\Gamma_n}{ \mt{X}{n-1}} +\gamma_n \m{C} + \gamma_n \mt{D}{n}$.
Solving the recursion yields
\begin{align}
\label{eq:main:recursion}
\mt{X}{n} = & \linearMapOne{\Gamma_n} \circ 
					\linearMapOne{\Gamma_{n-1}} \circ 
					\cdots \linearMap{\Gamma_{1}}{\mt{X}{0}} + \gamma_n \m{C} + \gamma_n \mt{D}{n}  \nonumber  \\
	&  + a_{n-1} \linearMap{\Gamma_n}{\m{C}} +  a_{n-1}\linearMap{\Gamma_n}{\mt{D}{n-1}}     \nonumber  \\ 
	& + \cdots + \nn \\
	& + a_1 \linearMapOne{\Gamma_n} \circ 
					\linearMapOne{\Gamma_{n-1}} \circ 
					\cdots \linearMap{\Gamma_{2}}{\m{C}} +  
					 a_1 \linearMapOne{\Gamma_n} \circ 
					\linearMapOne{\Gamma_{n-1}} \circ 
					\cdots  \linearMap{\Gamma_{2}}{\m{D}_1}
					\nn \\ 
	 =   &\quad  \P{1}{n}\{\mt{X}{0}\} + \mt{S}{n} \{\m{C} \}+ 
		\widetilde{\m{D}}_n,
\end{align}
where we have defined the linear map $\m{S}_n= \sum_{i=1}^n \gamma_i \P{i+1}{n}$ and the matrix
$\widetilde{\m{D}}_n = \sum_{i=1}^n \gamma_i \P{i+1}{n} \{\m{D}_i\}$. 
Since $\P{1}{n} \to \linearMapOne{0}$, our goal is to prove that $\m{S}_n \to \invLinearMapOne{B}$ and $\widetilde{\m{D}}_n \to \m{0}$.
\newcommand{\Binv}[1]{\m{B}_{#1}^{-1}}
By definition, 
\begin{equation}
\label{eq:limit}
	\sum_{i=1}^n \gamma_i \P{i+1}{n} = \invLinearMapOne{B_n} + \sum_{i=2}^n \P{i}{n} (\invLinearMapOne{B_{i-1}} - \invLinearMapOne{B_{i}}) - 
		\P{1}{n} \invLinearMapOne{B_1}.
\end{equation}
To see this, first note that $\gamma_n \I = (\I - \G{n}) \mt{B}{n}^{-1}$ for every $n$, and thus 
\begin{align}
\label{eq:map:intermediate}
\gamma_n \linearMapIdentityOne = \linearMapOne{I - \Gamma_n} \circ \invLinearMapOne{B_n}.
\end{align}
Therefore, if we collect the coefficients of the terms $\invLinearMapOne{B_n}$ in the right-hand side of \eqref{eq:limit}, we get
\begin{align}
\label{eq:last}
 \invLinearMapOne{B_n} + & \sum_{i=2}^n \P{i}{n} (\invLinearMapOne{B_{i-1}} - \invLinearMapOne{B_{i}}) - 
		\P{1}{n} \invLinearMapOne{B_1}  \nn \\
= &\quad  (\P{2}{n}-\P{1}{n}) \invLinearMapOne{B_1} + 
	(\P{3}{n}-\P{2}{n})  \invLinearMapOne{B_2} +\cdots +
	 (\P{n+1}{n}- \P{n}{n})  \invLinearMapOne{B_n} \nn \\
= &\quad   \P{2}{n} \circ \linearMapOne{I - \Gamma_1}  \circ \invLinearMapOne{B_1} + 
\P{3}{n} \circ \linearMapOne{I - \Gamma_2} \circ \invLinearMapOne{B_2} +\cdots +
		\P{n+1}{n} \circ  \linearMapOne{I - \Gamma_n} \circ \invLinearMapOne{B_n} & \nn \\
= & \quad \P{2}{n} (\gamma_1 \linearMapIdentityOne)  + \P{3}{n} (\gamma_2 \linearMapIdentityOne) + \cdots + 
	\P{n+1}{n} (\gamma_n \linearMapIdentityOne)  \quad \commentEq{\text{by } Eq. \eqref{eq:map:intermediate}}  \nn \\
= & \quad \sum_{i=1}^n \gamma_i \P{i+1}{n}, \nn
\end{align}
where we used the identity $\P{i+1}{n} - \P{i}{n} = \P{i+1}{n} \circ (\linearMapIdentityOne - \linearMapOne{\Gamma_i}) = \P{i+1}{n}  \circ  \linearMapOne{I - \Gamma_i}$. 
Furthermore, since $\m{B}_i$ are bounded,
\begin{align}
||\invLinearMapOne{B_{i-1}} - \invLinearMapOne{B_{i}}|| & =
|| |\invLinearMapOne{B_{i}} \circ (\linearMapOne{B_{i}}- \linearMapOne{B_{i-1}}) \circ \invLinearMapOne{B_{i-1}} || = 
\bigO{||\linearMapOne{B_i} - \linearMapOne{B_{i-1}}||} \nn \\
& = \bigO{||B_{i} - B_{i-1} ||} = \bigO{\gamma_i^2}. 
\commentEq{By assumption of Lemma \ref{lemma:recursions}} \nn
\end{align}
In addition, $||\sum_{i=2}^n \P{i}{n} \circ (\deltaB{i})|| \le
 K_0 e^{- \Gamma(n)} \sum_{i=2}^n e^{ \Gamma(i)} \bigO{\gamma_i^2}$.  Since $\sum_i \bigO{\gamma_i^2} < \infty$ and $e^{\Gamma(i)}$ is positive, increasing and diverging,
we can invoke  Kronecker's lemma and obtain $\sum_{i=2}^n  e^{\Gamma(i)}\bigO{\gamma_i^2} = o(e^{\Gamma(n)})$.
Therefore
\begin{align}
\sum_{i=2}^n \P{i}{n} \circ (\deltaB{i}) \to \linearMapOne{0},
\end{align}
and since $\P{1}{n} \to \linearMapOne{0}$, we conclude from Equation \eqref{eq:map:intermediate} that
\begin{equation}
\label{eq:limit2}
	\lim_{n \to \infty} \sum_{i=1}^{n} \gamma_i \P{i+1}{n} = \lim_{n \to \infty} \invLinearMapOne{B_n} = \invLinearMapOne{B}.
\end{equation}

Thus, $\m{S}_n \to \invLinearMapOne{B}$, as desired.
For $\widetilde{\m{D}}_n$ we have 
\begin{align}
\widetilde{\m{D}}_n =  \sum_{i=1}^n \gamma_i \P{i+1}{n} \{ \m{D}_i \}= &  
\invLinearMap{B_n} {\m{D}_n}  + 
 \sum_{i=2}^n \P{i}{n} \circ (\invLinearMap{B_{i-1}}{ \m{D}_{i-1} }-      	\invLinearMap{B_i}{\m{D}_i}) \nn \\ 
 	& +  \P{1}{n} \circ \invLinearMap{B_1}{\m{D}_1} \nn.
\end{align}
Since $||\m{D}_n|| \to 0$ it follows that $||\invLinearMap{B_n}{ \m{D}_n}|| \to 0$ and $
|| (\invLinearMap{B_{i-1}}{ \m{D}_{i-1} }-      	\invLinearMap{B_i}{\m{D}_i})|| =  \bigO{\gamma_i^2}$. Recall that $\P{1}{n} \to \linearMapOne{0}$, and thus $\widetilde{\m{D}}_n  \to \b{0}$. 
Finally, we substitute this result in Equation \eqref{eq:map:intermediate} to get $\mt{X}{n} \to \invLinearMapOne{B}\{\m{C}\}$.

For the second recursion of the lemma,
\begin{equation}
\label{eq:implicit:recursion}
\mt{Y}{n} = \invLinearMap{I + \gamma_n B_n} {\mt{Y}{n-1} +\gamma_n (\m{C} + \mt{D}{n})},
\end{equation}
the proof is similar. 
First, we  make the following definitions.
\begin{align}
& \m{\Gamma}_n   =  \I + \gamma_n \m{B}_n, \nn \\
& \Q{i}{n}  =  \invLinearMapOne{\Gamma_n} \circ 
					\invLinearMapOne{\Gamma_{n-1}} \circ 
					\cdots \invLinearMapOne{\Gamma_{i}}. \nn
\end{align}
As before, $\Q{i}{n} \to \linearMapNullOne$. Solving the recursion 
\eqref{eq:implicit:recursion} yields
\begin{align}
\mt{Y}{n} = \quad  \Q{1}{n}\{\mt{Y}{0}\} + \mt{S}{n} \{\m{C} \}+ 
		\widetilde{\m{D}}_n,
\end{align}
where  we defined $\m{S}_n  =  \sum_{i=1}^n \gamma_i \Q{i}{n}$ and $\widetilde{\m{D}}_n  =  \sum_{i=1}^n \gamma_i \Q{i}{n} \{\m{D}_i\}$. 
  The following identities
can also be verified by the definition of the linear maps.
\begin{align}
\label{eq:decomposition}
& \invLinearMapOne{B_n} \circ (\linearMapIdentityOne - \invLinearMapOne{\Gamma_n}) = \gamma_n \invLinearMapOne{\Gamma_n},\\
\label{eq:commutativity}
& \invLinearMapOne{B_n} \invLinearMapOne{\Gamma_n} = 
\invLinearMapOne{\Gamma_n} \invLinearMapOne{B_n}.
\end{align}
It holds,
\begin{align}
\invLinearMapOne{B_n} + \sum_{i=1}^n \Q{i}{n} \circ (\deltaB{i})  = &
	\invLinearMapOne{B_n} \circ (\linearMapIdentityOne - \invLinearMapOne{\Gamma_n}) + \invLinearMapOne{\Gamma_n} \circ \invLinearMapOne{B_{n-1}} \circ (\linearMapIdentityOne - \invLinearMapOne{\Gamma_n}) + \cdots  \nn \\
	= & \gamma_n \invLinearMapOne{\Gamma_n} + \gamma_{n-1}
	\invLinearMapOne{\Gamma_{n}}
	  \invLinearMapOne{\Gamma_{n-1}} + \cdots = \m{S}_n,\nn
\end{align}
where the first line is obtained by Eq. \eqref{eq:decomposition} and the second line by Eq. \eqref{eq:commutativity}. 
Thus, 
similar to the previously analyzed recursion, $\m{S}_n \to \invLinearMapOne{B}$ and $\widetilde{\m{D}}_n \to \m{0}$.
Therefore, $\m{Y}_n \to \invLinearMap{B}{\m{C}}$.

For both cases, if $\m{B}, \m{C}$ commute then 
$\invLinearMapOne{B}\{\m{C} \} =\m{X}$ such that 
$\m{B} \m{X} + \m{X} \m{B} = 2\m{C}$. Setting $\m{X} = \m{B}^{-1} \m{C}$ is a solution since $\m{B} \m{B}^{-1} \m{C} + \m{B}^{-1} \m{C} \m{B} 
 = \m{C} + \m{B}^{-1} \m{B} \m{C} = 2\m{C}$. By the Lyapunov theorem, 
 this solution is unique.
\end{proof}

\begin{corollary}
\label{corollary:recursions}
Consider the matrix recursions
\begin{align}
\label{eq:corollary:main}
		\m{X}_n & = \linearMap{I-\gamma_n B_n}{\m{X}_{n-1}} + \gamma_n^2 (\m{C} + \Mn{D}{n}), \\ 
\m{Y}_n & = \invLinearMap{I + \gamma_n B_n}{\m{Y}_{n-1} + \gamma_n^2 (\m{C} + \m{D}_n)},
\end{align}
where $\Mn{B}{n}, \m{B}, \m{C}, \Mn{D}{n}$ satisfy the assumptions of 
Lemma \ref{lemma:recursions}. 
Moreover, suppose $\gamma_n=\gamma_1 n^{-1}$.
If the matrix $\m{B}- \m{I}/\gamma_1$ is positive definite, then 
\begin{align}
& (1/\gamma_n) \Mn{X}{n} \to \invLinearMap{B-I/\gamma_1}{\m{C}} 
\text{ and } 
(1/\gamma_n) \m{Y}_n  \to \invLinearMap{B-I/\gamma_1}{\m{C}} 
\text{i.e.},\nn
\end{align}
 both matrices $(1/\gamma_n) \m{X}_n$ and $(1/\gamma_n) \m{Y}_n$ approximate the matrix $\invLinearMap{B-I/\gamma_1}{\m{C}}$.
If, in addition, $\m{B}$ and $\m{C}$ commute then 
$(1/\gamma_n) \m{X}_n \to (\m{B}-\m{I}/\gamma_1)^{-1} \m{C}$
and $ (1/\gamma_n) \m{Y}_n \to (\m{B}-\m{I}/\gamma_1)^{-1} \m{C}$.
\end{corollary}
\begin{proof}
\newcommand{\Xtilde}[1]{\Mn{\tilde{x}}{#1}}
\newcommand{\Ytilde}[1]{\Mn{\tilde{y}}{#1}}
Both $X_n, Y_n \to \m{0}$ by direct application of Lemma \eqref{lemma:recursions}. Let $\Xtilde{n} = (1/\gamma_n) X_n$. 
First, divide \eqref{eq:corollary:main} by $\gamma_n$ to obtain 
\begin{align}
\label{eq:cor1}
\Xtilde{n} & = \linearMap{I - \gamma_n B_n}{\Xtilde{n-1}}
 \frac{\gamma_{n-1}}{\gamma_n} + 
	\gamma_n (\m{C} + \Mn{D}{n}).
\end{align}
By \assumeMain\assumeGn, $\gamma_{n-1}/\gamma_n
 = 1 + \gamma_n/\gamma_1 + \bigO{\gamma_n^2}$.
 Then, 
 \begin{align}
 \linearMap{I - \gamma_n B_n}{\Xtilde{n-1}}
 \frac{\gamma_{n-1}}{\gamma_n} = 
  \linearMap{I - \gamma_n B_n}{\Xtilde{n-1}}
 + \gamma_n \Xtilde{n-1} + \bigO{\gamma_n^2}.
 \end{align}
Therefore, we can rewrite Eq. \eqref{eq:cor1} as
\begin{align}
\label{eq:cor2}
\Xtilde{n} = 
\linearMap{I - \gamma_n \Gamma_n}{ \Xtilde{n-1} }+ 	\gamma_n (\m{C} + \Mn{D}{n}),
\end{align}

where $\Gamma_n  =  B_n - I / \gamma_1 + \bigO{\gamma_n}$.
In the limit $\Gamma_n \to B - I/\gamma_1 > 0$.
Furthermore,  
$||\Gamma_{i-1} - \Gamma_{i}|| = \bigO{\gamma_i^2}$ 
by assumptions of Corollary \ref{corollary:recursions}.
Thus, we can apply Lemma~\ref{lemma:recursions} to conclude that $\Xtilde{n}  =  (1/\gamma_n) X_n  \to \invLinearMapOne{B - I/\gamma_1}\{\m{C}\}$. 
The proof for $Y_n$ follows the same reasoning since $(\I + \gamma_n B_n)^{-1} (\gamma_{n-1}/\gamma_n) = (\I + \gamma_n \Gamma_n)^{-1}$,
where $\Gamma_n  =  B_n - \I/\gamma_1 + \bigO{\gamma_n}$.
\end{proof}

\newcommand{\Wnoise}[1]{W_n(#1, \thetastar)}
\begin{customthm}{2.2}
\TheoremVariance
\end{customthm}
\begin{proof}
We begin with the implicit SGD procedure. 
For notational convenience we make the 
following definitions: 
$V_n  =  \Var{\thetaim{n}}$, 
$S_n(\theta)  =  \nabla \log f(Y_n; X_n, \theta)$.
Denote $\Ex{S_n(\theta)} =  h(\theta)$.
Let $J_h$ denote the Jacobian of 
function $h$, then, under typical regularity conditions of \assumeMains\assumeStrong\
and by Theorem~2.1:
\begin{align}
\label{var:regularity}
& \ExCond{S_n(\thetastar)}{X_n}=0 \nn\\
& \Var{S_n(\thetastar)} = \Ex{\VarCond{S_n(\thetastar)}{X_n}}  =  \Fisher{\thetastar}\nn\\
& J_{h}(\theta) = -\Fisher{\theta}, 
\commentEq{under regularity conditions} \nn \\
& h(\thetaim{n}) = -\Fisher{\thetastar} (\thetaim{n}-\thetastar) + \bigO{\gamma_n}
 \commentEq{by Theorem~2.1},\nn\\
& ||\Var{S_n(\theta)-S_n(\thetastar)}|| \le \Ex{||S_n(\theta)-S_n(\thetastar)||^2}
\le L_0^2 \Ex{||\theta-\thetastar||^2}.
\end{align}
We can now rewrite the definition of implicit SGD as follows,
\begin{align}
\label{var:eq1}
\thetaim{n} = \thetaim{n-1} + \gamma_n C_n S_n(\thetaim{n})
 = \thetaim{n-1} + \gamma_n \lambda_n C_n S_n(\thetaim{n-1}),
\end{align}
where $\lambda_n$  is defined in Theorem~3.1 and $\lambda_n = 1-\bigO{\gamma_n}$ by Eq. \eqref{eq:useful:lam}.
Then, taking variances on both sides of Eq. \eqref{var:eq1} yields
\begin{align}
\label{var:eq2}
V_n   =  V_{n-1} & + 
\gamma_n^2 C_n \Var{S_n(\thetaim{n}} C_n^\intercal + \gamma_n \Cov{\thetaim{n-1}}{S_n(\thetaim{n}} C_n^\intercal + 
 \gamma_n C_n \Cov{S_n(\thetaim{n})}{\thetaim{n-1}}.
\end{align}
We can  simplify all variance/covariance terms in Eq.
 \eqref{var:eq2} as follows.
\begin{align}
  C_n \Var{S_n(\thetaim{n})} C_n^\intercal 
 & =C_n \Var{S_n(\thetastar) + [S_n(\thetaim{n}) - S_n(\thetastar)]} C_n^\intercal \nn\\
	& = C \Fisher{\thetastar} C^\intercal + \littleO{1},
	\commentEq{by Eqs. \eqref{var:regularity}, 
	Theorem~(2.1), and \assumeMain\assumeCn}\nn\\
 \Cov{\thetaim{n-1}}{S_n(\thetaim{n})} & = 
	\Cov{\thetaim{n-1}}{S_n(\thetaim{n-1})}
	+  \Cov{\thetaim{n-1}}{(\lambda_n-1) S_n(\thetaim{n-1})}\nn\\
    &  =  \Cov{\thetaim{n-1}}{h(\thetaim{n-1})} + \bigO{\gamma_n} \nn\\
& = V_{n-1} \Fisher{\vthetastar} + \bigO{\gamma_n}.
	\commentEq{by Eq. \eqref{var:regularity}, Theorem~(2.1), 
Eq. \eqref{eq:useful:lam}}.\nn
\end{align}
Similarly, $\Cov{h(\thetaim{n})}{\thetaim{n-1}} = V_{n-1} \Fisher{\thetastar}
 + \bigO{\gamma_n}$.
We can now rewrite Eq.~\eqref{var:eq2} as
\begin{align}
\label{var:eq3}
V_n & =\linearMap{I - \gamma_n B_n}{\m{V}_{n-1}} + \gamma_n^2 [C \Fisher{\vthetastar} C^\intercal + \littleO{1}],
\end{align}
where $B_n  =  2 C_n \Fisher{\thetastar}$ and 
$B_n \to 2 C \Fisher{\thetastar}$.
Corollary~\ref{corollary:recursions} on recursion 
\eqref{var:eq3} yields the following closed-form,
since $\m{B}$ and $\m{C}$ commute and $C$ is symmetric:
\begin{align}
(1/n) V_n \to \VarAsymp. \nonumber
\end{align}
The regularity conditions \eqref{var:regularity} 
and the convergence rates of Theorem~2.1 
that are crucial for this proof 
also hold for the explicit procedure. 
\end{proof}

\begin{customthm}{2.3}
\TheoremAveraging
\end{customthm}
\begin{proof}
By Theorem~2.1 and \assumeMains\ \assumeLip, \assumeStrong, we have 
\begin{align}
\label{aisgd:eq1}
\nabla \log f(Y_n; X_n, \thetaim{n}) = 
	\nabla \log f(Y_n; X_n, \thetastar) - \Fisher{\thetastar} (\thetaim{n}-\thetastar) + \bigO{\gamma_n}.
\end{align}
Define, for convenience $\varepsilon_n =  \nabla \log f(Y_n; X_n, \thetastar)$, $F =  \Fisher{\thetastar}$. Then, the first-order implicit SGD iteration becomes 
\begin{align}
\thetaim{n}-\thetastar = (I + \gamma_n F)^{-1} (\thetaim{n-1}-\thetastar + \gamma_n \varepsilon_n + \bigO{\gamma_n^2}).
\end{align}
We make the following definitions.
\begin{align}
\label{aisgd:defs}
e_i &  =  \gamma_i (I+\gamma_i F)^{-1} (\varepsilon_i + \bigO{\gamma_i^2}),\nn\\
B_i^j & =  \prod_{k=j}^i (I+\gamma_k F)^{-1},\nn \\
D_j^n &  =   \prod_{k=n-1}^i B_{j+1}^k = I + B_{j+1}^{j+1} + B_{j+1}^{j+2} + \ldots + B_{j+1}^{n-1}.	
\end{align}
Then, we can solve the recursion for $\thetaBar{n}-\thetastar$ to obtain
\begin{align}
\label{aisgd:main_eq}
\thetaBar{n}-\thetastar = (1/n) D_0^n (\thetaBar{0}-\thetastar)
+ (1/n) \sum_i^{n-1} D_i^n e_i.
\end{align}
Our proof is now split into proving the following two lemmas.
\newcommand{\D}[2]{D_{#1}^{#2}}
\newcommand{\mrm}[1]{#1}
\begin{lemma}
\label{lemma:aisgd1}
Under \assumeMain\assumeGn\ $\D{0}{n} = \mrm{o}(n)$.
\end{lemma}
\begin{proof}
Matrix $F$ is positive definite by \assumeMain\assumeStrong.
Thus, if $\lambda$ is some eigenvalue of $\mrm{F}$ then the corresponding eigenvalue of $\D{0}{n}$ is $1 + \frac{1}{1+\gamma_{1} \lambda} +  \frac{1}{1+\gamma_{1} \lambda} \frac{1}{1+\gamma_{2} \lambda} + \cdots \le \sum_{i=0}^n \exp(-K \lambda \sum_{k=1}^i \gamma_k)$, where the last inequality is obtained by Lemma \ref{lemma:decay_factor}.
Because $\sum \gamma_i \to \infty$, the summands are $\mrm{o}(1)$, and thus $\D{0}{n}$ is $\mrm{o}(n)$.
\end{proof}

\begin{lemma}
\label{lemma:aisgd2}
Suppose \assumeMain\assumeGn\ and Eq. \eqref{aisgd:eq1} hold.
Then, 
\begin{align}
\gamma_i \D{i}{n} (I+\gamma_i F)^{-1} = \Omega_i^n + F^{-1}, 
\end{align}
such that
$\sum_{i=0}^{n-1} \Omega_i^n = o(n)$.
\end{lemma}
\begin{proof}
Our goal will be to compare the eigenvalues of $\gamma_i \D{i}{n}$
and $F$. Any matrix $\D{i}{n}$ shares the same eigenvectors with
$\mrm{F}$ because $F$ is positive definite, 
and thus a relationship on eigenvalues will automatically establish
a relationship on the matrices.
For convenience, define $q_{i}^j  =  \prod_{k=i}^j (1+\gamma_k \lambda)^{-1}$ for $\lambda>0$; by convention, $q_{i-1}^i = 1$. Also let $s_i^j  =  \sum_{k=i}^j \gamma_k$ be the function of partial sums.
By Lemma \ref{lemma:decay_factor} 
$q_i^j = \bigO{\exp(-K \lambda s_i^j)}$, for some $K>0$. 
For an eigenvalue $\lambda > 0$ of $\mrm{F}$ the corresponding eigenvalue, 
say $\lambda'$, of matrix $\gamma_i \D{i}{n}  (I + \gamma_i \mrm{F})^{-1}$ is equal to
\begin{align}
\label{eq1}
\lambda' = \frac{\gamma_i}{1+\gamma_i \lambda} (q_{i+1}^i + q_{i+1}^{i+1} + \ldots + q_{i+1}^{n-1}).
\end{align}
Thus,
\begin{align}
\lambda' (1+\gamma_i \lambda) =  \sum_{k=i}^{n-1} \gamma_i q_{i+1}^k.
\end{align}

Our goal will be to derive the relationship between $\lambda$ and $\lambda'$. By definition
\begin{align}
& \gamma_{i+1} \lambda q_{i+1}^{i+1} + q_{i+1}^{i+1} = 1 \nonumber \\
& \gamma_{i+2}  \lambda q_{i+1}^{i+2} + q_{i+1}^{i+2} = q_{i+1}^{i+1} \nonumber \\
& \ldots \ldots \nonumber \\
& \gamma_{n-2}  \lambda q_{i+1}^{n-2} + q_{i+1}^{n-2} =q_{i+1}^{n-3} \nonumber \\
& \gamma_{n-1} \lambda  q_{i+1}^{n-1} + q_{i+1}^{n-1} =q_{i+1}^{n-2} .
\end{align}
By summing over the terms we obtain:
\begin{align}
\label{eq2}
\lambda \sum_{k=i+1}^{n-1} \gamma_k  q_{i+1}^k + q_{i+1}^{n-1} = 1.
\end{align}
If we combine with \eqref{eq1} we obtain
\begin{align}
\label{eq3}
& \lambda \sum_{k=i}^{n-1} \gamma_i q_{i+1}^k +
\lambda \sum_{k=i}^{n-1} (\gamma_k - \gamma_i)  q_{i+1}^k
+ q_{i+1}^{n-1} = 1 + \gamma_i \lambda  ~\hbox{ or } \\
& (1+\gamma_i \lambda) \lambda \lambda' + \lambda \sum_{k=i}^{n-1} (\gamma_k - \gamma_i) q_{i+1}^k
+ q_{i+1}^{n-1} = 1 + \gamma_i \lambda.
\end{align}
We now focus on the second term. By telescoping the series we obtain
\begin{align}
\label{eq4}
 \lambda \sum_{k=i}^{n-1}  (\gamma_k - \gamma_i) q_{i+1}^k & =
\lambda \sum_{k=i}^{n-1} \left [ \sum_{j=i}^{k} (\gamma_{j+1}-\gamma_j) \right ]  q_{i+1}^k =  \lambda \sum_{k=i}^{n-1} \left [ \sum_{j=i}^{k} \gamma_j o(\gamma_j) \right ]  q_{i+1}^k \nonumber \\
& \le \lambda \mrm{o}(\gamma_i) \sum_{k=i}^{n-1} s_i^k   q_{i+1}^k \triangleq q_i^n.
\end{align}
In Eq. \eqref{eq4} we used $(\gamma_{j+1}-\gamma_j)/\gamma_j =\bigO{n^{-1-\gamma}}/n^{-\gamma} = \bigO{n^{-1}} = o(\gamma_j)$,
by \assumeMain\assumeGn. 
Our goal is now to show $\sum_{i=0}^{n-1} q_i^n = \littleO{n}$. 
Since $q_{i+1}^k = \bigO{\exp(-K\lambda s_{i+1}^k)}$ by \citep[p845, see A6 and A7]{polyak1992} we obtain that
$q_i^n \to 0$ for fixed $i$ as $n \to  \infty$. 
Therefore we can rewrite Eq.~\eqref{eq3} as
\begin{align}
\lambda' \lambda + q_i^n + \bigO{q_{i+1}^n} = 1,
\end{align}
where $\sum_{i=0}^n q_{i+1}^n = \littleO{n}$ and $\sum_{i=0}^{n-1} q_i^n = \littleO{n}$.
\end{proof}

Our proof is now complete. By Eq. \eqref{aisgd:main_eq} and 
Lemmas \ref{lemma:aisgd1} and \ref{lemma:aisgd2} we have 
\begin{align}
\thetaBar{n}-\thetastar = F^{-1} \sum_{i=1}^n \varepsilon_i +(1/n) \littleO{n}.\nn
\end{align}
Because $\Var{\varepsilon_i} = \Fisher{\thetastar}$, we finally obtain
\begin{align}
n \Var{\thetaBar{n}-\thetastar} = \Fisher{\thetastar}^{-1}.\nn
\end{align}

\end{proof}

\begin{customthm}{2.4}
\TheoremNormality
\end{customthm}
\begin{proof}
Let $S_n(\theta) =  \nabla \log f(Y_n; X_n, \theta)$ 
as in the proof of Theorem (2.2).
The conditions for Fabian's theorem---see \citet[Theorem 1]{fabian1968}---hold also for the implicit procedure. The goal is to show that 
\begin{align}
\label{thm5:eq1}
\thetaim{n}-\thetastar = (I - \gamma_n A_n) (\thetaim{n-1} - \thetastar) 
+ \gamma_n \xi_n(\thetastar) + \bigO{\gamma_n^2},
\end{align}
where $A_n \to A\succeq 0$, and $\xi_n(\theta) = S_n(\theta) - h(\theta)$,
and $h(\theta) = \Ex{S_n(\theta)}$; note, $\xi_n(\thetastar) = S_n(\thetastar)$.
Indeed, by a Taylor expansion on $S_n(\thetaim{n})$ 
and considering that $\thetaim{n} = \thetaim{n-1} + \gamma_n S_n(\thetaim{n})$, by 
definition, we have
\newcommand{\obsFisher}{\hat{\mathcal{I}_n}(\thetastar)}
\begin{align}
\label{normal:eq2}
(I + \gamma_n \obsFisher) (\thetaim{n}-\thetastar) = \thetaim{n-1}-\thetastar
 + \gamma_n S_n(\thetastar),
\end{align}
where $\obsFisher = -\nabla^2 S_n(\thetastar)$; 
note, $\Ex{\obsFisher} = \Fisher{\thetastar}$.
Because $(I + \gamma_n \hat{\mathcal{I}_n}(\thetastar))^{-1} 
 = I - \gamma_n \hat{\mathcal{I}_n}(\thetastar) + \bigO{\gamma_n^2}$, 
we can rewrite Eq. \eqref{normal:eq2} as
\begin{align}
\label{normal:eq3}
\thetaim{n}-\thetastar = (I - \gamma_n \obsFisher) (\thetaim{n-1} - \thetastar) 
+ \gamma_n S_n(\thetastar) + \bigO{\gamma_n^2}.
\end{align}
We can now apply Fabian's Theorem to derive asymptotic normality of 
$\thetaim{n}$. 
The variance matrix of the asymptotic normal distribution is derived in 
Theorem 2.4 under weaker conditions.
\end{proof}

\newtheorem{innercustomlemma}{Lemma}
\newenvironment{customlemma}[1]
  {\renewcommand\theinnercustomlemma{#1}\innercustomlemma}
  {\endinnercustomlemma}

\subsection{Stability}
Here, we prove Lemma~(2.1) in the main paper.
\begin{customlemma}{2.1}
\LemmaStability
\end{customlemma}
\begin{proof}
We will use the following intermediate result:
\begin{align}
	\max_{n > 0} | \prod_{i=1}^n (1-b/i)| \approx
				\begin{cases} 1-b  &\mbox{if } 0<b<1 \\ 
						         \frac{2^b}{\sqrt{2 \pi b}} & \mbox{if } b> 1 
				\end{cases} \nn
\end{align}
The first case is obvious. For the second case, $b>1$, assume without loss of generality that 
$b$ is an even integer. Then the maximum is given by 
\begin{align}
(b-1) (b/2-1) (b/3-1) \cdots (2-1) = \frac{1}{2}{b \choose b/2} = \Theta(2^b / \sqrt{2\pi b}),
\end{align}
where the last approximation follows from Stirling's formula. The stability result on the explicit SGD 
updates of Lemma~2.1 follows immediately by using the largest eigenvalue $\lmax$ of $\Fisher{\thetastar}$.
 For the implicit SGD updates, 
 we note that the eigenvalues of $(\I + \gamma_n \Fisher{\thetastar})^{-1}$ are less than one, for any $\gamma_n > 0$ and any Fisher matrix.
\end{proof}

\section{Appendix: Applications}
\label{appendix:applications}
\begin{customthm}{3.1}
\LinearTheorem
\end{customthm}

\begin{proof}
From the chain rule $\nabla \loglik{n}{\theta} = \linearScore{n}{\theta}$, 
and thus $\nabla \loglik{n}{\thetaim{n}} = \linearScore{n}{\thetaim{n}}$
and $\nabla \loglik{n}{\thetaim{n-1}} = \linearScore{n}{\thetaim{n-1}}$, and thus
the two gradients are colinear. 
Therefore there exists a scalar $\lambda_n$ such that
\begin{align}
\label{implicit_algo:eq1}
\nabla \loglik{n}{\thetaim{n}} & = \lambda_n \nabla \loglik{n}{\thetaim{n-1}} ~\hbox{ or } \nn\\
 \linearScore{n}{\thetaim{n}} & = 
\lambda_n \linearScore{n}{\thetaim{n-1}}.
\end{align}
We also have, 
\begin{align}
\label{implicit_algo:eq2}
\thetaim{n} & = \thetaim{n-1} + \gamma_n C_n \loglik{n}{\thetaim{n}}
\commentEq{by definition of implicit SGD in Eq.~(4)}\nn\\
 & = \thetaim{n-1} + \gamma_n \lambda_n C_n \loglik{n}{\thetaim{n-1}}.
 \commentEq{by Eq. \eqref{implicit_algo:eq1}}
\end{align}
Substituting the expression for $\thetaim{n}$ in Eq.\eqref{implicit_algo:eq2} into Eq. \eqref{implicit_algo:eq1}
we obtain the desired result of the Theorem in Eq. \eqref{algo:implicit:scale}.

We now prove the last claim of the theorem regarding 
the search bounds for $\lambda_n$.
For notational convenience, define $a  =  X_n^\intercal \thetaim{n-1}$, 
$g(x)  =  \ell'(x; Y_n)$, and $c = X_n^\intercal C_n X_n$, 
where $c > 0$ because $C_n$ are positive definite.
Also let $x_\star =  \gamma_n \lambda_n g(a)$, then the fixed-point equation \eqref{algo:implicit:fp} can be written as
\begin{align}
\label{algo:eq1}
x_\star  = \gamma_n  g(a + x_\star c).
\end{align}
where $g$ is decreasing by Assumption \assumeLinear.
If $g(a)=0$ then $x_\star=0$.
If $g(a) > 0$ then $x_\star>0$ and $\gamma_n  g(a + x c) < \gamma_n g(a)$ for all $x > 0$, since $ g(a + x c)$ is decreasing;
taking $x=x_\star$ yields $\gamma_n g(a) > \gamma_n g(a + x_\star c) = x_\star$, by the fixed-point equation \eqref{algo:eq1}. Thus, $0 < x_\star < \gamma_n g(a)$.
Similarly, if $g(a) < 0$ then $x_\star< 0$ and $\gamma_n  g(a + x c) > \gamma_n g(a)$ 
for all $x < 0$, since $ g(a + x c)$ is decreasing; taking $x=x_\star$ yields $\gamma_n g(a) < \gamma_n g(a + x_\star c) = x_\star$, by the fixed-point equation. 
Thus, $\gamma_n g(a)  < x_\star < 0$.
In both cases $0 < \lambda_n < 1$.
A visual proof is given Figure \ref{figure:proof}.
\begin{figure}[h!]
\centering
\includegraphics[width=\textwidth]{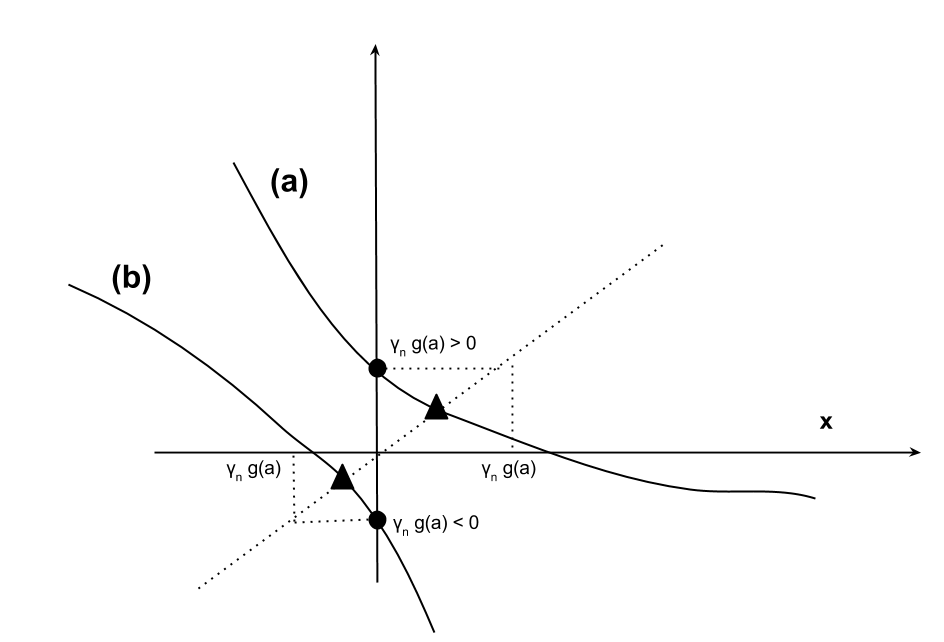}
\caption{(Search bounds for solution of Eq. \eqref{algo:eq1}) {\bf Case $g(a) > 0$:} Corresponds to Curve (a) defined as $\gamma_n g(a+x c), c>0$.
The solution $x_\star$ of fixed point equation \eqref{algo:eq1} 
(corresponding to right triangle) is between 0 and $\gamma_n g(a)$ since Curve (a) is decreasing.
{\bf Case $g(a) <0$: } Corresponds to Curve (b) also defined as $\gamma_n g(a+x c)$. The solution $x_\star$ of fixed point equation \eqref{algo:eq1} (left triangle) is between $\gamma_n g(a)$ and 0 since Curve (b) is also decreasing.
}
\label{figure:proof}
\end{figure}

\end{proof}

\section{Appendix: Additional experiments}
\label{appendix:additional_experiments}

\subsection{Normality experiments with implicit SGD}
\label{section:experiment:normality}
\begin{figure}[t]
 \centering
 \includegraphics[scale=0.55]{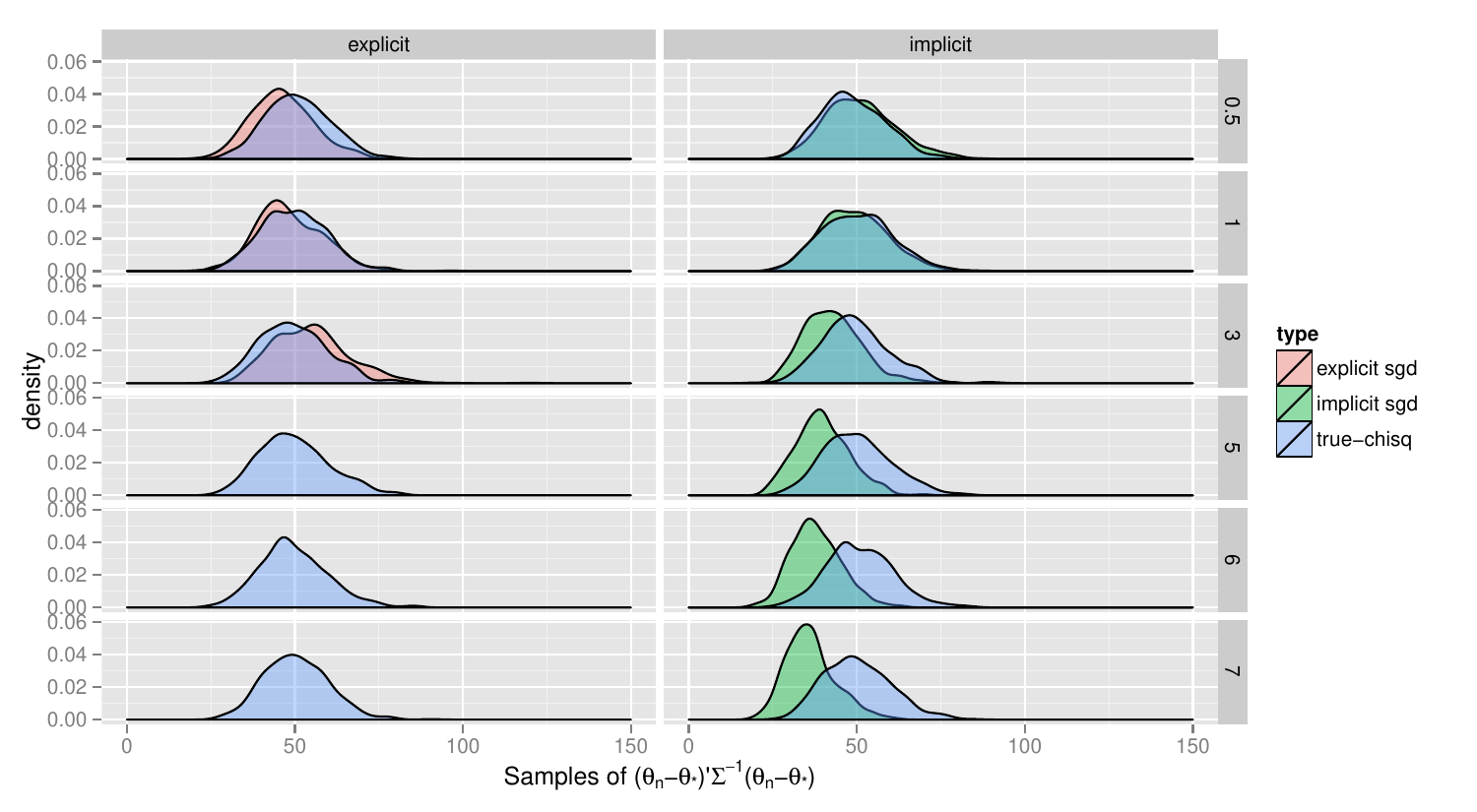}
 \caption{Simulation with normal model for $p=50$ parameters.
 Implicit SGD is stable and follows the nominal chi-squared distribution well, 
regardless of the particular learning rate.
Explicit SGD becomes unstable at larger $\gamma_1$ and its distribution does 
not follow the theoretical distribution chi-squared distribution well. In particular, 
the distribution of $N (\thetasgd{N}-\thetastar)^\intercal \Sigma^{-1} (\thetasgd{N}-\thetastar)$  quickly becomes unstable 
for larger values of the learning rate parameter, 
and eventually diverges when $\gamma_1>3$.}
\label{figure:normality_normal50}
 \end{figure}

\begin{figure}[t]
 \centering
 \includegraphics[scale=0.55]{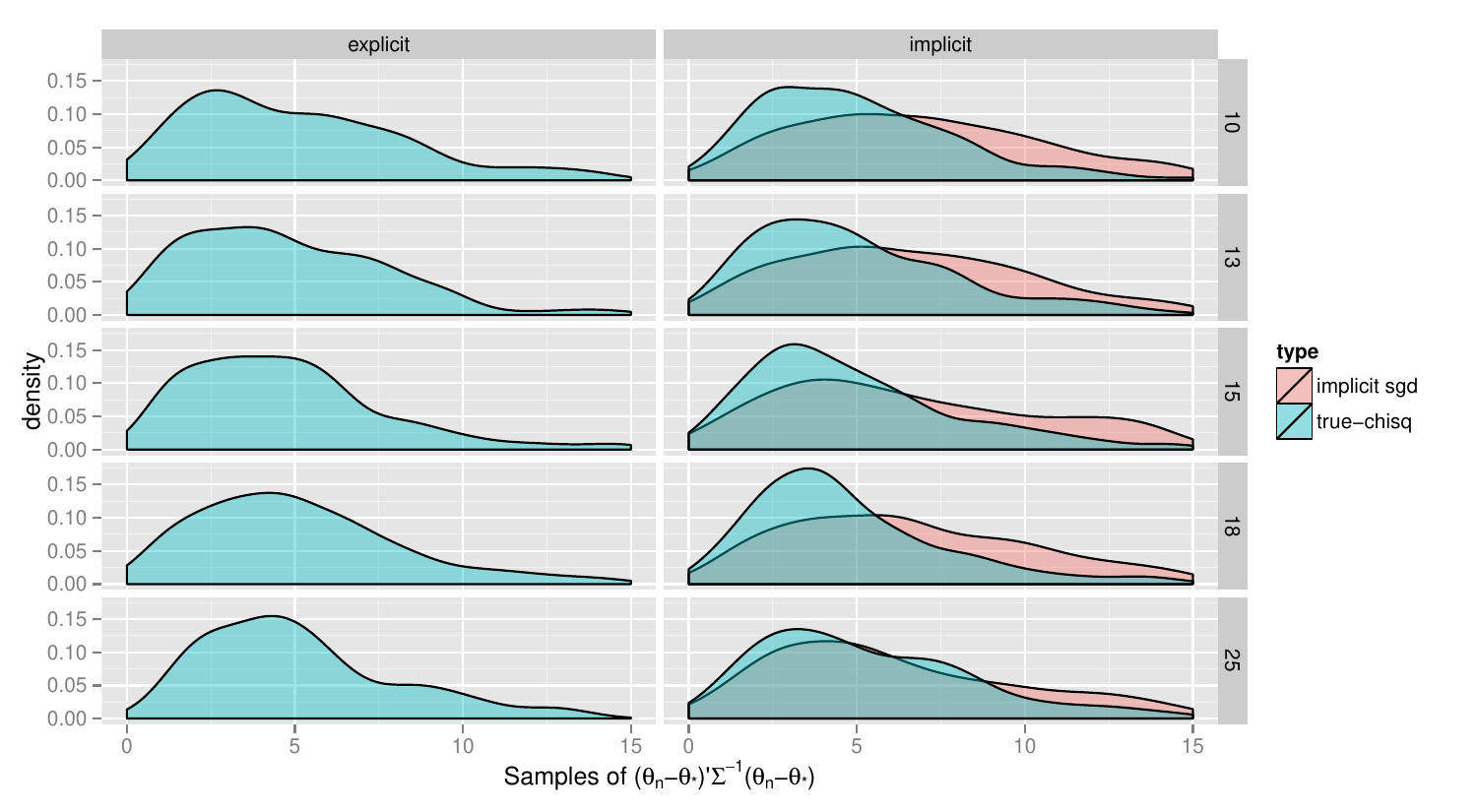}
 \caption{Simulation with logistic regression model for $p=5$.
 Learning rates are larger than in the linear normal model 
 to ensure the asymptotic covariance matrix of Theorem (2.2)
is positive definite.
 Implicit SGD is stable and follows the nominal chi-squared distribution regardless of the learning rate.
Explicit SGD is unstable at virtually all replications of this experiment.
\label{figure:normality_logistic5}}
 \end{figure}

In Figure \ref{figure:normality_normal50} we plot 
the experimental results of Section~4.1.2
for $p=50$ (parameter dimension). 
We see that explicit SGD becomes even more unstable in more dimensions 
as expected. In contrast, implicit SGD remains stable 
and validates the theoretical normal distribution for small learning rates. 
In larger learning rates we observe a divergence from the asymptotic 
chi-squared distribution (e.g., $\gamma_1 = 6$) because when the 
learning rate parameter is large there is more noise in the 
stochastic approximations, and thus more iterations are required 
for convergence. In this experiment we fixed the number of iterations 
for each value of the learning rate, but subsequent experiments verified 
that implicit SGD reaches the theoretical chi-squared distribution if 
the number of iterations is increased.
Finally, in Figure \ref{figure:normality_logistic5} we make a similar plot 
for a logistic regression model. 
In this case the learning rates need to be larger 
because with the same distribution of covariates for $X_n$, 
the Fisher information is smaller than in the linear normal model.
In summary, in almost all experiments explicit SGD was unstable and 
could not converge whereas implicit SGD was stable and followed the theoretical 
chi-squared distribution.

\subsection{Poisson regression}
\label{section:experiment:poisson}
\newcommand{\etheta}[1]{e^{\theta_{#1}}}
\newcommand{\nsamples}{100}
\newcommand{\niters}{20000}
\newcommand{\qprob}{0.2}
Here,  we illustrate our method on a bivariate Poisson model which is simple enough
to derive the variance formula analytically. This example was first presented by~\citet{toulis2014statistical}.
We assume binary features such that,
for any iteration $n$, $X_n$ is either $(0, 0)^\intercal$,
$(1, 0)^\intercal$ or $(0, 1)^\intercal$ with probabilities 0.6, \qprob\ and
\qprob\ respectively.
We set $\thetastar = (\theta_1, \theta_2)^\intercal$ for some $\theta_1, \theta_2$,
 and assume 
$Y_n \sim \mathrm{Poisson}(\exp(X_n^\intercal \thetastar))$, 
where the transfer function $h$ is the exponential, i.e., $h(x) = \exp(x)$.
It follows,
\begin{align}
\Fisher{\vthetastar} = \Ex{h'(X_n^\intercal \thetastar) X_n X_n^\intercal} = \qprob
\left( \begin{array}{cc}
\etheta{1} & 0 \\
0 & \etheta{2} \end{array} \right).  \nn
\end{align}
We set $\gamma_n = 10 /3n$ and $C_n=I$.
Setting $\theta_1 = \log 2$ and $\theta_2 = \log 4$, the asymptotic variance 
$\m{\Sigma}$ in Theorem (2.2) is equal to 
\begin{align}
\label{eq:poisson:variance}
 \m{\Sigma} = \frac{2}{3} \left( \begin{array}{cc}
\frac{\etheta{1}}{(4/3) \etheta{1}-1} & 0 \\
0 & \frac{\etheta{2}}{(4/3) \etheta{2}-1} \end{array} \right) =
 \left( \begin{array}{cc}
0.8 & 0 \\
0 & 0.62 \end{array} \right).
\end{align}
Next, we obtain \nsamples\ independent samples of $\thetasgd{n}$ and $\thetaimpl{n}$ 
for $n=\niters$ iterations of procedures in Eq.~(4) and in Eq.~(4),
and compute their empirical variances. 
We observe that the implicit estimates are particularly stable and have an empirical variance satisfying
\begin{align}
(1/\gamma_{n}) \widehat{\mathrm{Var}}(\thetaim{n}) =
 \left( \begin{array}{cc}
0.86 & -0.06 \\
-0.06 & 0.64 \end{array} \right), \nn
\end{align}
and that is close to the theoretical value in Eq.~\eqref{eq:poisson:variance}.
In contrast, the standard SGD estimates are unstable and their $L_2$ distance to 
the true values $\vthetastar$ are orders of magnitude larger than 
the implicit ones (see Table \ref{table:poisson} for sample quantiles).
By Lemma 2.1 in the main paper, such deviations are expected for 
standard SGD because the largest eigenvalue of $\Fisher{\thetastar}$ is $\lambda_{(2)} = 0.8$
satisfying $\gamma_1 \lambda_{(2)} = 8/3>1$. 
Note that it is fairly straightforward to stabilize the standard SGD procedure in this problem,
for example by modifying the learning rate sequence to $\gamma_n = \min\{0.15, 10/3n\}$.
In general, when the optimization problem is well-understood, it is easy to determine the learning rate schedule that avoids out-of-bound explicit updates.
In practice, however, we are working with problems that are not so well-understood and determining the correct learning rate parameters may take substantial effort. The implicit method eliminates this overhead.

\renewcommand*{\arraystretch}{1.2}
\begin{table}[t]
\caption{Quantiles 
of $||\thetasgd{n}-\vthetastar||$ and 
$||\thetaimpl{n}-\vthetastar||$. Values larger than \texttt{1e3} are marked ``*".
}
\label{table:poisson}
\begin{center}
\begin{small}
\begin{sc}
\begin{tabular}{l | cccccc}
	\multicolumn{7}{c}{quantiles}  \\
method   & 25\% & 50\%     & 75\% & 85\% & 95\%  & 100\%  \\
SGD      & 0.01 & 1.3    & 435.8 & * & * & * \\
Implicit & 0.00 & 0.01     & 0.02 & 0.02 & 0.03  & 0.04
\end{tabular}
\end{sc}
\end{small}
\end{center}
\end{table}

\subsection{Experiments with \texttt{glmnet}}
\label{appendix:additional_experiments:glmnet}
In this section, we transform the outcomes in the original experiment $Y$ 
through the logistic transformation and then fit  a logistic regression model. The results are shown in Table \ref{table:experiment:glmnet:logistic},
which replicates and expands on Table 2 of \citet{friedman2010regularization}.
The implicit SGD method maintains a stable running time over different correlations
and scales sub-linearly in the model size $p$. In contrast, \texttt{glmnet} is affected
by the model size $p$ and covariate correlation, and remains 2x-10x slower across
experiments. We note that the implicit SGD method is slower in the 
logistic regression example compared to the normal case (Table 3 in main paper). This is because 
the implicit equation of Algorithm 1 (in the main paper) needs to be solved numerically, whereas a closed-form
solution is available in the normal case.

\renewcommand*{\arraystretch}{0.7}
\begin{table}[t]
\caption{Experiments comparing implicit SGD with \texttt{glmnet}. 
Covariates $\m{X}$ are sampled as normal, with cross-correlation $\rho$, 
and the outcomes are sampled as $\b{y} \sim \mathrm{Binom}(\b{p})$, 
$\mathrm{logit}(\b{p}) = \mathcal{N}(\m{X} \vthetastar, \sigma^2 \m{I})$.
Running times (in secs) are reported for different values of $\rho$ averaged over 10 repetitions.}
\label{table:experiment:glmnet:logistic}
\begin{center}
\begin{small}
\begin{sc}
 \begin{tabular}{l ccccc }
 method & metric &  \multicolumn{ 4 }{c}{correlation ($\rho$)} \\
 & &  0 & 0.2 & 0.6 & 0.9 \\
 & &  \cline{1-4} \\
 & &  \multicolumn{4}{c}{$N=1000, p=10$}  \\
 & &  \cline{1-4} \\
 \multirow{2}{*}{\texttt{glmnet}}  & time(secs) & 0.02 & 0.02 & 0.026 & 0.051 \\
                 & mse & 0.256 & 0.257 & 0.292 & 0.358 \\
  &  &  &  &  &  \\
 \multirow{2}{*}{\texttt{sgd}}  & time(secs) & 0.058 & 0.058 & 0.059 & 0.062 \\
                 & mse &  0.214 & 0.215 & 0.237 & 0.27 \\
 & &  \cline{1-4} \\
 & &  \multicolumn{4}{c}{$N=5000, p=50$}  \\
 & &  \cline{1-4} \\
 \multirow{2}{*}{\texttt{glmnet}}   & & 0.182 & 0.193 & 0.279 & 0.579 \\
                 & &  0.131 & 0.139 & 0.152 & 0.196 \\
  & &   &  &  &  \\
 \multirow{2}{*}{\texttt{sgd}}   & & 0.289 & 0.289 & 0.296 & 0.31 \\
                 & &  0.109 & 0.108 & 0.116 & 0.14 \\
 & &  \cline{1-4} \\
 & &  \multicolumn{4}{c}{$N=100000, p=200$}  \\
 & &  \cline{1-4} \\
 \multirow{2}{*}{\texttt{glmnet}}  &  & 8.129 & 8.524 & 9.921 & 22.042 \\
                 & &  0.06 & 0.061 & 0.07 & 0.099 \\
  &  &  &  &  &  \\
 \multirow{2}{*}{\texttt{sgd}}  & &  5.455 & 5.458 & 5.437 & 5.481 \\
                 & &  0.045 & 0.046 & 0.048 & 0.058 \\
 \end{tabular}
\end{sc}
\end{small}
\end{center}
\end{table}

\subsection{Experiments with machine learning algorithms}
\label{appendix:experiments}
In this section we perform additional experiments with related methods 
from the machine learning literature. 
We focus on averaged implicit SGD defined in Eq.~(14) of the main paper, which was shown 
to be optimal under suitable conditions, because most machine learning 
methods are also designed to achieve optimality in the context of maximum-likelihood (or maximum a-posteriori) computation 
with a finite data set.
In summary, our experiments include the following procedures:

\begin{itemize}
\item Explicit SGD procedure in Eq.~(1) of the main paper.
\item Implicit SGD procedure in Eq.~(4) of the main paper.
\item Averaged explicit SGD: Averaged stochastic gradient descent with explicit updates of
the iterates \citep{xu2011towards, shamir2012stochastic, bach2013non}. This is equivalent to 
the procedure in Eq.(14) of the main paper, where the implicit update is replaced by an explicit one,
$\thetasgd{n} = \thetasgd{n-1} + \gamma_n \nabla \log f(Y_n; X_n, \thetasgd{n-1})$.
\item \proxsvrg: A proximal version of the stochastic gradient descent with progressive variance reduction (SVRG) method \citep{xiao2014proximal}.
\item \proxsag:  A proximal version of the stochastic average gradient (SAG)
method \citep{schmidt2013minimizing}.  While its theory has not been formally
established, \proxsag\ has shown similar convergence properties to \proxsvrg.\footnote{We note that the linear convergence rates for \proxsvrg\ and \proxsag\ refer to convergence to
the empirical minimizer (e.g., MLE), and not to ground truth $\theta_\star$.}
\item Adagrad~\citep{duchi2011} as defined in Eq.~(12).
 We note that \adagrad\ and similar adaptive methods effectively approximate the natural gradient by using a larger-dimensional learning rate. It has the
added advantage of being less sensitive than first-order methods to tuning of hyperparameters.
\end{itemize}

We test the performance of \aisgd\ on standard benchmarks of large-scale linear classification with real data sets against the aforementioned methods. Some of these test comparisons were recently published by~\citet{toulis2016stability}.
Our datasets are summarized in Table~\ref{table:datasets}.
The COVTYPE data~\citep{blackard1998comparison}
consists of forest cover types in which the task is to classify class 2 among 7
forest cover types.
{DELTA} is synthetic data offered in the PASCAL Large Scale Challenge
\citep{sonnenburg2008pascal} and we apply the default processing offered by the
challenge organizers.
The task in {RCV1} is to classify documents belonging to class {CCAT} in the
text dataset \citep{lewis2004rcv1}, where we apply the standard preprocessing
provided by \citet{bottou2012stochastic}.  In the MNIST data set
\citep{lecun1998gradient} of images of handwritten digits, the task is to
classify digit 9 against all others.

For \aisgd\ and \asgd, we use the learning
rate $\gamma_n = \eta_0(1+\lambda\eta_0 n)^{-3/4}$ prescribed in
\citet{xu2011towards}, where the constant $\eta_0$ is determined using a small
subset of the data.  Hyperparameters for other methods are set based on a
computationally intensive grid search over the entire hyperparameter space: for
\proxsvrg, this includes the step size $\eta$ in the proximal update and the
inner iteration count $m$, and for \proxsag, the same step size $\eta$.

\begin{center}
\begin{table}[tb]
\caption{Summary of data sets and the $L_2$ regularization parameter
$\lambda$ used}
  \centering
\begin{tabular}{|l|l|l|l|l|l|l|l|l|l|l|}
\hline
       & description          & type   & features& training set & test set & $\lambda$\\
\hline
covtype& forest cover type    & sparse & 54      & 464,809 & 116,203 & $10^{-6}$\\
delta  & synthetic data       & dense  & 500     & 450,000 & 50,000     & $10^{-2}$\\
rcv1   & text data            & sparse & 47,152  & 781,265 & 23,149  & $10^{-5}$\\
mnist  & digit image features & dense  & 784     & 60,000  & 10,000  & $10^{-3}$\\
\hline
\end{tabular}
\label{table:datasets}
\end{table}
\end{center}

\begin{figure}[ht!]
\vskip -0.1in
\begin{center}
\centerline{\includegraphics[scale=0.7]{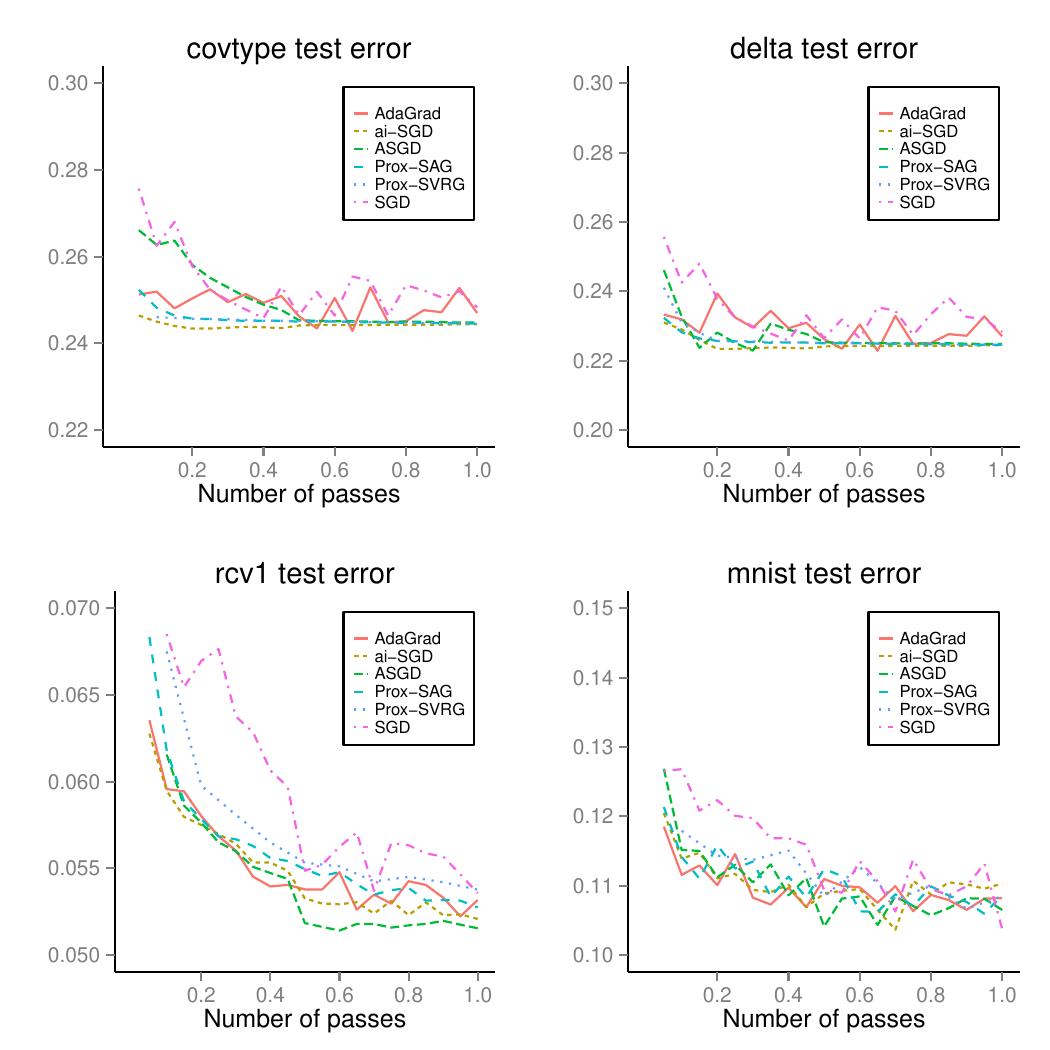}}
\caption{Large scale linear classification with log loss on four data sets. Each
plot indicates the test error of various stochastic gradient methods over a single
pass of the data.}
\label{figure:classification}
\end{center}
\vskip -0.1in
\end{figure}

The results are shown in Figure \ref{figure:classification}.
We see that \aisgd\ achieves comparable performance with the tuned proximal
methods \proxsvrg\ and \proxsag, as well as \adagrad.
All methods have a comparable convergence
rate and take roughly a single pass in order to converge.
\adagrad\ exhibits a larger variance in its estimate than the proximal methods, 
which can be explained from our theoretical results in Section 2.2.1.
We also note that as \aisgd\ achieves comparable results to the
other proximal methods, it also requires no tuning while \proxsvrg\ and \proxsag\ do 
require careful tuning of their hyparameters. This was confirmed 
from separate sensitivity analyses (not reported in this paper), which indicated that aisgd\ is robust to
fine-tuning of hyperparameters in the learning rate, whereas small perturbations
of hyperparameters in \asgd\ (the learning rate), \proxsvrg\ (proximal step size
$\eta$ and iteration $m$), and \proxsag\ (proximal step size $\eta$), can
lead to arbitrarily bad error rates.

\subsubsection{Averaged explicit SGD}
\begin{figure}[t]
\vskip -0.1in
\begin{center}
\centerline{\includegraphics[width=.7\textwidth]{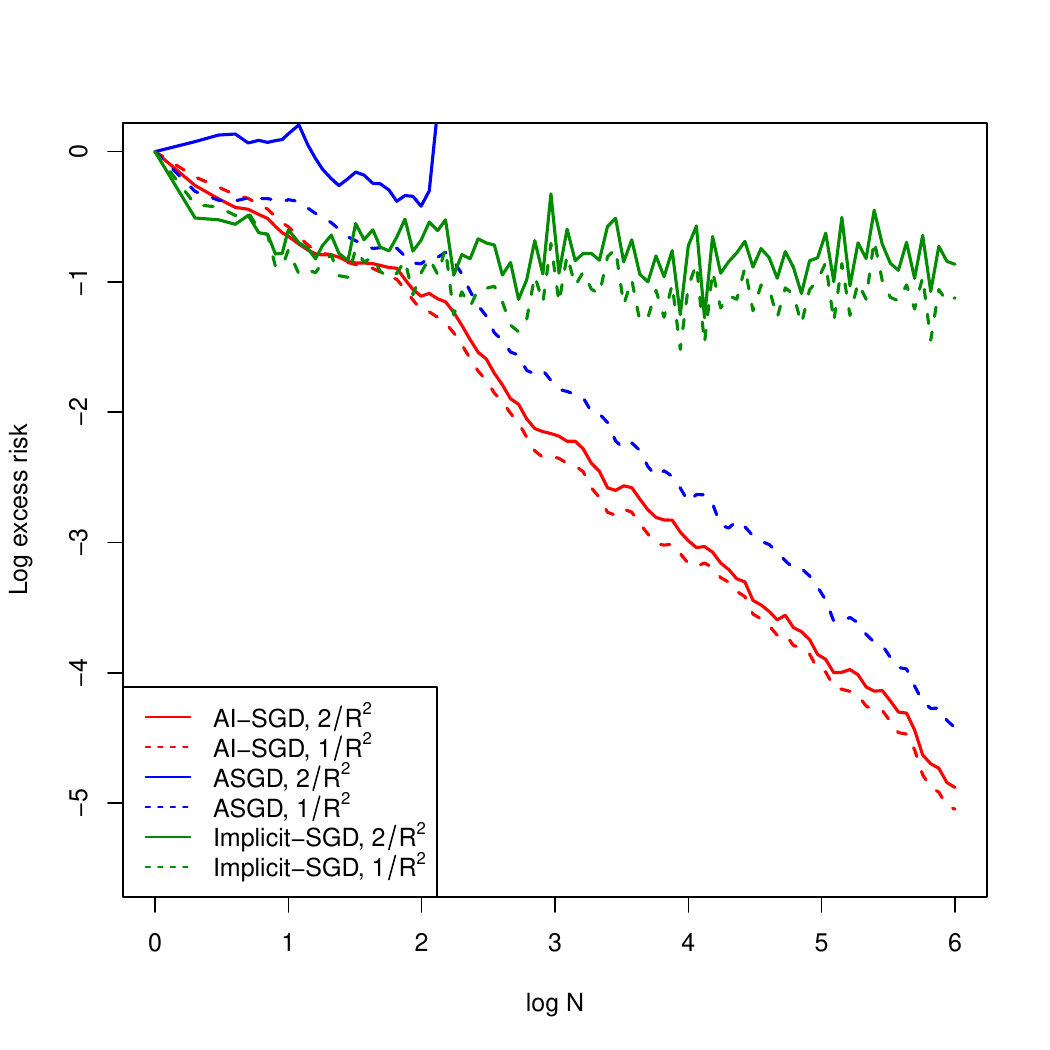}}
\caption{Loss of averaged implicit SGD, averaged explicit SGD, and plain implicit SGD in Eq.~(4) ($C_n=I$), on simulated multivariate normal data with $N=1\e6$ observations $p=20$ features. The plot shows that averaged implicit SGD is stable regardless of the specification of the learning rate $\gamma$ and without sacrificing performance. In contrast, explicit averaged SGD is 
very sensitive to misspecification of the learning rate.}
\label{figure:normal1}
\end{center}
\vskip -0.1in
\end{figure}

In this experiment we validate the theory of statistical efficiency and stability 
of averaged implicit SGD. To do so, we follow a simple normal linear regression example from \citet{bach2013non}.
We set $N=1\e6$ as the number of observations, and $p=20$ be the number of covariates. 
We also set  $\thetastar=(0,0,\ldots,0)^\intercal \in\mathbb{R}^{20}$ as
the true parameter value. 
The random variables  $X_n$ are distributed i.i.d. as 
$X_n \sim \mathcal{N}_p(0, H)$, where $H$ is a randomly generated symmetric matrix with eigenvalues $1/k$, for $k=1,\ldots, p$.
The outcome $Y_n$ is sampled from a normal distribution as $Y_n\:\vert\: X_n\sim \mathcal{N}(X_n^\intercal\theta_*, 1)$, for $n=1,\ldots,N$.
We choose a constant learning rate $\gamma_n \equiv \gamma$ according to the average radius of the data $R^2=\mathrm{trace}(H)$,
and for both averaged explicit and implicit SGD we collect iterates $\theta_n$ for $n=1,\ldots,N$,
and keep the average $\bar{\theta}_n$.
In Figure \ref{figure:normal1}, we plot $(\theta_n-\thetastar)^\intercal H (\theta_n-\thetastar)$ for each iteration for a maximum of $N$ iterations, i.e., a full pass over the data, in log-log space.

Figure \ref{figure:normal1} shows that \aisgd\ performs on par with \asgd\ for
the rates at which \asgd\ is known to be optimal. Thus, averaged implicit SGD is also optimal.
However, the benefit of the
implicit procedure in \aisgd\ becomes clear as the learning rate deviates;
notably, \aisgd\ remains stable for learning rates that are above the
theoretical threshold, i.e., $\gamma > 1/R^2$, whereas \asgd\ diverges in the case of $\gamma=2/R^2$. This stable behavior is also exhibited in \implicit, but
it converges at a slower rate than \aisgd, and thus cannot effectively combine stability with statistical efficiency. We note that
stability of \aisgd\ is also observed in the same experiments using decaying
learning rates.

\end{document}